%% file: main.tex
\documentclass[10pt,journal,compsoc,nofonttune]{IEEEtran}

\usepackage{amsmath,amsfonts}
\usepackage{array}
\usepackage[font=footnotesize,labelfont=sf,textfont=sf]{subfig}
\usepackage[font=small]{caption}
\usepackage{textcomp}
\usepackage{stfloats}
\usepackage{url}
\usepackage{verbatim}
\usepackage{graphicx}

\usepackage{url}
\usepackage[utf8]{inputenc}
\usepackage{xcolor}
\usepackage{xspace}
\usepackage{epsfig}
\usepackage{balance}
\usepackage{booktabs}
\usepackage{tabularx}
\usepackage{cite}
\usepackage{mathtools}
\usepackage{soul}
\usepackage{lipsum}
\usepackage{bm}
\usepackage{enumerate}
\usepackage{enumitem}
\usepackage[sort&compress,numbers]{natbib}

\usepackage[acronyms,nonumberlist,nopostdot,nomain,nogroupskip,acronymlists={hidden}]{glossaries}
\newglossary[algh]{hidden}{acrh}{acnh}{Hidden Acronyms}
\glsdisablehyper

\usepackage{tikz}
\usepackage{pgfplots}
\pgfplotsset{compat=newest}
\pgfplotsset{plot coordinates/math parser=false}
\newlength\fheight
\newlength\fwidth
\usetikzlibrary{plotmarks,patterns,decorations.pathreplacing,backgrounds,calc,arrows,arrows.meta,spy,matrix,scopes}
\usepgfplotslibrary{patchplots,groupplots}
\usepackage{tikzscale}

\input{acronyms.tex}

\usepackage{algorithm}
\usepackage{algorithmicx}
\usepackage{algcompatible}
\usepackage[noend]{algpseudocode}

\newcommand{\cast}{\gls{cast}\xspace}

\newcommand{\blue}[1]{{#1}}

\begin{document}

\title{Colosseum as a Digital Twin:\\
Bridging Real-World Experimentation\\
and Wireless Network Emulation}

\author{\IEEEauthorblockN{Davide Villa,
Miead Tehrani-Moayyed,
Clifton Paul Robinson,\\
Leonardo Bonati,
Pedram Johari,
Michele Polese,
Tommaso Melodia}\\
\IEEEauthorblockA{Institute for the Wireless Internet of Things, Northeastern University, Boston, MA, U.S.A.}\\
\IEEEauthorblockA{E-mail: \{villa.d, tehranimoayyed.m, robinson.c, l.bonati, p.johari, m.polese, melodia\}\\@northeastern.edu}\\
\thanks{This is a revised and substantially extended version of~\cite{villa2022cast}, which appeared in the Proceedings of ACM WiNTECH~2022%
.}
%
%
\thanks{This work was partially supported by the U.S.\ National Science Foundation under grant CNS-1925601, by the U.S.\ Department of Transportation, Federal Highway Administration, and by OUSD(R\&E) through Army Research Laboratory Cooperative Agreement Number W911NF-19-2-0221. The views and conclusions contained in this document are those of the authors and should not be interpreted as representing the official policies, either expressed or implied, of the Army Research Laboratory or the U.S.\ Government. The U.S.\ Government is authorized to reproduce and distribute reprints for Government purposes notwithstanding any copyright notation herein.}
}



\maketitle

\glsunset{3gpp}
\glsunset{usrp}

\begin{picture}(0,0)(10,-285)
\put(0,0){
\put(0,0){\footnotesize This paper has been accepted for publication on IEEE Transactions on Mobile Computing.}
\put(0,-10){
\scriptsize \textcopyright~2024 IEEE. Personal use of this material is permitted. Permission from IEEE must be obtained for all other uses, in any current or future media, including}
\put(0, -17){
\scriptsize reprinting/republishing this material for advertising or promotional purposes, creating new collective works, for resale or redistribution to servers or lists,}
\put(0, -24){
\scriptsize or reuse of any copyrighted component of this work in other works.}
}
\end{picture}

\begin{abstract}
Wireless network emulators are being increasingly used for developing and evaluating new solutions for \gls{nextg} wireless networks.
However, the reliability of the solutions tested on emulation platforms heavily depends on the precision of the emulation process, model design, and parameter settings. 
To address, obviate, or minimize the impact of errors of emulation models, in this work, we apply the concept of \gls{dt} to large-scale wireless systems.
%
%
Specifically, we demonstrate the use of Colosseum, the world's largest wireless network emulator with hardware-in-the-loop, as a \gls{dt} for \gls{nextg} experimental wireless research at scale. 
%
As proof of concept, we leverage the \gls{cast} to create the \gls{dt} of a publicly available over-the-air indoor testbed for sub-6\:GHz research, namely, Arena.
Then, we validate the Colosseum \gls{dt} through experimental campaigns on emulated wireless environments, including scenarios concerning cellular networks and jamming of Wi-Fi nodes, on both the real and digital systems.
Our experiments show that the \gls{dt} is able to provide a faithful representation of the real-world setup, obtaining an average similarity of up to \blue{0.987} in throughput and \blue{0.982} in \gls{sinr}.
\end{abstract}



\glsresetall
\glsunset{3gpp}
\glsunset{usrp}


\section{Introduction}
\label{sec:intro}



Current wireless technologies are the key enablers of the digital world. They meet the ever-growing demand of connecting larger and larger groups of people, vehicles, wearables, robots---virtually anything; they enable hosts of applications previously unthinkable: from autonomous vehicles roaming space and the oceans to life-saving telemedicine. 
A lead actor in this wireless revolution is the \gls{5g}, which has taken decisive steps toward redefining the cellular architecture. \gls{5g} provides unprecedented connectivity rates, capacity, and latency by opening and streamlining access to the \gls{ran}, using new frequency bands and advanced spectrum management techniques~\cite{BonatiPDBM20}.
The development of~6G
is already undergoing and is expected to achieve even higher capacity, lower latency, and increased bandwidth compared to currently deployed 
\gls{5g} systems~\cite{zhang20196g, giordani2020toward}, enabling the true realization of the \gls{iot} and connecting over~30 billion devices by 2030~\cite{statistaIoT}. 

%
%
%
%
%

In this context, powerful experimental wireless platforms have been recently developed to provide an ecosystem for advanced wireless research through repeatable and reproducible experimentation and the creation of large datasets. 
%
These platforms are becoming the nexus of \gls{ai}-enabled wireless research, where researchers can design, develop, train, and test new solutions for \gls{nextg} wireless systems.
Examples 
include the US~NSF \gls{pawr} program with its four at-scale, outdoor programmable platforms~\cite{pawr}, 
and indoor testbeds including the Drexel Grid~\cite{dandekar2019grid}, ORBIT \cite{kohli2021openaccess}, and Arena \cite{bertizzolo2020comnet}. 

While these testbeds provide good examples of indoor and outdoor wireless propagation environments, their scale can hardly capture the dynamics of real-world deployments. Also, their scope is limited to the physical environment where they are deployed. 
Alternatively, 
for site-independent wireless 
experimentation,
researchers can use
large-scale wireless emulation platforms.
%
By emulating a virtually unlimited variety of scenarios, these instruments are becoming a key resource to design, develop, and validate networking solutions in quasi-realistic environments, at scale, and with a diverse set of fully-customizable \gls{rf} channel conditions, traffic scenarios, and network topologies~\cite{bonati2021colosseum,sichitiu2020aerpaw,dandekar2019grid}. 
An exemplary large-scale emulation-based wireless platform is Colosseum---the world's largest wireless network emulator with hardware-in-the-loop~\cite{bonati2021colosseum}.


Solution development and testing for \gls{nextg} networks are in fact evolving toward integrating actual networked systems with a digital model that provides a replica of the physical network to be used for continuous prototyping, testing, and self-optimization of the living network.
These \emph{\glspl{dt}}~\cite{BarricelliCF19} are trending at the forefront of wireless research testing and prototyping~\cite{Saracco22}.
Similar to the \glspl{dt} used for some time in the industrial sector, a digital replica of a telecommunication network enables researchers to design optimized network architectures and to develop new \gls{ai}-based features to further expand and enhance the capabilities of \gls{nextg} systems. In contrast to the currently used simulation-based network planning tools, \gls{dt}-based systems will be connected to real-world deployed physical subsystems with real-time feedback loops, providing high-fidelity design and planning platforms.


While recent research focuses on \gls{nextg} telecommunication systems as a technology for reliable two-way communication between specific physical objects and their digital models, the realization of high-fidelity emulation-based \glspl{dt} for wireless systems as a whole, namely, a \gls{dtmn}~\cite{ericssondtmn, spirent5G}, is still a challenge largely untackled.
In this work, we provide the first demonstration of the capabilities of Colosseum as a \gls{dtmn}.
Particularly, we develop and test a comprehensive set of tools to: (i)~create an emulated \gls{dt} of virtually any real-world wireless scenario in Colosseum;
(ii)~validate the emulated environment through channel sounding; and 
(iii)~twin a standardized protocol stack through a \gls{cicd} framework.
The first two elements are carried out by using \textit{CaST}, a \emph{channel emulation generator and sounder toolchain}, preliminarily presented in~\cite{villa2022cast}, which we extend to include capabilities to realize the third element.
As a use-case, we leverage the extended version of \textit{CaST} to create and deploy a realistic \gls{dtmn} of Arena~\cite{bertizzolo2020comnet}, an over-the-air wireless testbed, on Colosseum.

The main contributions of this work are:

\begin{itemize}

\item We extend \acrshort{cast} to develop the first \gls{dtmn} on Colosseum, using Arena as a use case.
%
This use case demonstrates the scope and capabilities of Colosseum as a \gls{dt}, providing the research community with a set of tools to twin real-world environments.

\item 
We develop a \gls{cicd} pipeline for real-time twinning of selected protocol stacks, e.g., cellular and Wi-Fi.
This shows the flexibility of our tool by enabling researchers to test the latest version of open-source protocols as they are released, efficiently and automatically.

\item We compare key network performance metrics, namely, throughput and \gls{sinr}, of the Arena/Colosseum \gls{dtmn} 
to validate the fidelity of our twinning process. This is performed through a cellular network scenario with the open-source srsRAN software suite implementation~\cite{gomez2016srslte}, and via an adversarial jamming scenario using a GNU Radio-based Wi-Fi protocol stack~\cite{bloessl_ieee_2022}.

\end{itemize}

Results show that the twinning process is able to faithfully create a digital version of a real-world \gls{rf} environment,
achieving an average normalized cross-correlation similarity of \blue{0.987} in throughput and \blue{0.982} in \gls{sinr} using the cellular stack, and the Wi-Fi stack in the jamming scenario.

The remainder of this paper is organized as follows. 
In Section~\ref{sec:dt} we provide a brief primer on the concept of \glspl{dt} for \gls{nextg} systems in the context of wireless network emulators. 
Section~\ref{sec:dtplatforms} presents the platforms that we use to develop and implement our \gls{dtmn}. 
Section~\ref{sec:digitizing} defines the steps required for digitizing a real-world environment.
Section~\ref{sec:results} describes our experimental setup and results.
Section~\ref{sec:relatedwork} surveys related previous works.
Finally, Section~\ref{sec:conclusion} concludes the paper.



\section{Digital Twins}
\label{sec:dt}



\begin{figure}[th]
    \centering
    \includegraphics[width=\columnwidth]{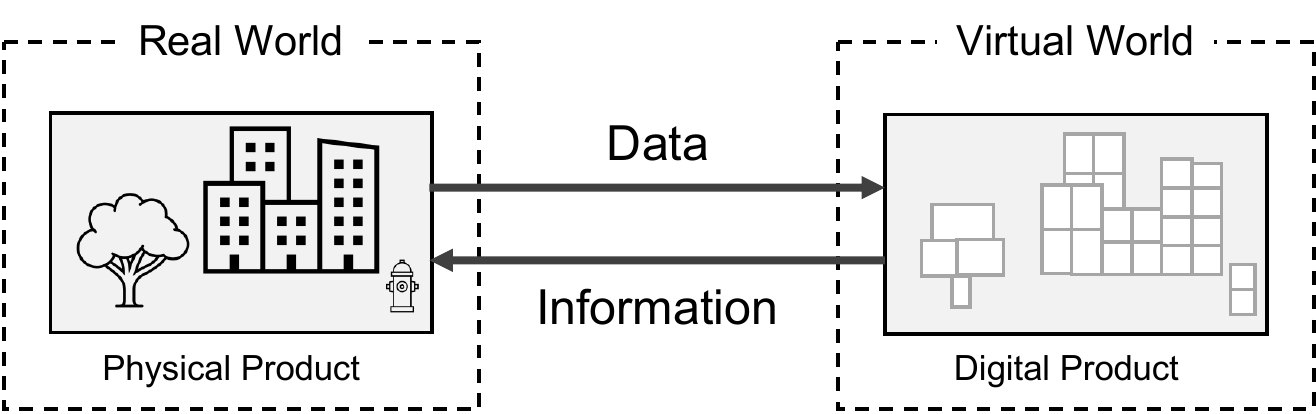}
    \caption{High-level representation of digital twin components.}
    \label{fig:dt-architecture}
\end{figure}

The \gls{dt} concept is finding increasing momentum
as a means of enhancing the performance of physical systems by using their virtual counterparts~\cite{Jones2020characteristing}.
The origin of this name is universally credited to Grieves and Vickers~\cite{grieves2015dt}, who 
define a \gls{dt} as a system consisting of three primary elements (Figure~\ref{fig:dt-architecture}): (i)~a physical product in the real world; (ii)~a virtual representation of the product in the virtual world; and (iii)~a connection of data and information tying the two.

Over the years, the \gls{dt} idea has extended into new domains, starting from its description and adding different flavors to this concept.
For example, some works consider \gls{dt} as an enabler for Industry~4.0 applications, as detailed in~\cite{rolle2020architecture}, while others suggest its use in areas such as product design, assembly, or production planning~\cite{Tao2018dt}.
Moreover, the continuous evolution of \gls{dt}s and their applications ushered the concept of \glspl{dtn}, as systems interconnect multiple \glspl{dt}~\cite{wu2021dtn}.
%
Finally, \glspl{dt} have been adopted in the context of the wireless communications ecosystem and cellular networks.

\begin{figure*}[t]
    \centering
    \includegraphics[width=\textwidth]{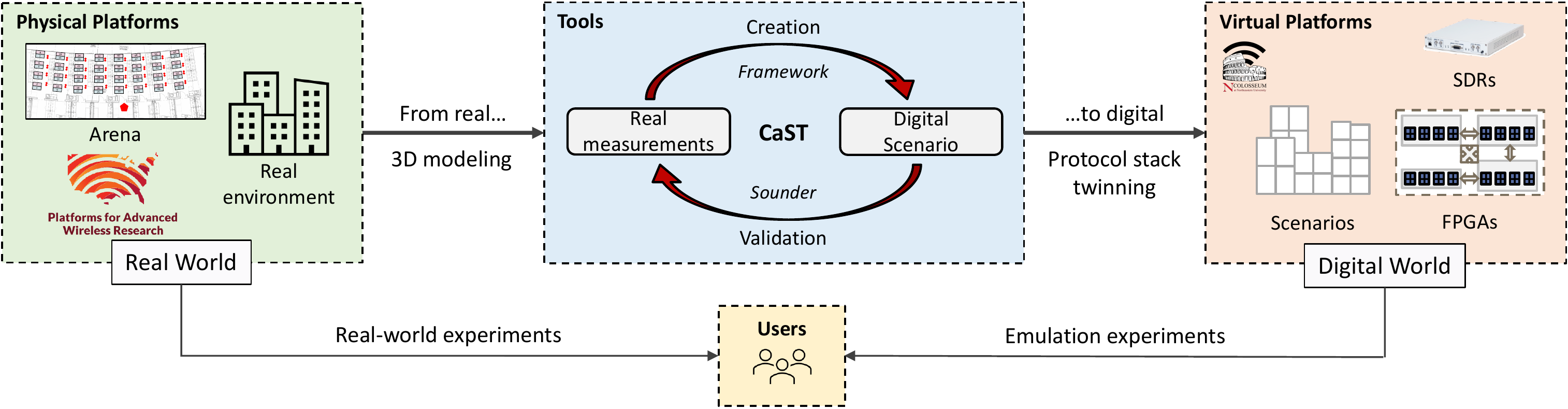}
    \caption{Main components of our high-level representation of a \gls{dt}.}
    \label{fig:dt-colosseum}
\end{figure*}

In this work, we apply the concept of \gls{dt} to experimental wireless research, and, to the best of our knowledge, in what is the \textit{first example of \gls{dtmn} for real-world applications.}
Figure~\ref{fig:dt-colosseum} shows a high-level diagram of all main components of our \gls{dt} representation.
Specifically, we develop a set of tools to create and validate a comprehensive digital representation of a particular real-world system inside a virtual environment. 
%
%
%
This would enable researchers to run wireless experiments inside a \gls{dt} of virtually any type of physical environment; develop and test new algorithms; and derive results as accurately and as close as possible to the behavior that they would obtain in the real-world environment.

To this aim, we propose Colosseum, the world's largest wireless network emulator~\cite{bonati2021colosseum}, as a \gls{dt} for real-world wireless experimental testbeds and environments.
Thanks to its large-scale emulation capabilities, Colosseum twins both the real and digital worlds by capturing conditions of real environments and reproducing them in a high-fidelity emulation. 
This is done through so-called \gls{rf} scenarios that model the characteristics of the physical world (e.g., channel effects, propagation environment, mobility, etc.) and convert them into digital emulation terrains to be used for wireless experimentation. 
%
%
Colosseum can also twin the protocol stack itself, i.e., it allows the deployment of the same generic software-defined stacks that can replicate the functionalities of real-world wireless networks, e.g., O-RAN managed cellular protocol stacks~\cite{bonati2021intelligence,polese2022coloran}, orchestrators~\cite{doro2022orchestran}, and different \gls{tcp} congestion control algorithms~\cite{pinto2023wintech}.

Colosseum is not limited to \gls{sdr} devices, but it also supports the integration of \gls{cots} devices---\blue{as long as they expose \gls{rf} connectors compatible with those of the \gls{mchem} \glspl{usrp}, e.g., SMA connectors or equivalent ones}---as demonstrated in~\cite{baldesi2022charm}, where \gls{soc} boards running OpenWiFi are used.
%
Additionally, custom equipment and waveforms can be integrated within Colosseum,
as demonstrated in prior work where we integrated a proprietary jammer~\cite{robinson2023esword}, and a radar waveform~\cite{villa2023wintech}, within the system.

Through the utilization of these scenarios and the twinning of protocol stacks and generic waveforms, users can collect data and test solutions in many different environments representative of real-world deployments, and fine-tune their solutions before deploying them in production networks to ensure they perform as expected.
An example of this is provided by~\cite{bonati2022openrangym-pawr}, where data-driven solutions for cellular networking were prototyped and tested on Colosseum before moving them to other platforms.
%
Overall, this allows users to retain full control over the digitized virtual world, to reproduce all---and solely---the desired channel effects, and to repeat and reproduce experiments at scale.
This is particularly important for \gls{ai}/\gls{ml} applications~\cite{bonati2022openrangym-pawr}, where: (i)~access to a large amount of data is key to designing solutions as general as possible; and (ii)~\gls{ai} agents need to be thoroughly tested and validated in different conditions to be sure they do not cause harm to the commercial infrastructure.

To enable RF twinning between physical and digital worlds in Colosseum, we utilize our recently developed tool \cast, an end-to-end toolchain to create and characterize realistic wireless network scenarios with a high degree of fidelity and accuracy~\cite{villa2022cast}.
\gls{cast} is composed of two main parts: (i)~a streamlined framework to create realistic mobile wireless scenarios from real-world environments (thus digitizing them); and (ii)~a \gls{sdr}-based channel sounder to characterize emulated \gls{rf} channels.
The protocol stack twinning is enabled by a \gls{ci} and \gls{cd} platform that can deploy in the Colosseum system the latest, or a specifically desired, version of a wireless protocol stack. We support any software-defined stack that has been designed for real-world experiments and have implemented a specific version of a \gls{cicd} framework for the OpenAirInterface 5G cellular implementation~\cite{KALTENBERGER2020107284}.


As proof of concept, we use \gls{cast} to create the \gls{dt} of a publicly available over-the-air indoor testbed for sub-6\:GHz research, namely Arena~\cite{bertizzolo2020comnet}.
This allows us to showcase the capabilities of Colosseum as a \gls{dt} platform, as well as the level of fidelity that can be achieved by the twinning process and operations.

%

\section{Digital Twin Platforms}
\label{sec:dtplatforms}

In this section, we describe the two platforms that are part of our \gls{dt} ecosystem: (i)~Colosseum, for large-scale emulation/digitization of physical environments, is described in Section~\ref{sec:colosseum}; and (ii)~Arena, for over-the-air real-world experimentation, in Section~\ref{sec:arena}.

\subsection{Large-scale Emulation: Colosseum}
\label{sec:colosseum}

\begin{figure*}[t]
    \centering
    \includegraphics[width=\textwidth]{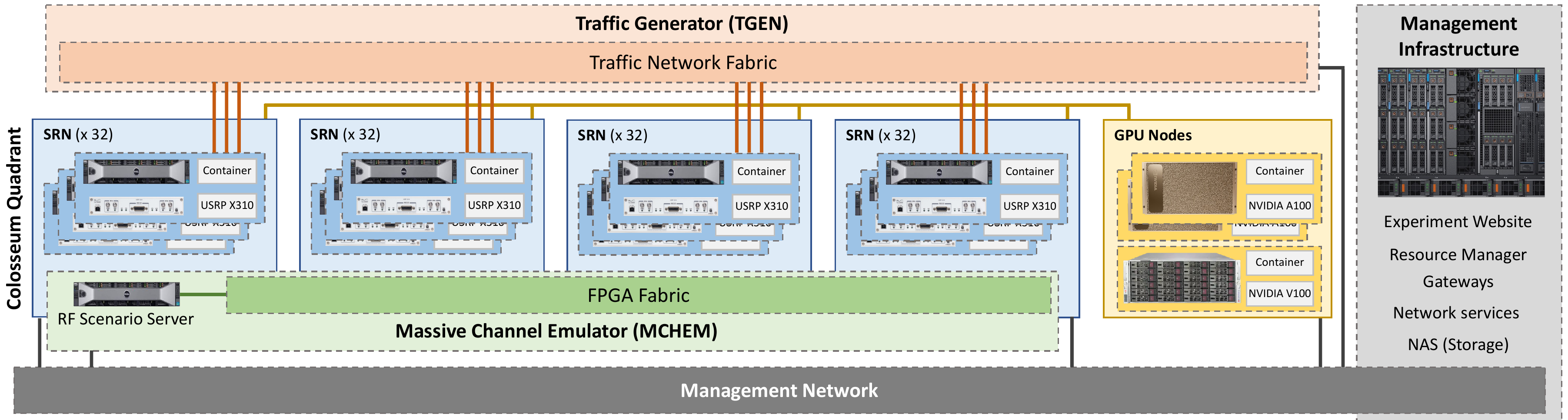}
    \caption{Colosseum architecture, adapted from~\cite{bonati2021colosseum}.}
    \label{fig:colosseum-architecture}
\end{figure*}

Colosseum is the world's largest publicly available wireless network emulator with hardware-in-the-loop.
At a high level, Colosseum consists of five main components, depicted in Figure~\ref{fig:colosseum-architecture}~\cite{bonati2021colosseum}: (i)~128 \acrfullpl{srn}; (ii)~the \acrfull{mchem}; (iii)~the \acrfull{tgen}; (iv)~the GPU nodes; and (v)~the management infrastructure.

The \glspl{srn}, which are divided into four quadrants, comprise 128~high-performance Dell PowerEdge R730 compute servers, each driving a dedicated USRP X310 \gls{sdr}---able to operate in the $[10\:\mathrm{MHz}, 6\:\mathrm{GHz}]$ frequency range---through a 10~Gbps fiber cable.
These servers are equipped with Intel Xeon E5-2650 CPUs with 48~cores, as well as NVIDIA Tesla K40m GPUs, to support heavy computational loads (e.g., \gls{ai}/\gls{ml} applications) and be able to properly drive their dedicated \gls{sdr}.
Users of the testbed can reserve \glspl{srn} for their experiments through a web-based \gls{gui}, as well as specify the date/time, and amount of time they need these resources for.
At the specified reservation time, Colosseum exclusively allocates the requested resources to the users and instantiates on them a softwarized protocol stack---also specified by the user when reserving resources---in the form of a \gls{lxc}.
After these operations have been carried out, users of the testbed can access via SSH to the allocated \glspl{srn}, and use the softwarized protocol stack instantiated on them (e.g., cellular, Wi-Fi, etc.) to drive the \glspl{sdr} and test solutions for wireless networking in a set of diverse environments emulated by Colosseum.

\begin{figure}[ht]
    \centering
    \includegraphics[width=\columnwidth]{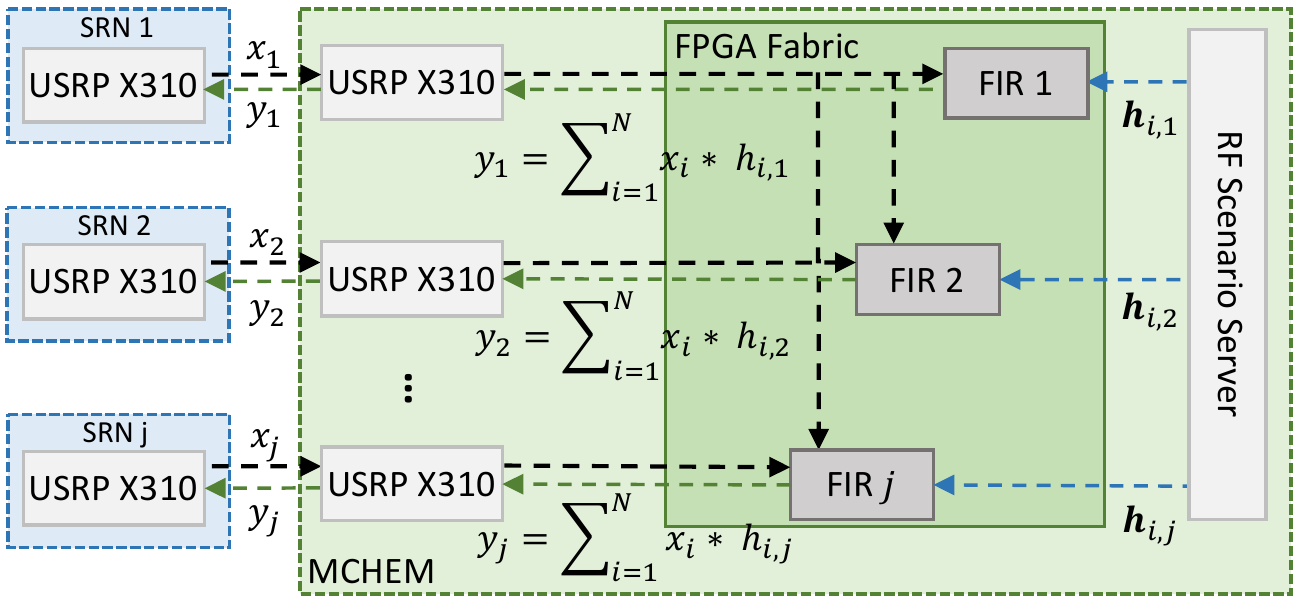}
    \caption{FPGA-based RF scenario emulation in Colosseum, from~\cite{bonati2021colosseum}.}
    \label{fig:mchem_diagram}
\end{figure}

These environments---called \gls{rf} scenarios in the Colosseum jargon---are emulated by Colosseum \gls{mchem}.
\gls{mchem} is formed of 16~NI ATCA 3671~\gls{fpga} distributed across the four quadrants of Colosseum.
Each ATCA module includes 4~Virtex-7 690T \glspl{fpga} that process through \gls{fir} filters the signals from/to an array of \glspl{usrp} X310 (32~USRPs per \gls{mchem} quadrant, for a total of 128~USRPs across the four quadrants of Colosseum) connected one-to-one, through SMA cables, to the \glspl{usrp} driven by the \glspl{srn} controlled by the users (see Figure~\ref{fig:mchem_diagram}).

Instead of being transmitted over the air, signals generated by the \gls{srn} USRPs are sent to the corresponding USRP on the \gls{mchem} side.
From there, they are converted in baseband and to the digital domain, and processed by the \gls{fir} filters of the \gls{mchem} \glspl{fpga} that apply the \gls{cir} corresponding to the \gls{rf} scenario chosen by the user of the testbed (see Figure~\ref{fig:mchem_diagram}).

Specifically, these \gls{fir} filters comprise 512~complex-valued taps that are set to reproduce the conditions and characteristics of wireless channels in real-world environments, i.e., the \gls{cir} among each pair of \gls{srn}.
As an example, and as depicted in Figure~\ref{fig:mchem_diagram}, signal $x_i$ generated by one of the \glspl{srn} is received by the USRP of \gls{mchem} and transmitted to its \glspl{fpga}.
Here, the \gls{fir} filters load the vector ${h}_{i,j}$ corresponding to the 512-tap \gls{cir} between nodes $i$ and $j$ (with $i, j \in \{1,...,N\}$ set of \glspl{srn} active in the user experiment) from the \gls{rf} scenario server, which contains a catalog of the scenario available on Colosseum.
Then, they apply these taps to $x_i$ through a convolution operation.
The signal $y_{j} = \sum_{i=1}^{N} x_i \ast h_{i,j}$ resulting from this operation, i.e., the originally transmitted $x_i$ signal with the \gls{cir} of the emulated channel, is finally sent to \gls{srn} $j$.
Analogous operations also allow Colosseum to perform superimposition of signals from different transmitters, and to consider interfering signals (besides the intended ones), as it would happen in a real-world wireless environment~\cite{ashish2018scalable}.
In this way, thus, Colosseum can emulate effects typical of real and diverse wireless environments, including fading, multi-path, and path loss, in terrains up to $1\:\mathrm{km^2}$ of emulated area, and with up to $80$\:MHz bandwidth, and can support the simultaneous emulation of different scenarios from multiple users.
Furthermore, Colosseum is capable of emulating node mobility discretely.
Every millisecond, the RF Scenario Server loads different pre-defined channel taps into the Colosseum FPGAs, effectively mimicking changes in channel conditions resulting from node position changes.

Similarly to the emulation of \gls{rf} environments, the \gls{tgen} allows users of the testbed to emulate different IP traffic flows among the reserved nodes.
This tool, which is based on the U.S.\ Naval Research Laboratory's \gls{mgen}~\cite{mgen}, enables the creation of flows with specific packet arrival distributions (e.g., Poisson, uniform, etc.), packet size, and rate.
These traffic flows, namely \textit{traffic scenarios}, are sent to the \glspl{srn} of the user experiment that, then, handles them through the specific protocol stack instantiated on the \glspl{srn} (e.g., Wi-Fi, cellular, etc.).

Recently, Colosseum added various GPU nodes to the pool of resources that can be reserved by users.
These include two NVIDIA DGX servers, state-of-the-art computing solutions with 8~NVIDIA A100 GPUs each and interconnected through a Tbps internal NVlink switching interface, and one large memory node (Supermicro SuperServer 8049U-E1CR4T) with 6~NVIDIA V100 GPUs, 128-core Intel Xeon Gold 6242 CPUs, and 3\:TB of RAM.
These resources, which can be reserved from the same web-based \gls{gui} used for the \glspl{srn}, can stream data in real-time from/to the \glspl{srn} through high-speed links and have the capability of powering computational-intensive workloads, such as those typical of \gls{ai}/\gls{ml} applications.

Finally, Colosseum includes a management infrastructure---not accessible by the users---that is used to maintain the rest of the system operational (see Figure~\ref{fig:colosseum-architecture}).
Some of the services offered by this include: (i)~servers that run the website used to reserve resources on the testbed; (ii)~resource managers to schedule and assign \glspl{srn} and GPU nodes to users; (iii)~multiple \gls{nas} systems to store experiment data and container images; (iv)~gateways and firewalls to enable user access and isolation throughout experiments; and (v)~precise timing servers and components to synchronize the \glspl{srn}, the GPU nodes, and the \glspl{sdr}.

\subsection{Over-the-Air Experimentation: Arena}
\label{sec:arena}

\begin{figure*}[t]
    \centering
    \includegraphics[width=\textwidth]{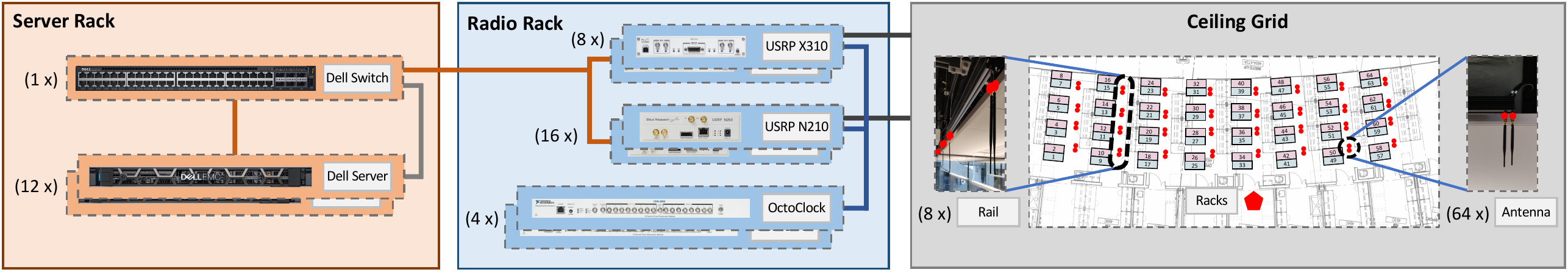}
    \caption{Arena architecture.}
    \label{fig:arena-architecture}
\end{figure*}

Arena is an over-the-air wireless testbed deployed on the ceiling of an indoor laboratory space~\cite{bertizzolo2020comnet}.
The architecture of Arena is depicted at a high level in Figure~\ref{fig:arena-architecture}.
Its main building blocks are: (i)~the ceiling grid; (ii)~the radio rack; and (iii)~the server rack.

The ceiling grid concerns~64 VERT2450 omnidirectional antennas hung off a $2450\:\mathrm{ft^2}$ indoor office space.
These are deployed on sliding rails and arranged in an $8 \times 8$ configuration to support \gls{mimo} applications.
The antennas of the ceiling grid are cabled through $100$\:ft low-attenuation coaxial cables to the radio rack.
This is composed of~24 USRP \glspl{sdr} (16~USRP N210 and 8~USRP X310) synchronized in phase and frequency through four OctoClock clock distributors.
Similarly to the USRPs on Colosseum, these \glspl{sdr} can be controlled through softwarized protocol stacks (e.g., cellular, Wi-Fi, etc.) deployed on the compute nodes of the server rack, to which they are connected through a Dell S4048T-ON \gls{sdn} programmable switch.
The server rack includes 12~Dell PowerEdge R340 compute nodes that are powerful enough to drive the \glspl{sdr} of the radio rack and use them for wireless networking experimentation in a real wireless propagation environment.
%

\smallskip
Because of the similarities offered by these two testbeds, software containers can be seamlessly transferred between the Colosseum and Arena testbeds with minimal modifications (e.g., specifying the network interface used to communicate with the \glspl{sdr}), as discussed in Section~\ref{sec:protocol-stack-twinning}.)
As we will show in Section~\ref{sec:results}, this allows users to design and prototype solutions in the controlled environment provided by the Colosseum \textit{digital twin}, to transfer them on Arena, and to validate these solutions in a real and dynamic wireless ecosystem.

\section{Digitizing Real-world Environments}
\label{sec:digitizing}

The process of digitizing real-world environments into their \gls{dt} representation is composed of different steps: (i)~\gls{rf} scenario twinning, in which the physical environment is represented into a virtual scenario and validated thereafter; and (ii)~protocol stack twinning, in which softwarized protocol stacks are swiftly transferred from the real world to the \gls{dt}, thus allowing users to evaluate their performance in the designed virtual scenarios.
We will describe these steps in the remainder of this section.

\subsection{RF Scenario Twinning}
\label{sec:scenario-twinning}

The \gls{rf} scenario twinning operations are performed by our \acrfull{cast}~\cite{villa2022cast}, which we made publicly available to the research community.\footnote{\url{https://github.com/wineslab/cast}}
This tool allows users to characterize a physical real-world \gls{rf} environment and to convert it into its digital representation, to be used in a digital twin, such as the Colosseum wireless network emulator.
\cast is based on an open \gls{sdr}-based implementation that enables: (i)~the creation of virtual scenarios from physical terrains; and (ii)~their validation through channel sounding operations to ensure that the characteristics of the designed \gls{rf} scenarios closely mirror the behavior of the real-world wireless environment.

\subsubsection{Scenario Creation}

\begin{figure}[ht]
    \centering
    \includegraphics[width=\columnwidth]{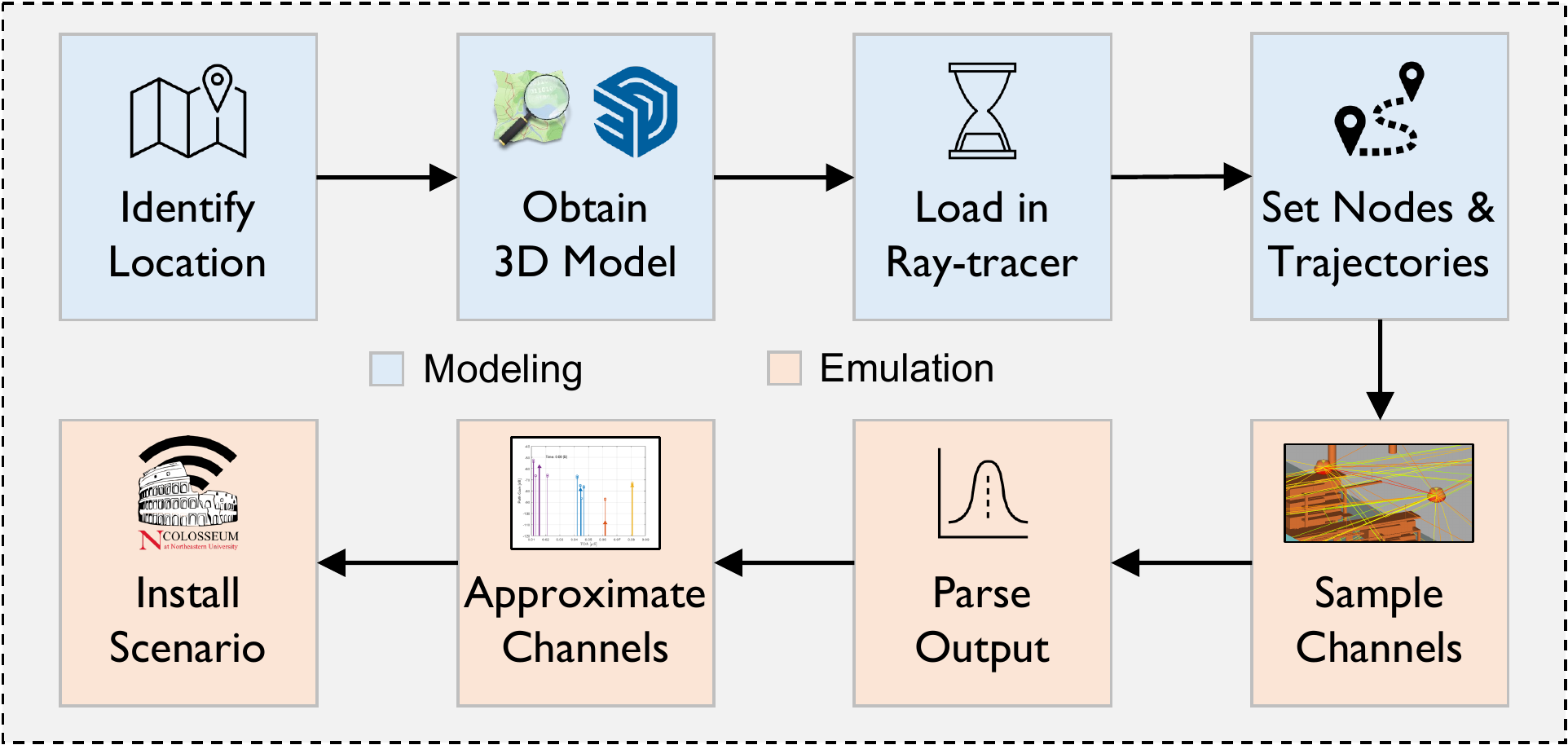}
    \caption{CaST scenario creation workflow.}
    \label{fig:scenariocreation}
\end{figure}

The scenario creation framework consists of several steps that capture the characteristics of a real-world propagation environment and model it into an \gls{rf} emulation scenario to install on Colosseum.
These steps, which are shown in Figure~\ref{fig:scenariocreation}, concern: (i)~identifying the wireless environment to emulate; (ii)~obtaining a 3D model of the environment; (iii)~loading the 3D model in a ray-tracing software; (iv)~modeling nodes and defining their trajectories; (v)~sampling the channels between each pair of nodes; (vi)~parsing the ray-tracing output of the channel samples; (vii)~approximating the obtained channels in a format suitable for the emulation platform (e.g., Colosseum \gls{mchem} \glspl{fpga}); and, finally, (viii)~installing the scenario on Colosseum.

\textbf{Identify the Wireless Environment.}
The first step consists of identifying the wireless environment, i.e., the physical location to twin in the channel emulator.
The area to model can be of different sizes, and representative of different environments, e.g., indoor (see Section~\ref{sec:arenadtscen}), outdoor (as shown in~\cite{villa2022cast}), urban, or rural.

\textbf{Obtain the 3D Model.}
The second step concerns obtaining the 3D model of the area to digitize.
This can be obtained from various databases, e.g., \gls{osm}, which is publicly available for outdoor environments or it needs to be designed using 3D modeling software, e.g., SketchUp.

\textbf{Load the Model in the Ray-tracer and Assign Material Properties.}
The 3D model obtained in the previous step needs to be converted into a file format (e.g., STL) suitable to be loaded into a ray-tracing software, e.g., the MATLAB ray-tracer or \gls{wi}, a commercial suite of ray-tracing models and high-fidelity \gls{em} solvers developed by Remcom~\cite{WI}.
Each object in the 3D model imported by the ray-tracing software consists of surfaces, and the material properties of these surfaces should be set to have reasonable ray-tracing results. The level of granularity in this step may depend on the ray-tracer platform, e.g., in the \gls{wi}, the material properties can be assigned to each surface. In the current version of MATLAB ray-tracer, this assignment is limited to the terrain and the buildings. The flexibility in assigning materials with a high level of detail leads to complex structures in the environment objects and accurate ray-tracing results.

\textbf{Model Nodes and Define Trajectories.}
Once the 3D model of the environment has been loaded in the ray-tracing software and the material properties are assigned, the radio nodes need to be modeled, which includes setting the nodes' radio parameters, modeling the antenna pattern, and defining locations of the nodes in the physical environment.
These nodes can be either static or mobile, in which case their trajectories and movement speeds need to also be defined.
The radio parameters of the nodes, e.g., carrier frequency, bandwidth, transmit power, receiver noise figure, ambient noise density, and antenna characteristics, need to be set as well.

\textbf{Sample the Channels.}
At this point, the channel is sampled through the ray-tracing software with a predefined sampling time interval $T_s$, which allows for capturing the mobility of the nodes in a discrete way.
To this aim, the node trajectories are spatially sampled with a spacing $D_i = V_i \cdot T_s$, where $V_i$ is the speed of node $i$.
Since spatial consistency plays a key role in providing a consisting correlated scattering environment in the presence of mobile nodes, we follow the \gls{3gpp} recommendations and consider a coherence distance of $15$\:m to guarantee an apt spatial consistency~\cite{3gppModel}.

\textbf{Parse the Output.}
The next step consists of parsing the ray-tracer output to extract a synchronized channel between each pair of nodes in the scenario for each discrete time instant $t$ spaced at least $1\:\mathrm{ms}$.
The temporal characteristic of the wireless channels is considered as a \gls{fir} filter, where the \gls{cir} is time-variant and expressed by:

\begin{equation}
\label{eq:time_variant_cir}
    h(t,\tau) = \sum_{i = 1}^{N_t} \Tilde{c_i}(t) \cdot \delta(t-\tau_i(t)),
\end{equation}

\noindent
where $N_t$ is the number of paths at time $t$, and $\tau_i$ and $c_i$ are the \gls{toa} and the path gain coefficient of the $i$-th path, respectively.
The latter is a complex number with magnitude $a_i$ and phase $\varphi_i$

\begin{equation}
\label{eq:cir_coefficient}
    \Tilde{c_i}(t) = a_i(t) \cdot e^{j\varphi_{i}(t)}
\end{equation}

\textbf{Approximate the Channels.}
The \gls{cir} characterized in the previous steps needs to be converted in a format suitable for \gls{mchem} \glspl{fpga}, e.g., 512~channel taps, 4~of which assume non-zero values, spaced with steps of $10\:\mathrm{ns}$ and with a maximum excess delay of $5.12\:\mathrm{\mu s}$.
To do this, we leverage a \gls{ml}-based clustering technique to reduce the taps found by the ray-tracing software, align the tap delays, and finalize their dynamic range, whilst ensuring the accuracy of the emulated scenario~\cite{tehrani2021creating}.

\textbf{Install the Scenario.}
Finally, the channel taps resulting from the previous steps are fed to Colosseum scenario generation toolchain, which converts them in \gls{fpga}-friendly format and installs the resulting \gls{rf} scenario on the \gls{dt}, ready to be loaded on-demand by the RF Scenario Server.

\subsubsection{Scenario Validation}

Now that the scenario has been created and installed in the \gls{dt}, we validate its correct functioning through the channel sounder embedded in \cast~\cite{villa2022cast}.
In doing this, we also ensure that the scenario installed in the \gls{dt} closely follows the behavior experienced in the real-world environment.
%

\begin{figure}[htbp]
    \centering
    \includegraphics[width=\columnwidth]{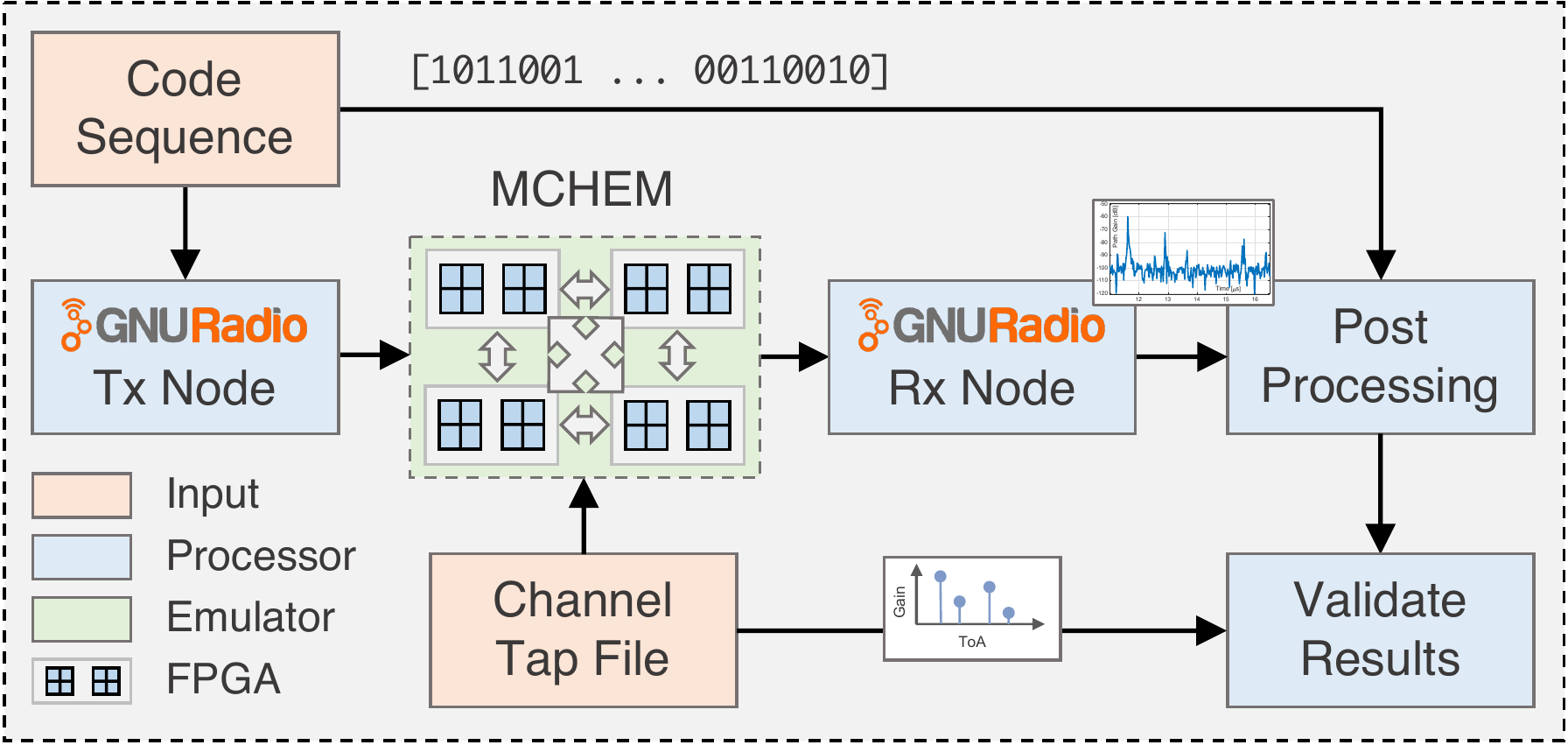}
    \caption{CaST channel sounding workflow.}
    \label{fig:soundingbd}
\end{figure}

The main steps of \cast channel sounder, shown in blue shades in Figure~\ref{fig:soundingbd}, are: (i)~the transmission of a known code sequence used as a reference for the channel sounding operations; (ii)~the reception of the transmitted code sequence, processed by \gls{mchem} through the channel taps of the emulated \gls{rf} scenario; (iii)~the post-processing of the received data and its correlation with the originally transmitted code sequence; and (iv)~the validation of the results with the modeled channel taps.

The \gls{cast} sounder uses a transmitter and a receiver node implemented through the GNU Radio open-source \gls{sdr} development toolkit~\cite{gnuradio}.
This software toolkit allows implementing and programming \glspl{sdr} through provided signal processing blocks that can be interconnected to one another.

In our sounding application, the transmitter takes as input a known code sequence---how to derive the specific code sequence is described in Section~\ref{sec:cast-tuning}---and transmits it to the receiver node through the wireless channel emulated by the Colosseum \gls{dt} through the \gls{rf} scenario to evaluate.
The transmitted signal is composed of sequential repetitions of the code sequence encoded through a \gls{bpsk} modulation.
While other modulation types are not restricted, we leverage \gls{bpsk} because it offers sufficient channel information for the sounding in Colosseum.
Additionally, it allows for simple data computations that are less susceptible to errors and approximations, resulting in a cleaner and less disrupted signal.
%
Data is streamed to the USRP controlled by the \gls{srn} that transmits it to the receiving node through \gls{mchem}.
For increased flexibility of the channel sounder, \cast allows users to set \blue{various USRP parameters, such as clock source, sample rate, and frequency.}

At the receiver side, the \gls{srn} USRP samples the signal sent by \gls{mchem}, i.e., the transmitted signal processed with the channel taps of the emulated scenario.
%
%
This signal is cross-correlated with the originally transmitted known code sequence to extract the \gls{cir} $h(t)$ of the emulated scenario, and the \gls{pl} $p(t)$.
The \gls{cir} is then used to obtain the \gls{toa} of each multi-tap component of the transmitted signal, which allows measuring the distance between taps, while the \gls{pl} allows measuring the intensity and attenuation of such components as a function of the time delay.
To perform the above post-processing operations, let $c(t)$ be the $N$-bit known code sequence, and $s^{IQ}(t)$ and $r^{IQ}(t)$ the \gls{iq} components of the transmitted ($s(t)$) and received ($r(t)$) signals, respectively.
The \gls{iq} components of the \gls{cir} are computed by separately correlating $r^{I}(t)$ and $r^{Q}(t)$ (i.e., the $I$ and $Q$ components of $r^{IQ}(t)$) with the $I$ and $Q$ components of $s(t)$ divided by the inner product of the transmitted known sequence with its transpose:

\begin{align}
    h^{I}(t) &= \frac{r^{I}(t) \otimes s^{I}(t)}{s^{I^T}(t) \times s^{I}(t)},\label{eq:hicorr} \\[1em]
    h^{Q}(t) &= \frac{r^{Q}(t) \otimes s^{Q}(t)}{s^{Q^T}(t) \times s^{Q}(t)},\label{eq:hqcorr}
\end{align}

\noindent
where $\otimes$ is the cross-correlation operation~\cite{buck2002computer} between two discrete-time sequences $x$ and $y$, which measures the similarity between $x$ and shifted (i.e., lagged) repeated copies of $y$ as a function of the lag $k$ \blue{following:}
\begin{equation}\label{eq:xcorrcast}
    \blue{\chi(k) = \sum_{n=1}^{N} x(n) \cdot y(n+k),}
\end{equation}
\blue{with $\chi$ denoting the cross-correlation and $N$ the length of the $x$ sequence.}
It is worth noticing that if the considered modulation is a \gls{bpsk}, the denominator is equal to the length $N$ of $c(t)$.
The amplitude of the \gls{cir} can be computed as:

\begin{equation}
    \label{eq:habs}
    |h(t)| = \sqrt{(h^{I}(t))^2 + (h^{Q}(t))^2}
\end{equation}

\noindent
and the path gains as:

\begin{equation}
    \label{eq:pgdb}
    G_p(t) [dB] = 20log_{10}(|h(t)|) - P_t - G_t - G_r,
\end{equation}

\noindent
where $P_t$ is the power of the transmitted signal, and $G_t$ and $G_r$ are the transmitter and receiver antenna gains expressed in $dB$.

\subsection{Protocol Stack Twinning}
\label{sec:protocol-stack-twinning}

The twinning of protocol stacks from real to virtual environments (and back) is key in the \gls{dt} ecosystem, as it allows users to swiftly transfer and evaluate real-world solutions in a controlled setup through automated tools.
Twinning at the protocol stack level, combined with the RF scenario twinning discussed in Section~\ref{sec:scenario-twinning}, makes it possible to seamlessly prototype, test, and transition end-to-end, full-stack solutions for wireless networks to and from digital and physical worlds.
After validation in the controlled environment of the \gls{dt}---to make sure whatever is tested works as expected---the protocol stack solutions can be transitioned back to real-world deployments where they are ultimately used on a production network. As an example, in our prior works~\cite{bonati2022openrangym-pawr,polese2022coloran}, we have shown how \gls{ai} solutions for 5G cellular networks trained and tested on the digital twin environment---Colosseum---can be effective and also work on real-world environments---Arena and the \gls{pawr} platforms~\cite{pawr}. 

At a high level, the twinning of the protocol stack involves: (i) tracking one or multiple remote, centralized version control systems that host the code of the protocol stack; and (ii) providing pipelines that can automatically replicate the same software build in the digital and physical domains. In addition, it is possible to embed automated steps for the performance validation (i.e., profiling of relevant performance metrics), similar to the scenario validation step of \gls{cast}.

\begin{figure}[h]
    \centering
    \includegraphics[width=\columnwidth]{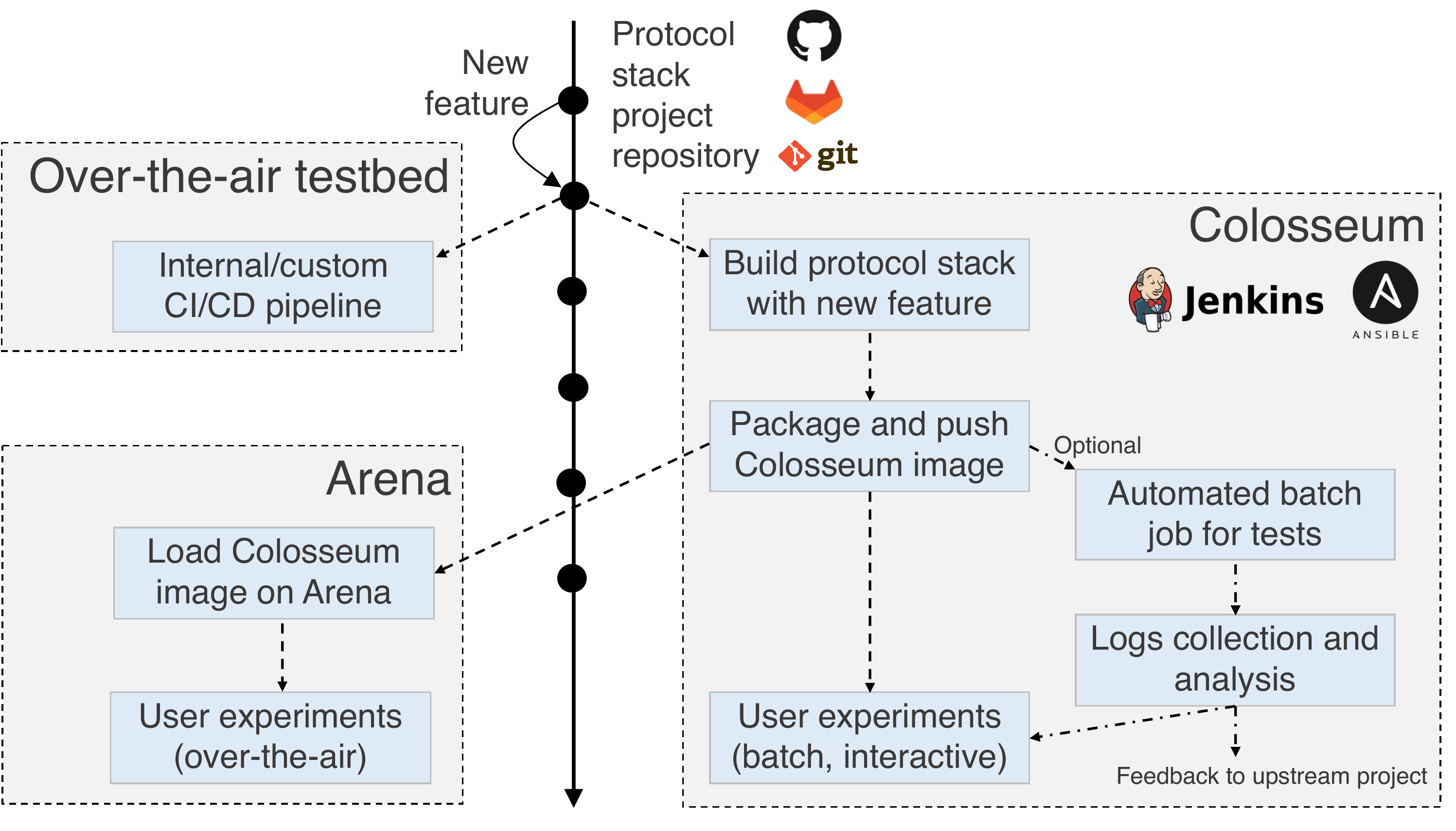}
    \caption{Protocol stack twinning workflow across a digital environment, i.e., Colosseum, and two physical environments, i.e., Arena and a generic over-the-air testbed.}
    \label{fig:protocol-twinning}
\end{figure}

Figure~\ref{fig:protocol-twinning} illustrates how the protocol stack twinning is implemented in Colosseum (right), with extensions to a generic over-the-air testbed (top left) and the specific integration with Arena (bottom left). The figure refers to a project repository for a sample protocol stack, hosted on a versioning platform that supports the \texttt{git} version control system (e.g., GitHub, GitLab, among others). In Colosseum, we implemented this pipeline for the OpenAirInterface reference stack for 5G base stations and \glspl{ue}, and are working toward the integration of additional components for O-RAN testing~\cite{bonati2022openrangym-pawr}. 

Whenever a new feature (i.e., a commit on selected branches, or a pull request) is pushed to the target project repository, 
a \gls{ci}/\gls{cd} framework implemented with Jenkins and Ansible triggers the automated process in the digital Colosseum domain. Specifically, Jenkins monitors the remote repository and orchestrates the kickoff of the build job, and Ansible applies the relevant configuration parameters to the machine that actually executes the build job (e.g., a Colosseum \gls{srn} or a dedicated virtual machine on AWS). Once the build is successful, the Jenkins job packages the output of the process into an LXC image which is stored on the Colosseum \gls{nas}.  

Once this is done, Colosseum can be further used to perform automated testing, e.g., to automatically test solutions and algorithms on the \gls{dt} and collect relevant metrics from such experiments. These can be shared with the relevant stakeholders, e.g., the developers of the protocol stack framework being deployed and tested. Moreover, since no over-the-air transmissions happen in Colosseum, as the channels are emulated through \gls{mchem} (see Section~\ref{sec:colosseum}), this \gls{dt} environment enables users to test networking solutions over frequencies and bandwidths that would normally require compliance with the \gls{fcc} regulations.

Finally, the image with the relevant components can be used by experimenters in Colosseum or moved to the physical domain, e.g., Arena, for validation on a real-world infrastructure. In addition, the protocol stack can be twinned in other over-the-air testbeds following their internal and custom \gls{cicd} pipelines, as long as the centralized repository that the different testbeds track provides shared specifications for the build environment (e.g., operating system, compiler versions, packages, etc.). As an example, the Colosseum protocol stack twinning process already replicates internal \gls{cicd} pipelines used in the Eurecom/OpenAirInterface facilities~\cite{eurecomCi}.
%
%
%
%
%
%

%

\section{Experimental Evaluation}
\label{sec:results}

This section discusses the capabilities of our \gls{dt} system through experimental evaluations, and it is organized as follows: (i)~we showcase \gls{cast} tuning process (Section~\ref{sec:cast-tuning}); (ii)~we leverage \gls{cast} to validate Colosseum scenarios, both with single and multiple taps (Section~\ref{sec:cast-colosseum-validation}); (iii)~we describe the Arena scenario designed as part of this paper (Section~\ref{sec:arenadtscen}); and (iv)~we compare some experimental use cases (e.g., for cellular networking and Wi-Fi applications) both in the Arena testbed and in its \gls{dt} representation (Section~\ref{sec:usecases}).

%

%
%
%

\subsection{\acrshort{cast} Tuning}
\label{sec:cast-tuning}

\begin{figure}[ht]
\centering
    \centering
    \includegraphics[width=\linewidth]{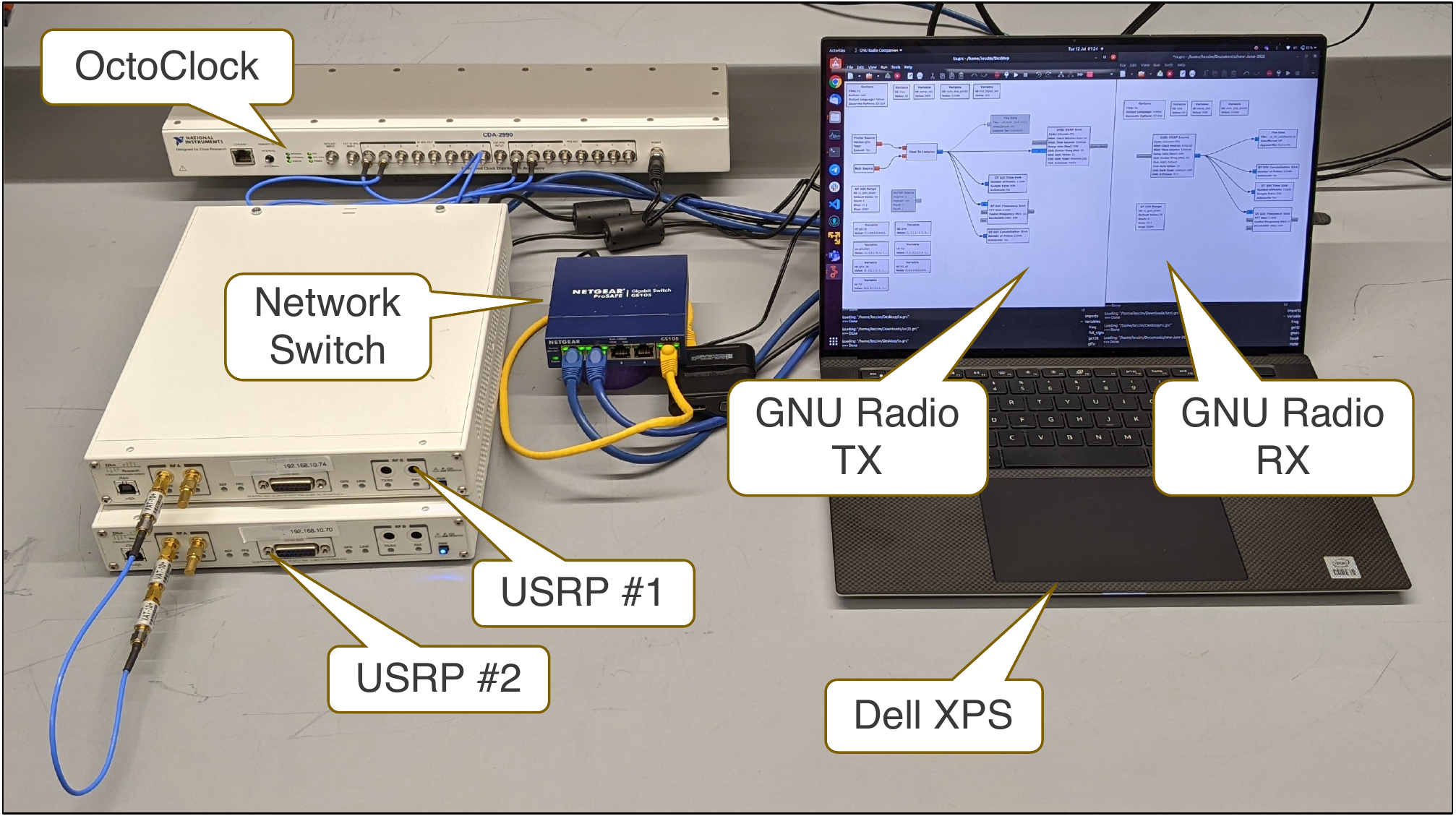}
    \vspace{-5pt}
    \caption{Controlled laboratory environment used for the \acrshort{cast} tuning process.}
    \label{fig:localtestbed}
\end{figure}

As a first step, we tune \gls{cast} parameters and configurations (see Section~\ref{sec:scenario-twinning}) outside the Colosseum channel emulator to identify a code sequence with high auto-correlation and low cross-correlation between transmitted code sequence and received signal which functions well within our combination of software and hardware.
It is worth noting that the tuning of CaST is a one-time operation, and it would need to be repeated only upon changes in the Colosseum hardware (e.g., radio devices and channel emulation system).
This step, which is key for \cast to be able to derive taps from arbitrary \glspl{cir}, is performed in the controlled environment shown in Figure~\ref{fig:localtestbed}.

This consists of two \gls{usrp} X310 \glspl{sdr} equipped with a UBX-160 daughterboard and synchronized in phase and frequency through an OctoClock clock distributor to mirror the same deployment used in Colosseum.
Differently from the Colosseum deployment, however, the two \glspl{usrp} are connected through a $12$\:inches SMA cable, and $30$\:dB attenuators (to shield the circuitry of the daughterboard from direct power inputs, as indicated in their datasheet).
This is done to derive the above-mentioned code sequences in a baseline and controlled setup without additional effects introduced by over-the-air wireless channels, or channel emulators.
The USRPs are connected through a network switch to a Dell XPS laptop, used to drive them.
The sounding parameters used in this setup are summarized in Table~\ref{table:localconfig}.
We consider different values for the gains of the \glspl{usrp} (i.e., in $[0, 15]$\:dB) to evaluate their effect on the sounding results.
The receiving period time and data acquisition are set to $3$\:s.
\begin{table}[ht]
    \centering
    \footnotesize
    \caption{Configuration parameters used in the controlled laboratory setup.}
    \label{table:localconfig}
    \begin{tabularx}{0.8\columnwidth}{
        >{\raggedright\arraybackslash\hsize=\hsize}X
        >{\raggedright\arraybackslash\hsize=\hsize}X }
        \toprule
        Parameter & Value \\
        \midrule
        Center frequency & $1$\:GHz \\
        Sample rate & $[1, 50]$\:MS/s \\
        USRP transmit gain & $[0, 15]$\:dB \\
        USRP receive gain & $[0, 15]$\:dB \\
        \bottomrule
    \end{tabularx}
\end{table}

\begin{figure}[ht]
    \centering
    \subfloat[Gold sequence]{\label{fig:cirlocalgold}\includegraphics[width=0.48\columnwidth]{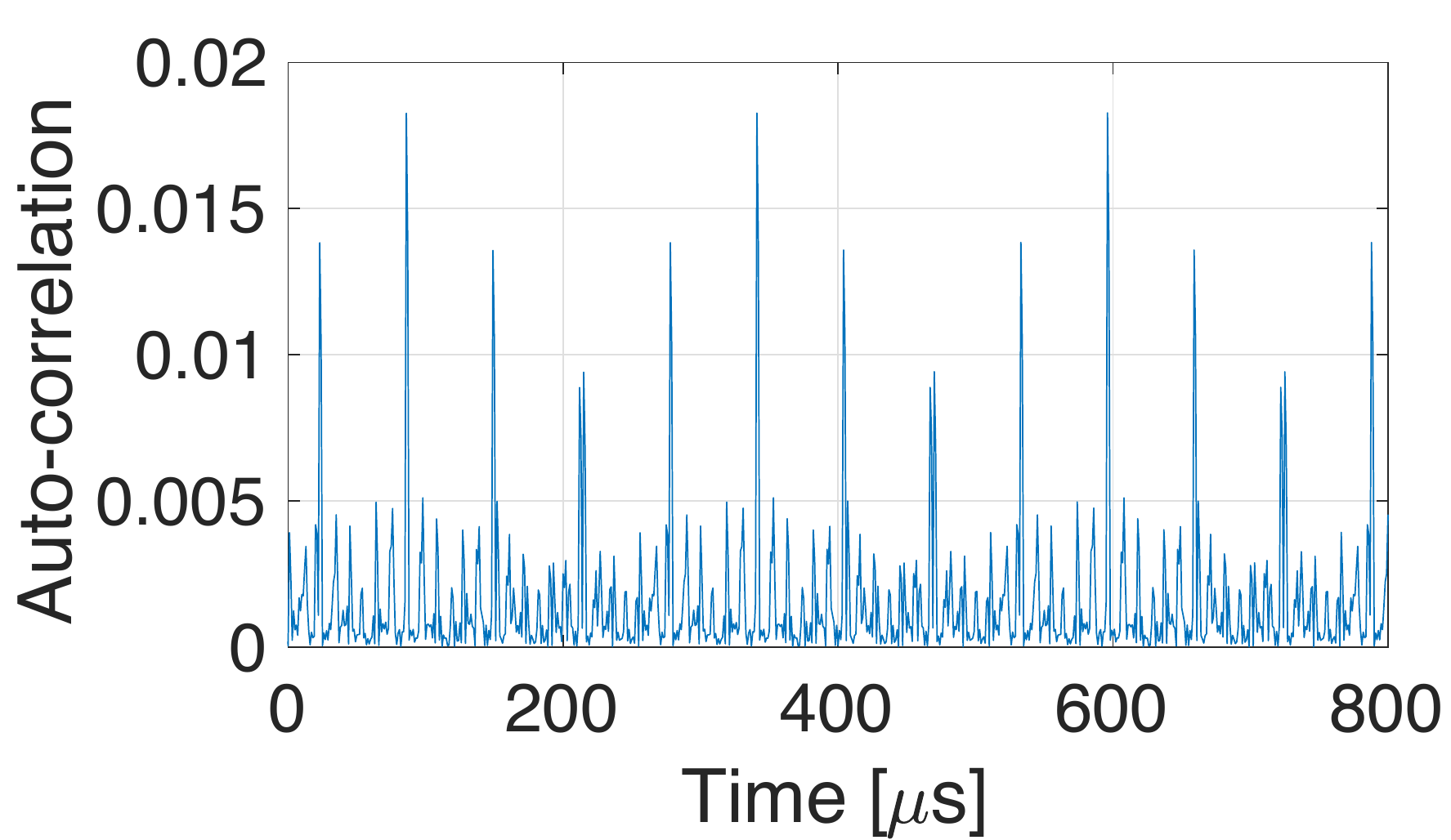}}%
    \hfill
    \subfloat[Ga$_{128}$ sequence]{\label{fig:cirlocalga}\includegraphics[width=0.48\columnwidth]{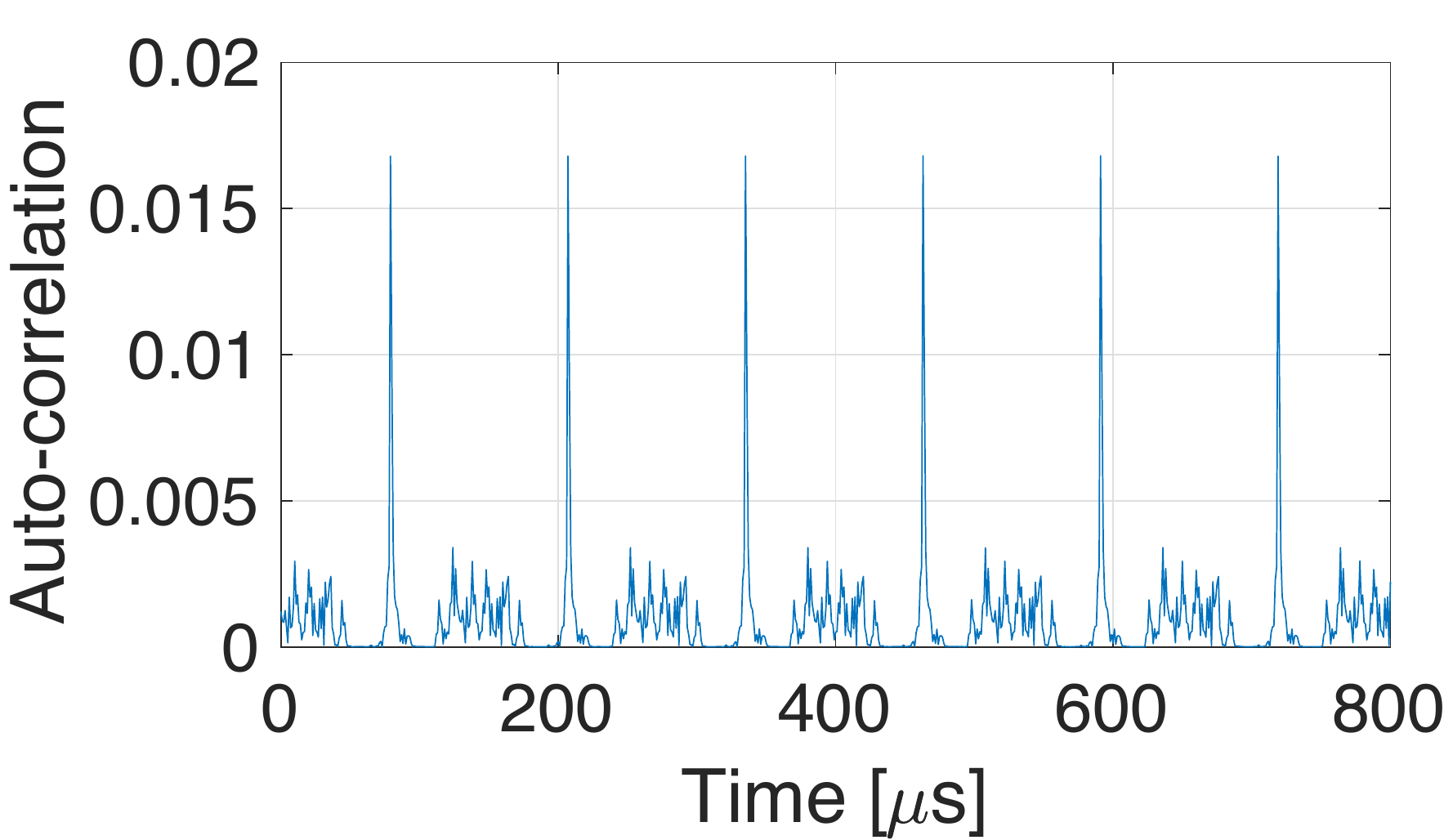}}
    \hfill
    \subfloat[\acrshort{ls} sequence]{\label{fig:cirlocalls1}\includegraphics[width=0.48\columnwidth]{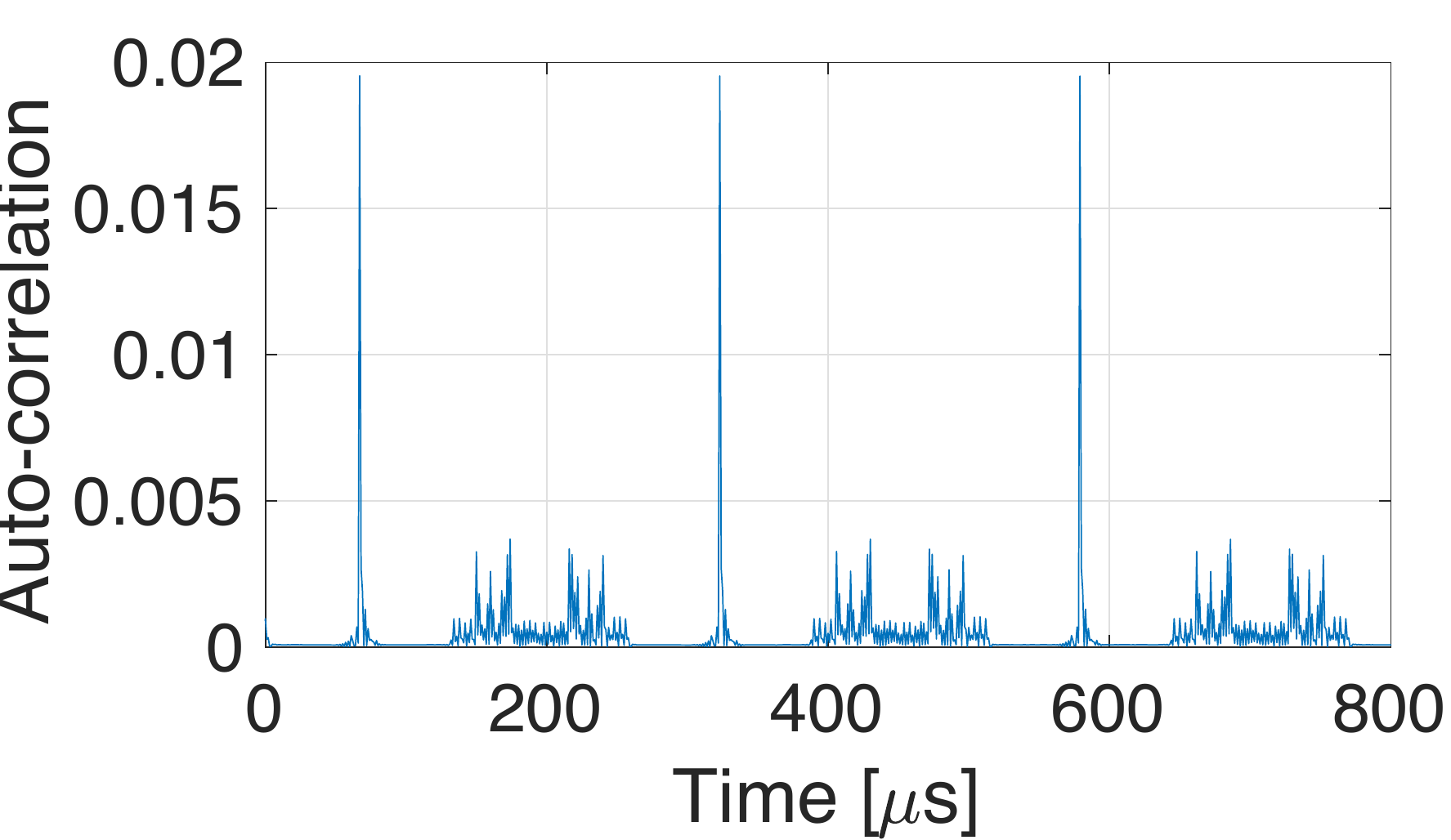}}
    \hfill
    \subfloat[\acrshort{glfsr} sequence]{\label{fig:cirlocalglfsr}\includegraphics[width=0.48\columnwidth]{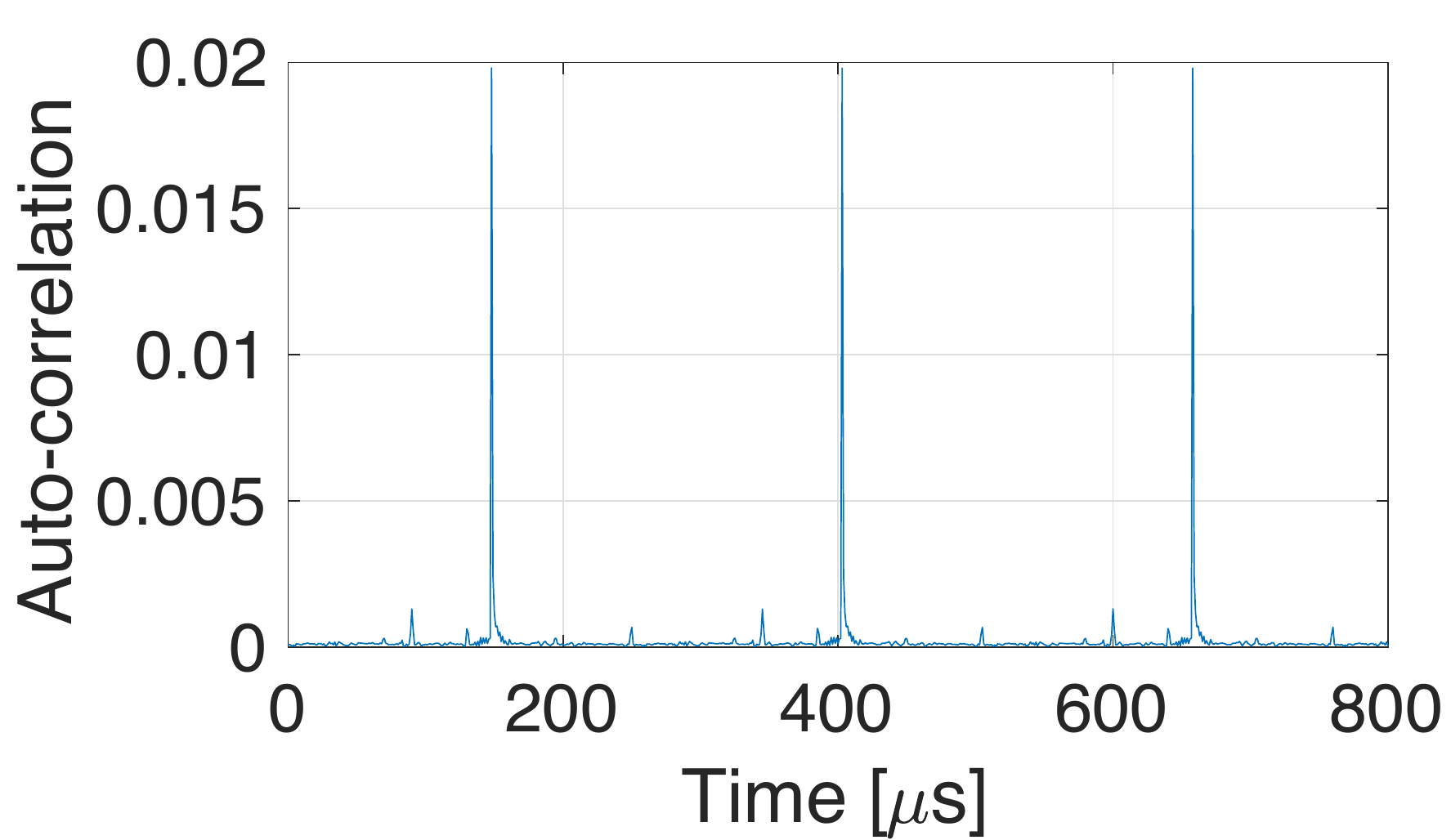}}
    \caption{Correlation of different code sequences in the controlled laboratory environment.}
    \label{fig:cirlocal}
\end{figure}

\textbf{Finding the Code Sequence.}
Code sequences have been widely investigated in the literature because of their role in many different fields~\cite{velazquez2016sequence,stanczak2001are}.
Good code sequences achieve a high auto-correlation (i.e., the correlation between two copies of the same sequence), and a low cross-correlation (i.e., the correlation between two different sequences).
For our channel-sounding characterization, we consider and test four different code sequences by leveraging the laboratory environment shown in Figure~\ref{fig:localtestbed}:
\begin{itemize}
    \item \textit{Gold sequence}. These sequences
    are created by leveraging the XOR operator in various creation phases applied to a pair of codes, $u$, and $v$, which are called a preferred pair. This pair of sequences must satisfy specific requirements to qualify as suitable for a gold sequence, as detailed in~\cite{zhang2011analysis}.
    Gold sequences have small cross-correlation within a set, making them useful when more nodes are transmitting in the same frequency range. They are mainly used in telecommunication (e.g., in \gls{cdma}) and in satellite navigation systems (e.g., in GPS).
    In this work, we use a Gold sequence of $255\:\mathrm{bits}$ generated with the MATLAB Gold sequence generator system object with its default first and second polynomials, namely $z^6+z+1$ and $z^6+z^5+z^2+z+1$, for the generation of the preferred pair sequences.
    
    \item \textit{Golay complementary sequence}. Being complementary, these sequences have the property that the sum of their out-of-phase aperiodic auto-correlation coefficients is equal to $0$~\cite{golay1961seq}. Their applications range from multi-slit spectrometry and acoustic measurements to Wi-Fi networking and \gls{ofdm} systems. In our tests, we use a $128$-bit type~A Golay Sequence (Ga$_{128}$) as defined in the IEEE 802.11ad-2012 Standard~\cite{ieee-802_11ad-standard}.
    
    \item \textit{\gls{ls} sequence}.
    These sequences exhibit the property of reaching very low auto-correlation and cross-correlation values in a certain portion of time, based on the maximum delay dispersion of the channel, called \gls{ifw}. This allows the mitigation of the interference if the maximum transmission delay is smaller than the \gls{ifw} length.
    In our experiments, we use an \gls{ls} sequence generated following the directions in~\cite{garcia2010generation}, and only leveraging the first codeset of $\{-1, 1\}$ without including the \gls{ifw}.
    
    \item \textit{\gls{glfsr} sequence}. These sequences add time offsets to \gls{lfsr} codes by leveraging extra XOR gates at the output of the \gls{lfsr}.
    %
    This allows to achieve a higher degree of randomness if compared to the classic \gls{lfsr}, making them more efficient and fast in detecting potential faults with increased auto-correlation results~\cite{pradhan1999glfsr}.
    %
    %
    In this paper, we leverage GNU Radio
    to generate a $255$-bits sequence with the following parameters: shift register degree 8, bit mask 0, and seed 1.
\end{itemize}
Each of these sequences has been separately used by the transmitter node to construct the sending signal and to send it to the receiver node with a sample rate of $1$\:MHz.
After that, the receiver node performs the post-processing operations.
Results of $800\:\mathrm{\mu s}$ \gls{cir} for each code sequence are shown in Figure~\ref{fig:cirlocal}.
We can notice that all code sequences are able to correctly identify the starting position of the transmitted signal, as shown by the peak values.
The distance $D_{peak}$ of each peak can be written as a function of the code length $N$ and the sampling rate $SR$.

\begin{equation}
\label{eq:dpeak}
    D_{peak} = \frac{N}{SR}
\end{equation}

\noindent
Therefore, $D_{peak}$ is equal to $255\:\mathrm{\mu s}$ for the Gold, \gls{ls}, and \gls{glfsr} codes, each showing 3~transmitted sequences in Figure~\ref{fig:cirlocal}, and to $128\:\mathrm{\mu s}$ for the Ga$_{128}$ code, which displays 6~sequences instead.
We notice that \gls{glfsr} shows the highest auto-correlation and lowest cross-correlation among the four considered code sequences.
This results in an overall cleaner \gls{cir}.
For these reasons, we adopt the \gls{glfsr} code sequence in our experimental evaluation through \gls{cast}.

\begin{figure}[ht]
    \centering
    \includegraphics[width=\columnwidth]{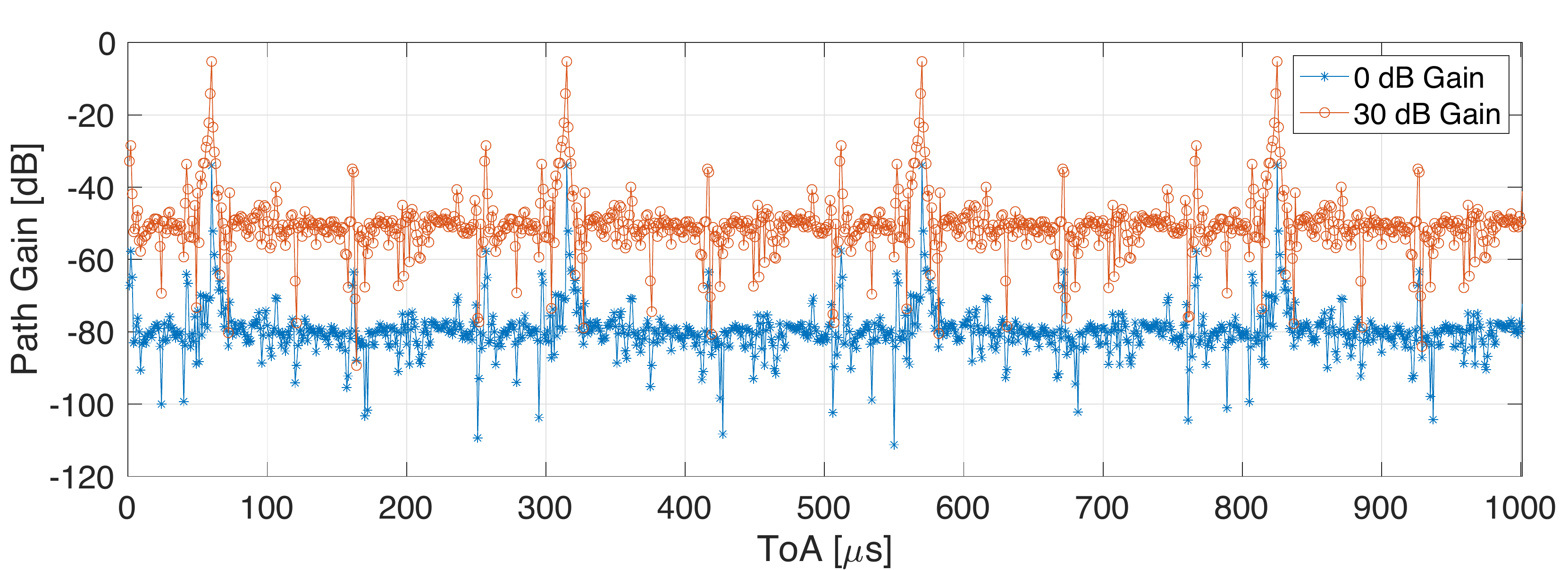}
    \vspace{-5pt}
    \caption{Received path gains in the controlled laboratory environment with $0$ and $30$\:dB total transmit and receive gains use cases ($15$\:dB at both transmitter and receiver sides).}
    \label{fig:pglocal}
\end{figure}

\textbf{\acrshort{cast} Validation in a Laboratory Environment.}
After identifying the code sequence for our application, we evaluate \gls{cast} in the laboratory setup shown in Figure~\ref{fig:localtestbed}.
To this aim, we test our sounder with a \gls{glfsr} code sequence and various configuration parameters, e.g., sample rate, center frequency, and antenna gains, to study its behavior and gather reference information to be leveraged in the Colosseum experiments.
Figure~\ref{fig:pglocal} shows a time frame of the received path gains for the case with $0$\:dB (blue line in the figure), and $30$\:dB total transmit and receive gains ($15$\:dB at both transmitter and receiver sides, orange line).

The figure shows signals that repeat based on the length of the transmitted code sequence, i.e., every 255 sample points (or equivalently every $255\:\mathrm{\mu s}$, since one point equals to $1/\mathrm{sample\_rate} = 1\:\mathrm{\mu s}$).
The peaks represent the path loss of the single tap of this experiment, which are equal to $34.06$\:dB for the $0$\:dB case, and $5.24$\:dB for the $30$\:dB case.
Since we have $30$\:dB attenuation in this validation setup, these results are in line with our expectations (with some extra loss due to the physical components of the setup, e.g., cable attenuation and noise).
We also notice that in the $30$\:dB case, the measured loss is slightly more severe due to imperfections in the power amplifiers of the \glspl{usrp}.
We use these results as a reference for our channel-sounding operations.

\subsection{Validation of Colosseum Scenarios through \acrshort{cast}}
\label{sec:cast-colosseum-validation}

After the tuning and validation in the controller laboratory environment, we can leverage \gls{cast} to validate the behavior of Colosseum \gls{mchem}.
We deploy the \gls{cast} sounder on the Colosseum wireless network emulator by creating an \gls{lxc} container from the open-source \gls{cast} source code.
This container, which has been made publicly available on Colosseum, contains all the required libraries and software to perform channel-sounding operations, as well as for the post-processing of the obtained results.
This enables the re-usability of the sounder with different \glspl{srn} and scenarios, as well as portability to different testbeds (e.g., to the Arena testbed described in Section~\ref{sec:arena}).
It also allows the automation of the channel sounding operations through automatic runs supported by Colosseum, namely \textit{batch jobs}.

To achieve our goal of characterizing \gls{mchem}, we test a set of synthetic \gls{rf} scenarios (i.e., single- and multi-tap \gls{rf} scenarios) on Colosseum, i.e., scenarios created specifically for the purpose of channel sounding.
These scenarios have been manually generated with specific channel characteristics to validate the behavior of \gls{mchem}, and have been made publicly available for all Colosseum users.
The parameters used in this evaluation are the same as the ones in Table~\ref{table:localconfig} with the only exception of the sample rate that is set at $50$\:MS/s to have a $20$\:ns resolution (thus being able to properly retrieve tap delays and gains), and the \gls{glfsr} code sequence found above.

\begin{figure}[ht]
    \centering
    \includegraphics[width=\columnwidth]{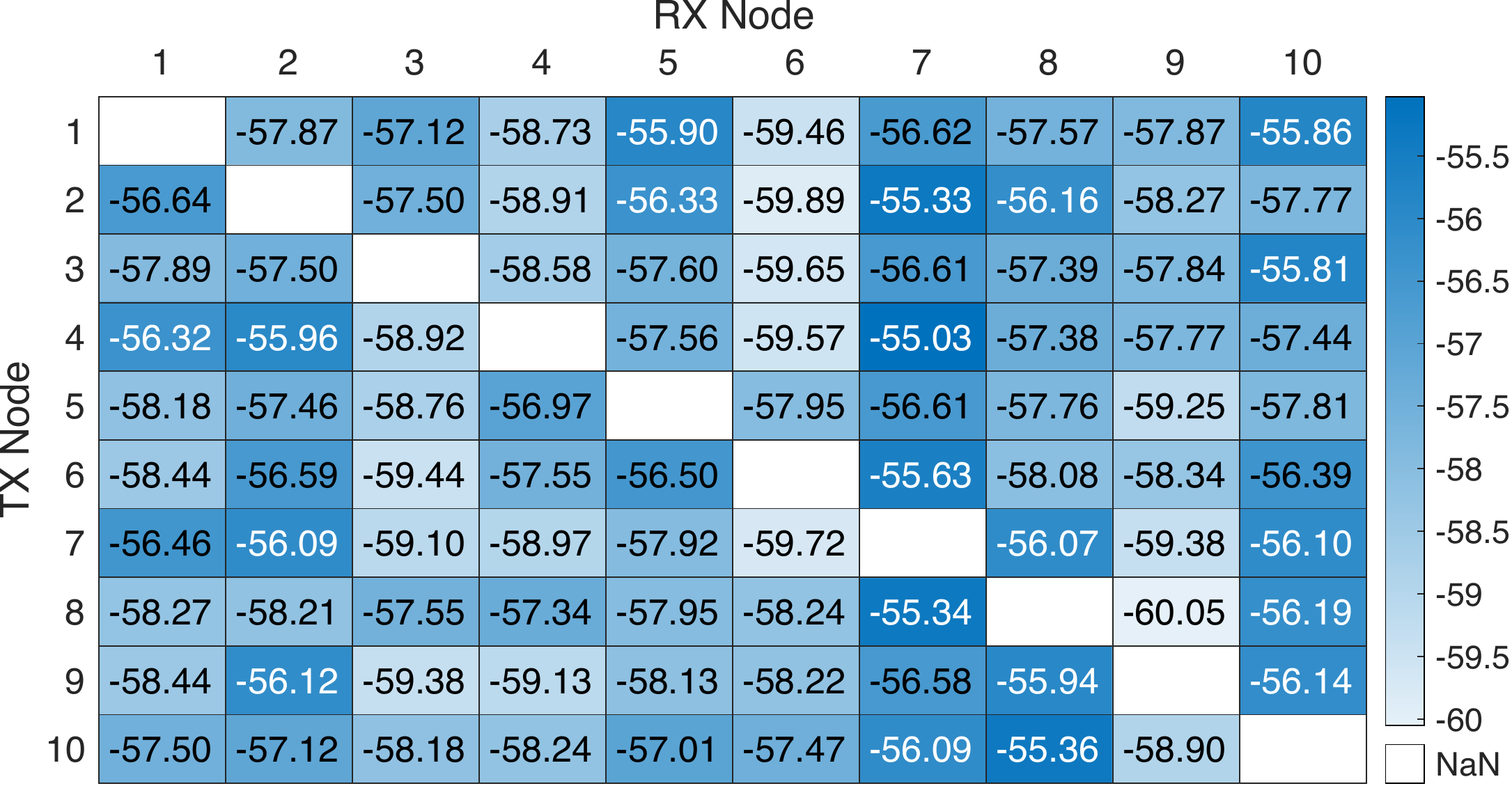}
    \caption{Path loss heatmap as measured by \acrshort{cast} in a $0$\:dB Colosseum \gls{rf} scenario with~10 SRNs.}
    \label{fig:heatmap0db}
\end{figure}

\textbf{Single-tap Scenario.}
The first synthetic \gls{rf} scenario that we consider is a single-tap scenario with nominal $0$\:dB path loss (i.e., $0$\:dB of path loss added to the inherent loss of the hardware components of the testbed).
To find the base loss of \gls{mchem}, i.e., the loss due to Colosseum hardware-in-the-loop, we instantiate \gls{cast} on 10~\glspl{srn}, and sound the channels among them, measuring the path loss of each link, shown in Figure~\ref{fig:heatmap0db}.
%

Each cell in the figure represents the average path loss for $2$\:s of reception time between transmitter (row) and receiver (column) nodes.
%
Results show an average Colosseum base loss of $57.55$\:dB with a \gls{sd} of $1.23$\:dB.
We also observe that the current dynamic range of Colosseum is approximately $43$\:dB, i.e., between the $57.55$\:dB base loss at $1$\:GHz and
the noise floor of $-100$\:dB.
%

\begin{figure}[htbp]
\centering
    \centering
    \includegraphics[width=\columnwidth]{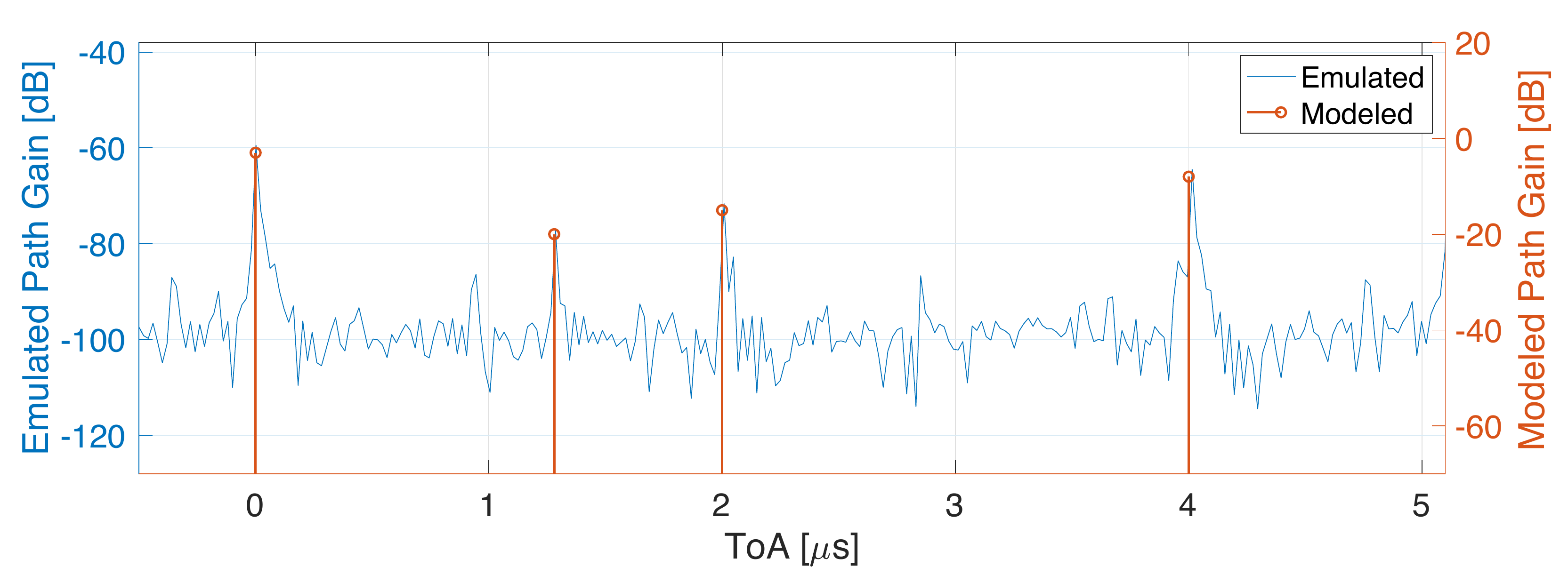}
    \caption{Comparison between emulated and modeled path gains in Colosseum for a single time frame.}
    \label{fig:cirpgcomparison}
\end{figure}

\textbf{Multi-tap Scenario.}
The second synthetic \gls{rf} scenario that we consider is a four-tap scenario in which taps have different delays and path gains.
%
We characterize such a scenario on Colosseum through \cast channel sounding operations.
Results for the emulated and modeled path gains for a single time frame are shown in Figure~\ref{fig:cirpgcomparison} in blue and orange, respectively.

We notice that the \glspl{toa} match between the modeled \gls{cir} and the taps emulated by the Colosseum \gls{rf} scenario, namely they occur at $0$, $1.28$, $2$, and $4\:\mathrm{\mu s}$.
We also notice that the received powers are in line with our expectations.
Indeed, by adding the Colosseum base loss computed in the previous step to the power measured by \gls{cast} (in blue in the figure), we obtain the modeled taps (corresponding to $-3$, $-20$, $-15$, and $-8$\:dB, shown in orange in the figure).

We now analyze the accuracy of the measurements performed with \gls{cast} by computing the relative difference between the emulated taps over time.
We do so by considering $1,500$ time frames.
Results show that the average difference between the strongest tap of each time frame is in the order of $10^{-6}$\:dB, with a \gls{sd} of $0.03$\:dB.
Analogous results occur for the second tap---which is the weakest tap in our modeled \gls{cir}---with a \gls{sd} of $0.17$\:dB, and for the third and fourth taps.
Finally, differences between the first and second taps of each time frame (i.e., between strongest and weakest taps in our modeled \gls{cir}) amount to $0.52$\:dB with a \gls{sd} of $0.18$\:dB.
These results are a direct consequence of the channel noise, which impacts weaker taps more severely.

Overall, results demonstrate \gls{mchem} accuracy in emulating wireless \gls{rf} scenarios in terms of received signal, tap delays, and gains.
This also shows \cast effectiveness in achieving a $20$\:ns resolution, thus sustaining a $50$\:MS/s sample rate, and a tap gain accuracy of $0.5$\:dB, which allows \cast to capture even small differences between the modeled and emulated \gls{cir}.

\subsection{Arena Digital Twin Scenario}
\label{sec:arenadtscen}

\begin{figure}[ht]
    \centering
    \subfloat[Real-world location]{\label{fig:arena-real-loc}\includegraphics[width=0.48\columnwidth]{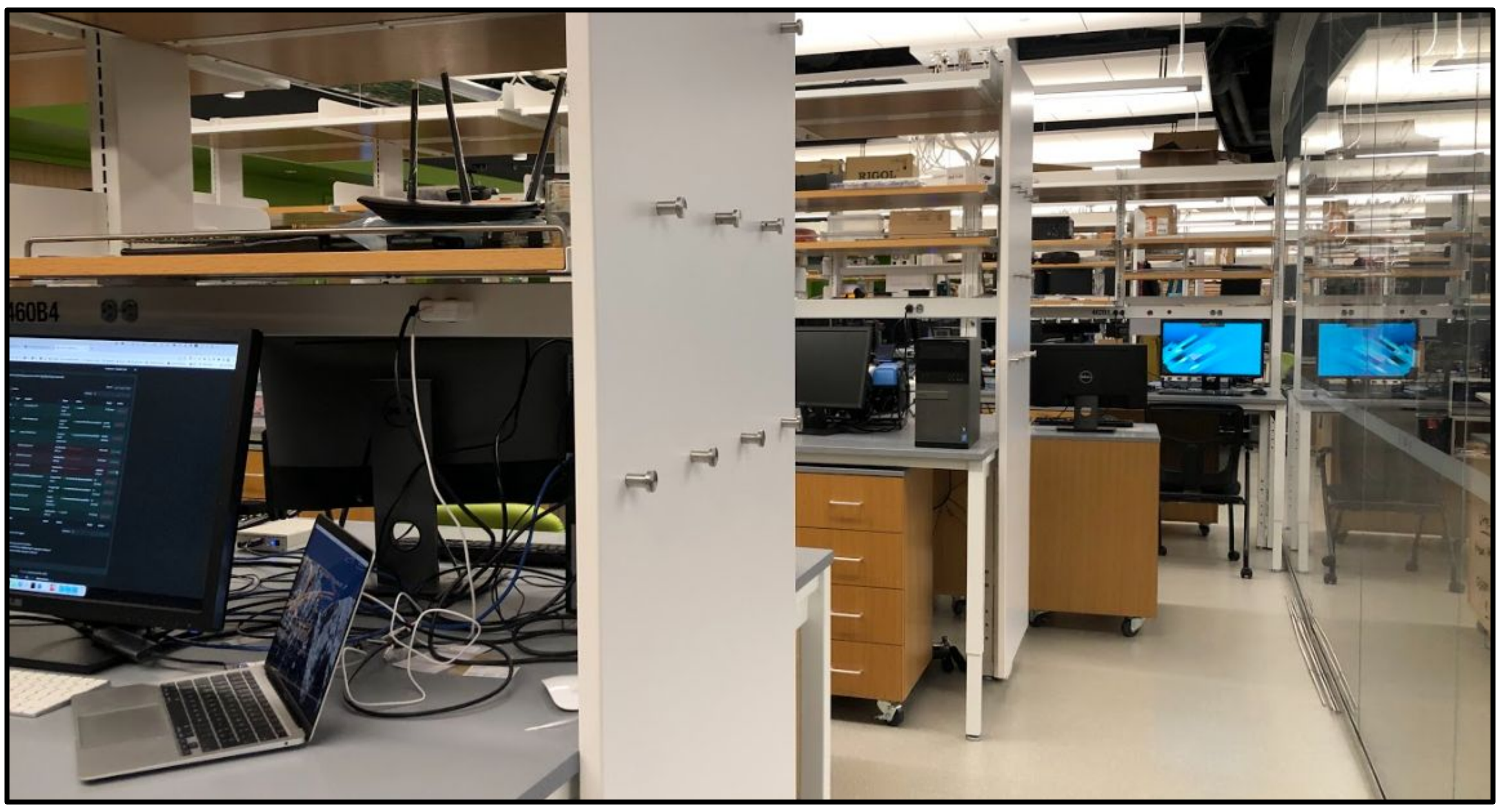}}%
    \hfill
    \subfloat[Digital-twin scenario]{\label{fig:arena-twin-loc}\includegraphics[width=0.48\columnwidth]{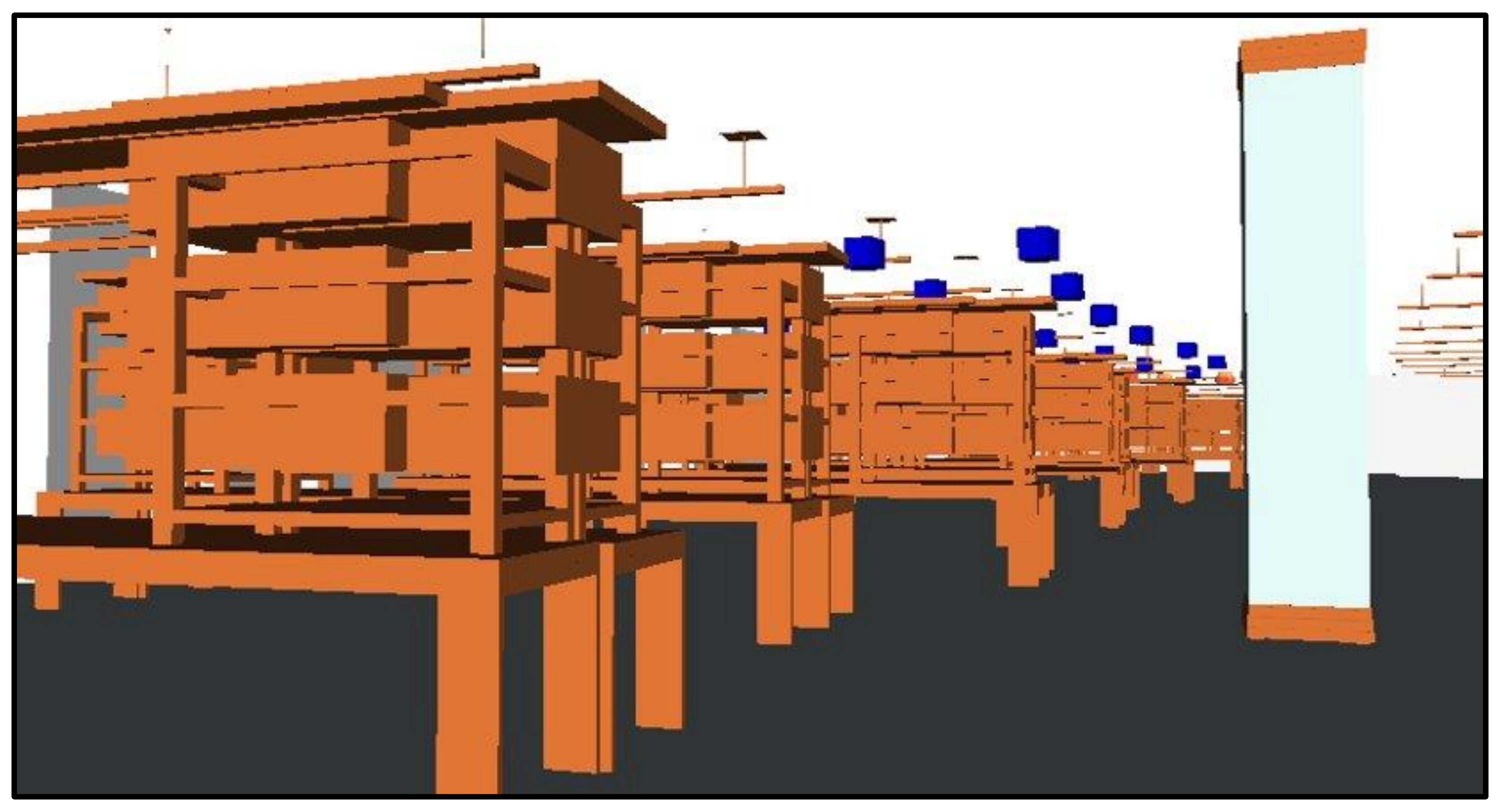}}
    \hfill
    \caption{The \blue{conversion} from a real-world location, into a digital medium scenario used to create the digital twin representation.}
    \label{fig:arena-real-to-dig}
\end{figure}

We use Sketchup~\cite{sketchup} software to create a 3D representation of the Arena testbed.
This software allows users to model a broad range of environments starting from an architectural layout (e.g., of the Arena testbed, a picture of which is shown in Figure~\ref{fig:arena-real-loc}), and with different surface renderings, e.g., glass walls and windows, wooden walls, carpeted floors~\cite{sketchup}.
%
%
%
The resulting 3D model (shown in Figure~\ref{fig:arena-twin-loc}) is then fed to the ray-tracing software, Wireless inSite~\cite{WI} in this case, to create a \gls{dt} scenario on Colosseum following the steps described in Section~\ref{sec:scenario-twinning}.
%

\begin{figure}[ht]
    \centering
    \includegraphics[width=\columnwidth]{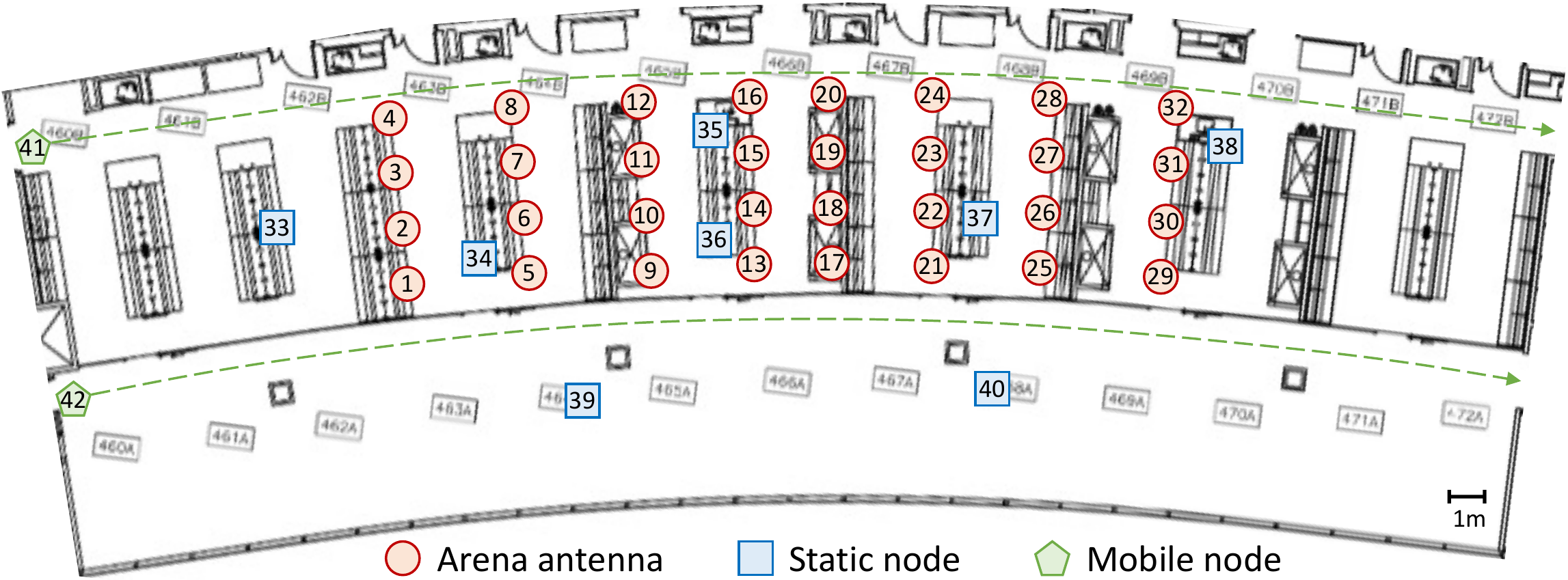}
    \caption{Location of the nodes in an Arena \gls{dt} scenario.}
    \label{fig:dt-arena-nodes}
\end{figure}

For the developed Arena scenario, we model the antenna points of the Arena testbed in 32 locations (one for each antenna pair), as well as 8~static nodes distributed in their surroundings, and 2~mobile nodes traversing the laboratory space at a constant speed of $1.2$\:m/s. The height of the nodes (both static and mobile) is set to $1$\:m, e.g., to emulate handheld devices, or devices lying on table surfaces.
The modeled locations and nodes are shown in Figure~\ref{fig:dt-arena-nodes}, where the red circles represent the antenna pairs of Arena, while the blue squares and green pentagons identify the static and mobile nodes, respectively.
The dashed green arrows denote the movement direction of the mobile nodes.
By using a Dell T630 machine with 2 Xeon E52660 $14$\:cores CPU, $128$\:GB RAM, and Tesla K40 GPU, the ray-tracing operations of this scenario took $14$\:h and $21$\:m, while the channel approximation process $2$\:h and $33$\:m.
Additionally, the installation process in Colosseum required around $19$\:h and $30$\:m by leveraging a virtual machine hosted on a Dell PowerEdge M630 Server with 24 CPU cores and $96$\:GB of RAM.
It is worth noting that these are one-time operations and the scenario is played on-demand right away afterward.

\begin{figure}[ht]
    \vspace{-10pt}
    \centering
    \includegraphics[width=\columnwidth]{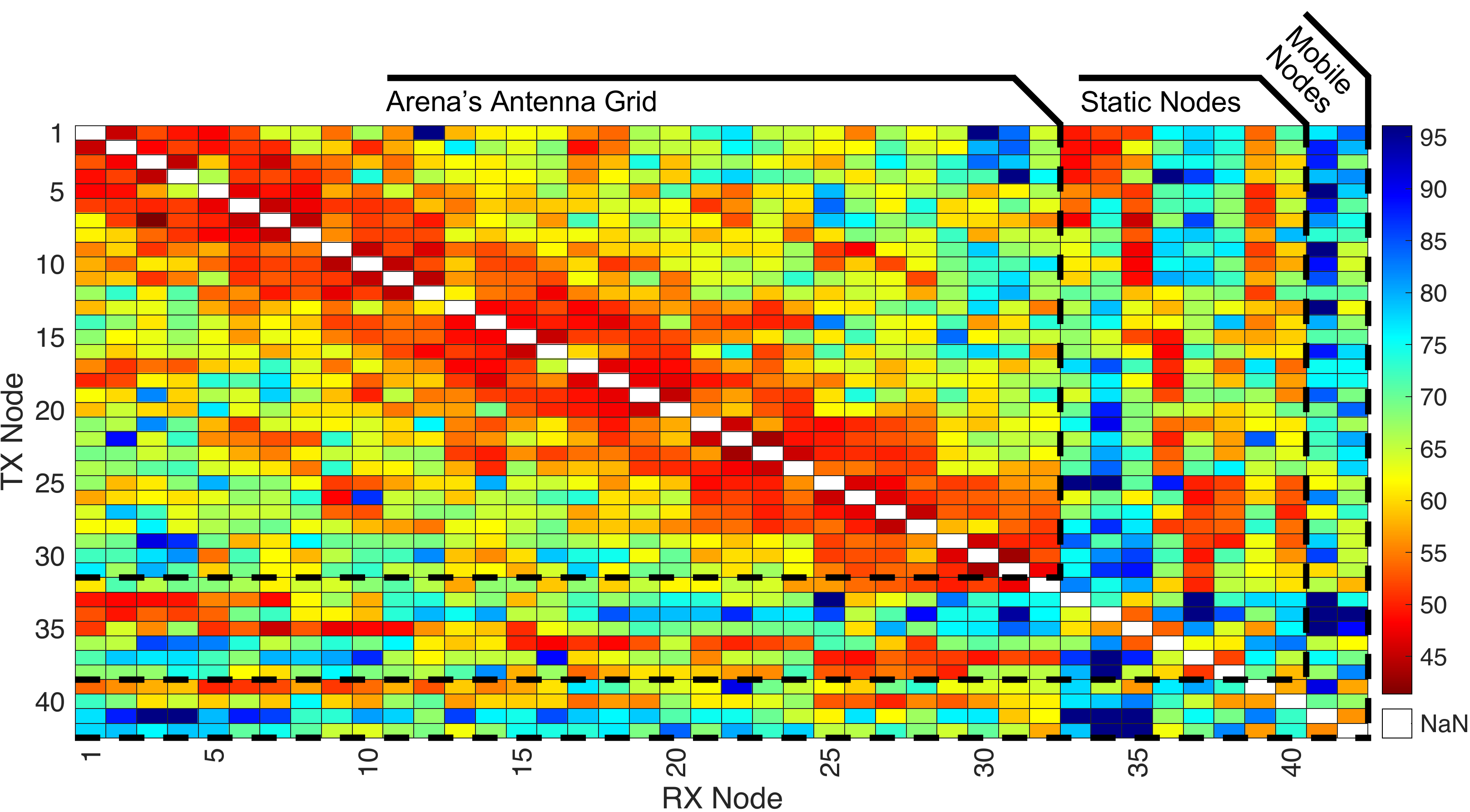}
    \caption{Heat map of the path loss among the nodes of Figure~\ref{fig:dt-arena-nodes}, with a line separator between antenna, static, and mobile. The mobile nodes are considered in the starting position on the left.}
    \label{fig:dt-arena-heatmap}
\end{figure}

Figure~\ref{fig:dt-arena-heatmap} shows the heat map of the path loss among the transmit-receive node pairs (the mobile nodes are considered in the starting position on the left).
As expected, closer nodes experience a lower path loss, which increases with the distance between the nodes.
A similar trend is also visible for the static nodes, even though this is less noticeable due to their scattered locations.
On the other hand, due to their remote starting locations on the side of the room, the mobile nodes exhibit a very high path loss against all nodes, as depicted in Figure~\ref{fig:dt-arena-heatmap}. These path losses decrease as they get closer to each node on their path, and increase again while reaching their end locations on the other side of the room.
%

\subsection{Experimental Use Cases}
\label{sec:usecases}

In this section, we show outcomes of relevant experimental use cases run on both the Arena testbed, as well as on its \gls{dt} representation.
The first use case involves the deployment of a cellular networking system using the srsRAN software suite, while the second one encompasses a Wi-Fi adversarial jamming use case built on GNU Radio.

\subsubsection{\blue{Cellular Networking}}
\begin{figure}[ht]
    \centering
    \includegraphics[width=\columnwidth]{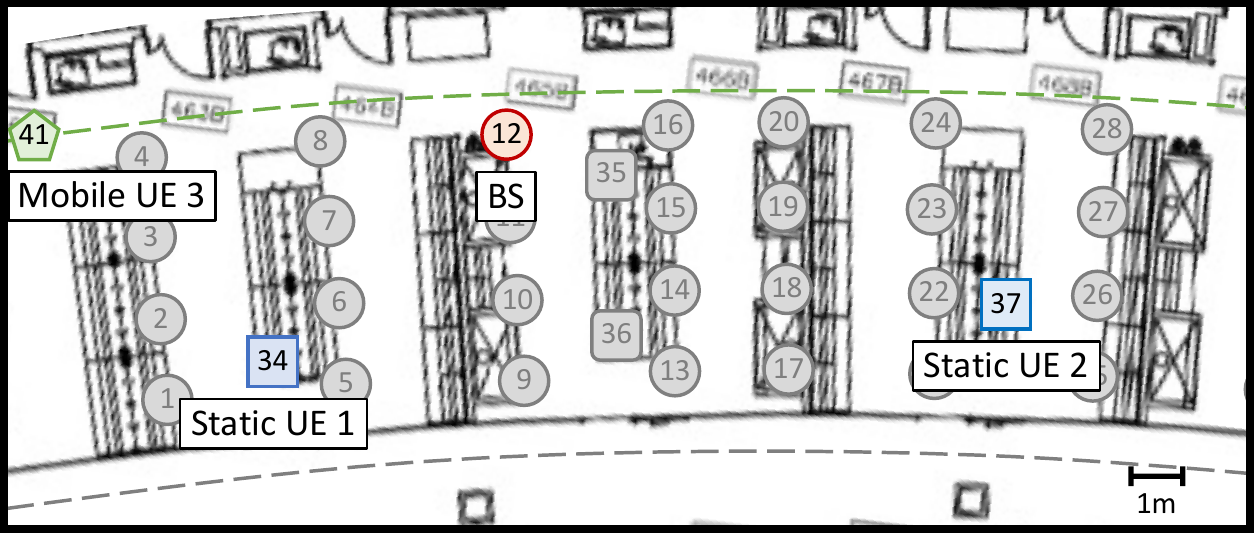}
    \caption{Location of the nodes in the cellular experiment.}
    \label{fig:cellular-nodes-singlebs}
\end{figure}

%
\blue{\textbf{Single \acrshort{bs}.}} In the first cellular networking use case, we leverage SCOPE~\cite{bonati2021scope}---an open-source framework based on srsRAN~\cite{gomez2016srslte} for experimentation of cellular networking technologies---to deploy a twinned \gls{ran} protocol stack with one \gls{bs} and three \glspl{ue} in the Arena over-the-air testbed and in the Colosseum emulation system.
The same node positions, shown in Figure~\ref{fig:cellular-nodes-singlebs}, are used in the two platforms: the \gls{bs}, which transmits over a $10$\:MHz spectrum, is located on node~12, two static \glspl{ue} on nodes~34 and~37, and one mobile \gls{ue} on node~41.
In Arena, \glspl{ue} are implemented through commercial smartphones (Xiaomi Redmi Go), while on Colosseum, they are deployed on the \glspl{sdr} of the testbed.

We conduct two experiments on each system: the first one involves a downlink \gls{udp} traffic sent at a $5$\:Mbps rate; the second one \gls{tcp} downlink traffic.
The traffic generation for each experiment is achieved using iPerf, a benchmarking tool designed for assessing the performance of IP networks~\cite{iperf}.
The following results show the average of at least 5 separate experiment realizations.
%

%
%

\begin{figure}[h]
    \centering
    \subfloat[Arena]{\label{fig:cellular-arena-results-udp}\includegraphics[width=0.49\columnwidth]{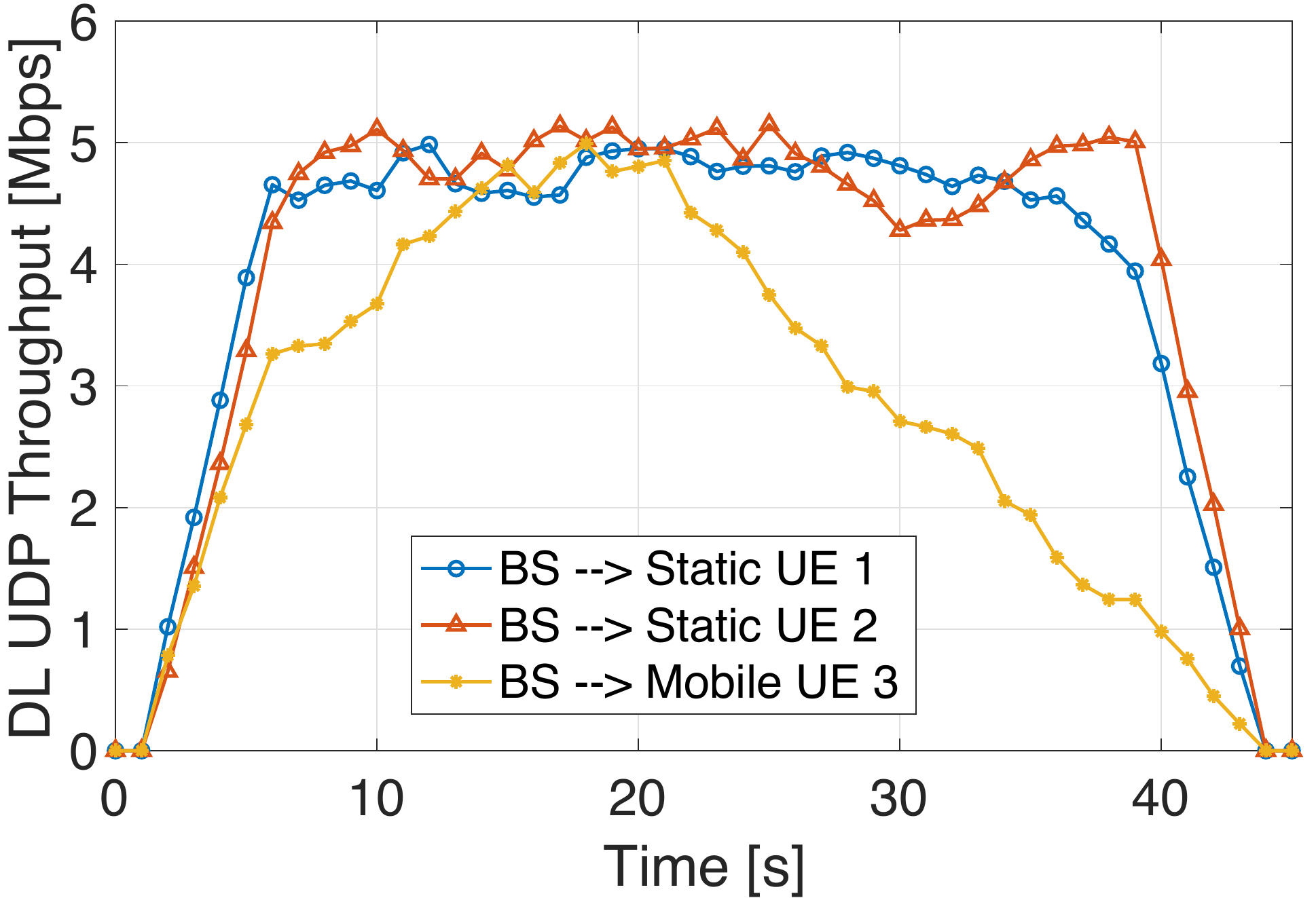}}
    \hfill
    \subfloat[Colosseum]{\label{fig:cellular-colosseum-results-udp}\includegraphics[width=0.49\columnwidth]{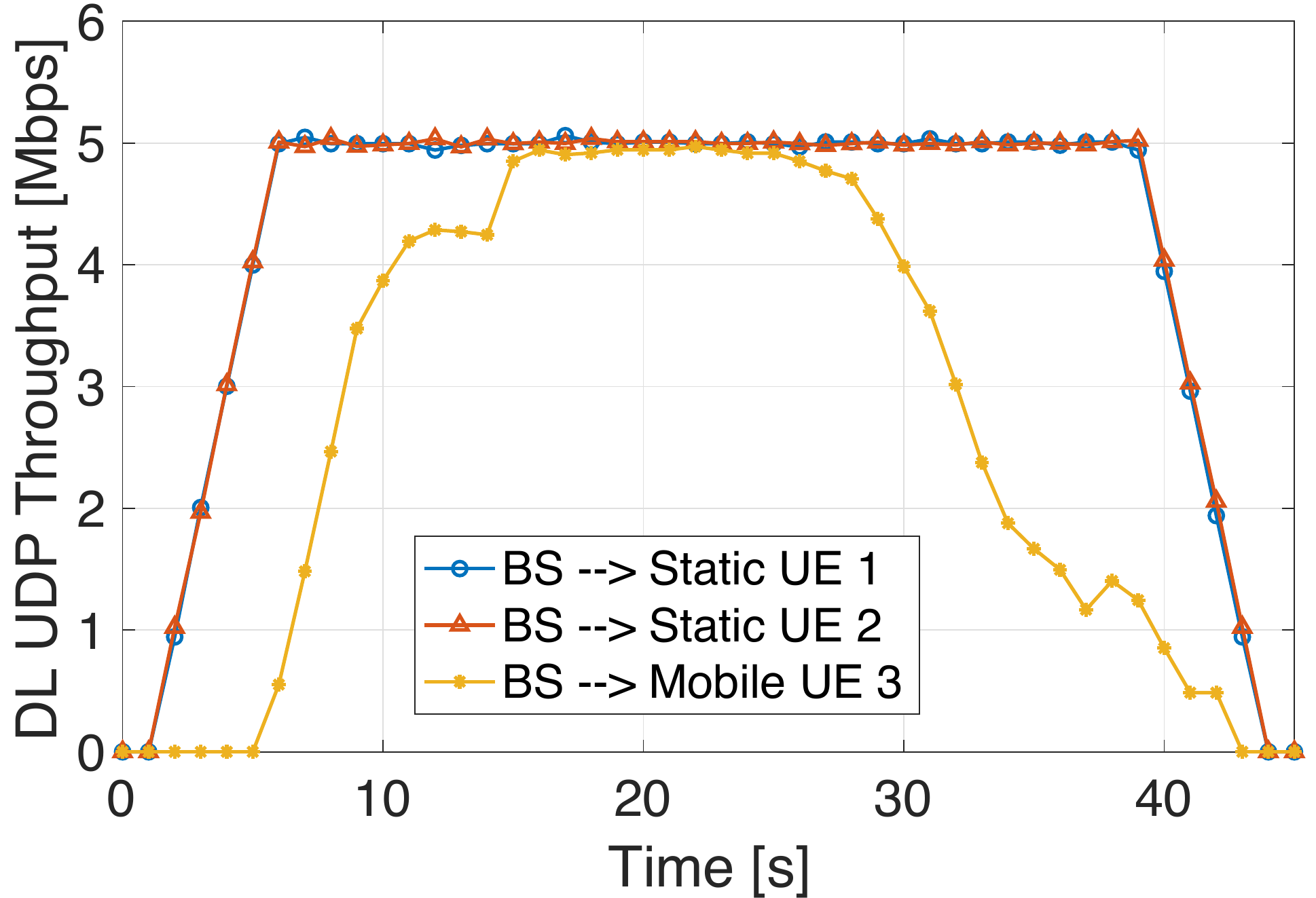}}
    \hfill
    \caption{\gls{udp} downlink throughput of the cellular use case on the Arena and Colosseum testbeds.}
    \label{fig:cellular-results-udp}
\end{figure}

Figure~\ref{fig:cellular-results-udp} shows the \gls{udp} downlink throughput for static (blue and orange lines), and mobile (yellow line) nodes on the Arena (Figure~\ref{fig:cellular-arena-results-udp}) and Colosseum (Figure~\ref{fig:cellular-colosseum-results-udp}) testbeds. 
We can notice similar trends and patterns exhibited on both testbeds. 
Specifically, the throughput of the static nodes remains stable around $5$\:Mbps in both Colosseum and Arena, where we notice a less stable behavior due to the use of over-the-air communications, and potential external interference.
%
%
As expected, the throughput of the mobile node---that starts from the top-left location shown in Figure~\ref{fig:cellular-nodes-singlebs} and travels to the right along the trajectory depicted with the green line in the figure---increases as the node gets closer to the \gls{bs} (where it reaches a $5$\:Mbps peak), and then decreases as the node gets farther away.

\begin{figure}[h]
    \centering
    \subfloat[Arena]{\label{fig:cellular-arena-results-tcp}\includegraphics[width=0.49\columnwidth]{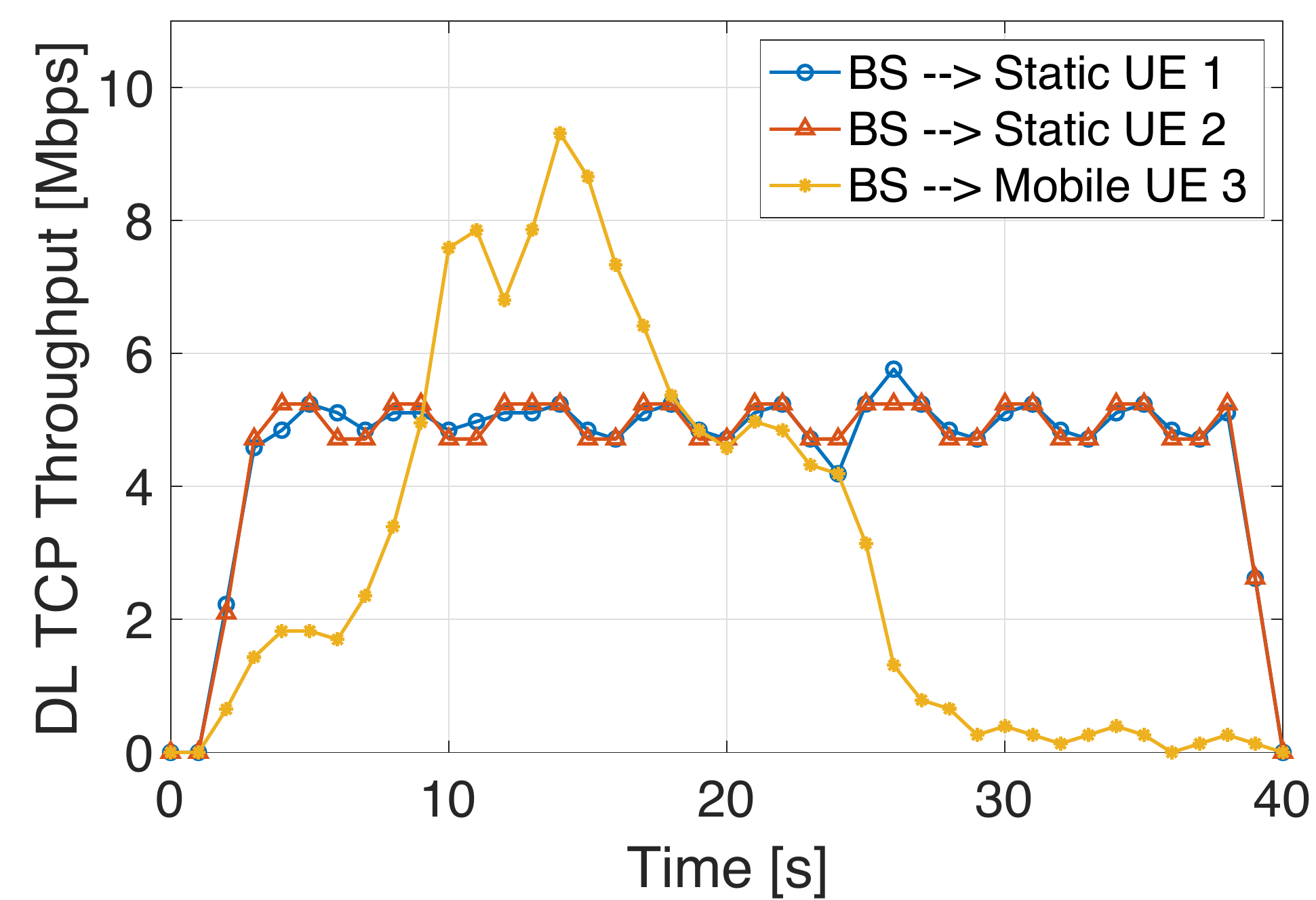}}
    \hfill
    \subfloat[Colosseum]{\label{fig:cellular-colosseum-results-tcp}\includegraphics[width=0.49\columnwidth]{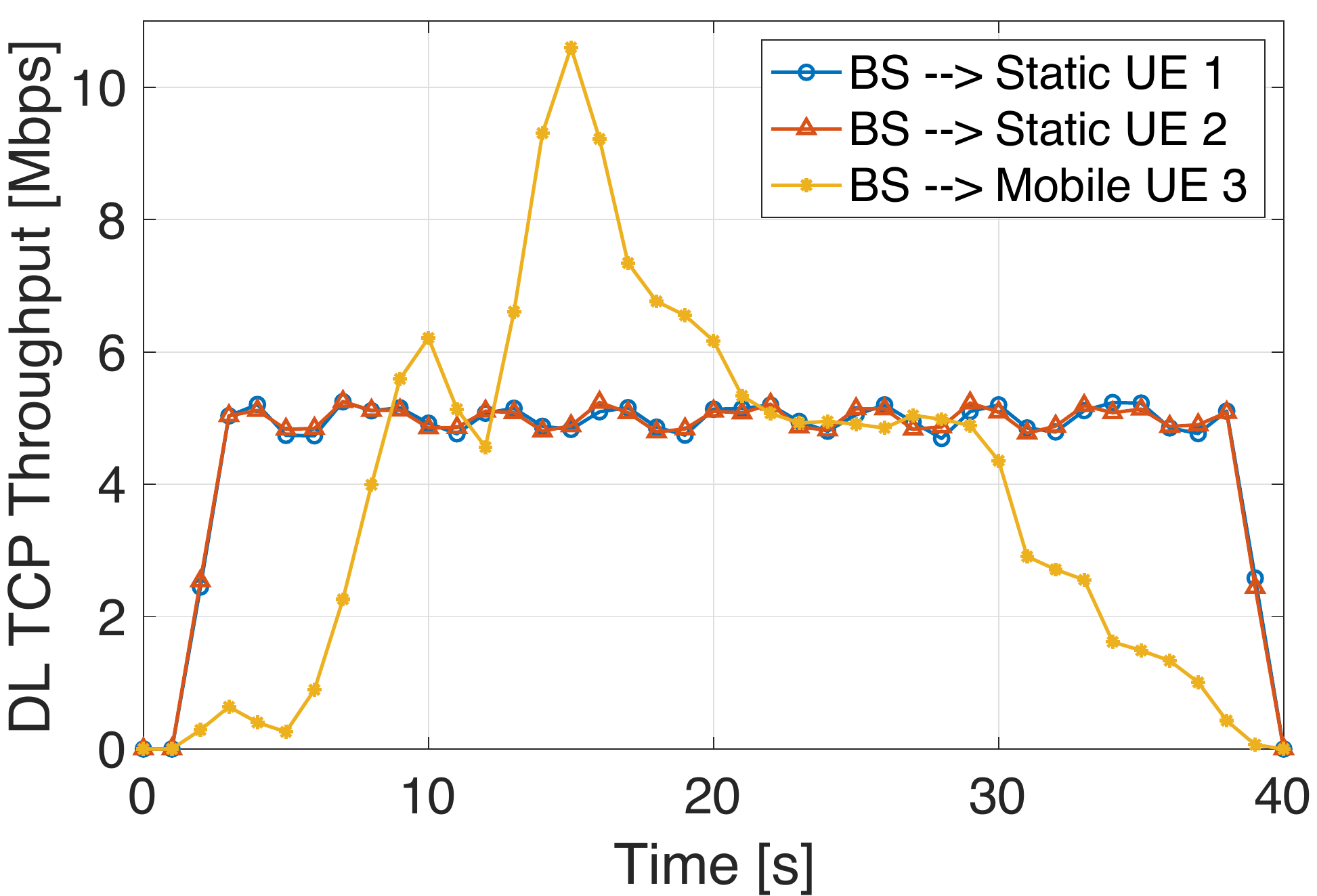}}
    \hfill
    \caption{\gls{tcp} downlink throughput of the cellular use case on the Arena and Colosseum testbeds.}
    \label{fig:cellular-results-tcp}
\end{figure}

\begin{figure}[h]
    \centering
    \subfloat[Arena]{\label{fig:sinr-cellular-arena-results-tcp}\includegraphics[width=0.49\columnwidth]{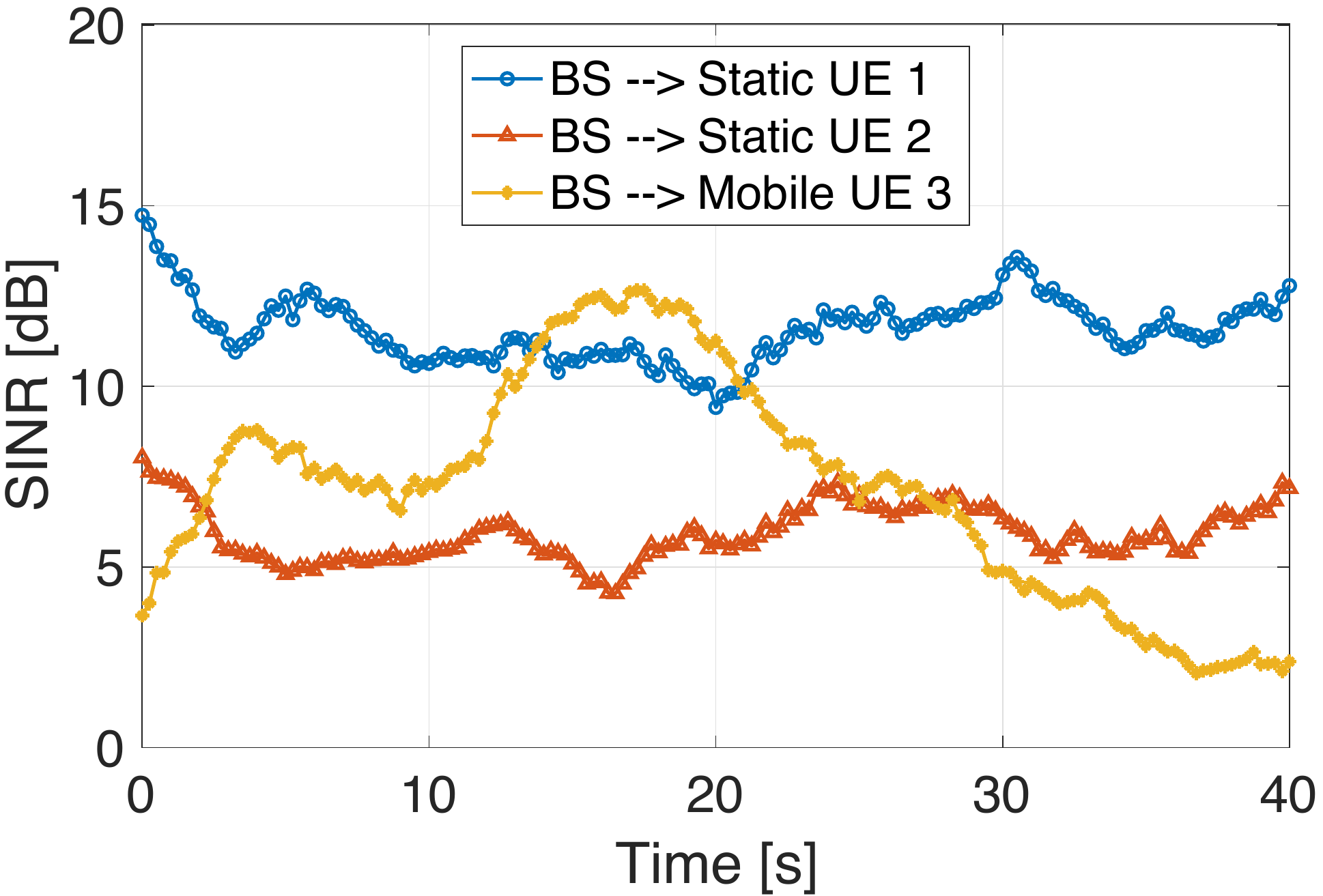}}
    \hfill
    \subfloat[Colosseum]{\label{fig:sinr-cellular-colosseum-results-tcp}\includegraphics[width=0.49\columnwidth]{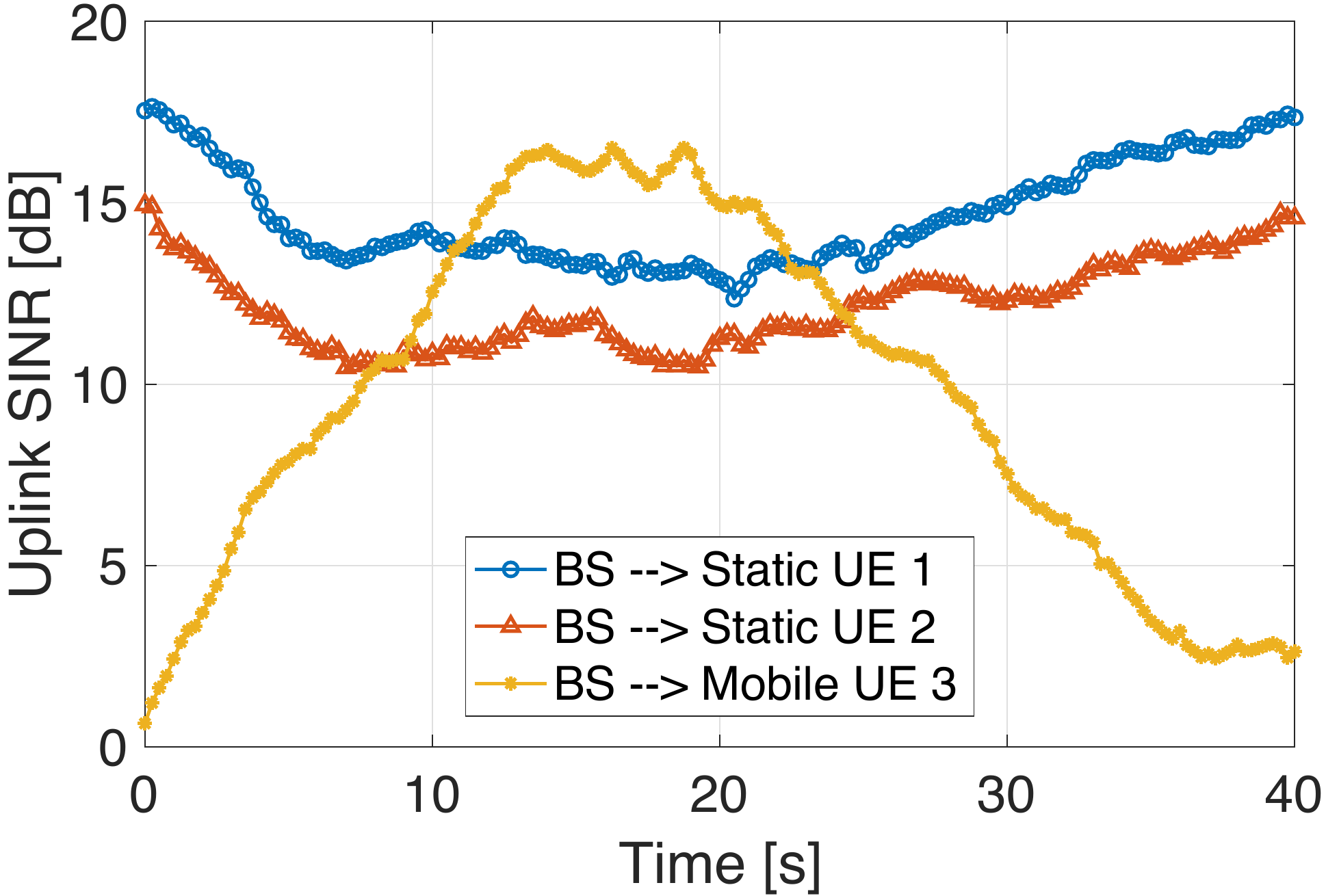}}
    \hfill
    \caption{\gls{tcp} \gls{sinr} of the cellular use case on the Arena and Colosseum testbeds.}
    \label{fig:sinr-cellular-results-tcp}
\end{figure}

Figure~\ref{fig:cellular-results-tcp} and Figure~\ref{fig:sinr-cellular-results-tcp} plot the \gls{tcp} downlink throughput and \gls{sinr} results of the second experiment for static (blue and orange) and mobile (yellow) nodes on Arena (Figure~\ref{fig:cellular-arena-results-tcp} and Figure~\ref{fig:sinr-cellular-arena-results-tcp}) and Colosseum (Figure~\ref{fig:cellular-colosseum-results-tcp} and Figure~\ref{fig:sinr-cellular-colosseum-results-tcp}).
%
Also in this use case, a similar pattern is clearly noticeable.
In particular, the two static nodes maintain a relatively stable throughput of the nominal $5$\:Mbps \gls{tcp} traffic on both testbeds with no apparent impact during the passage of the mobile \gls{ue} close to the static \glspl{ue}, which in the Arena case is moved manually.
This behavior is visible in the \gls{sinr} results, which show a decrease due to the created interference when the mobile node approaches each of the static \glspl{ue}.
The mobile node exhibits a similar trend with two high peaks in throughput
on both systems.
These peaks can be attributed to the \gls{tcp} protocol, which retransmits data to ensure delivery in case of packet failures as soon as the signal improves, resulting in higher application throughput values compared to the nominal $5$\:Mbps.
Specifically, each peak corresponds to the time right after the mobile device transitions close to static node 1 (around time $10$\:s), and then to static node 2 (around time $15$\:s).
Due to the increased interference, the mobile node loses more packets, which will eventually be retransmitted.
%

\begin{figure}[ht]
    \centering
    \includegraphics[width=\columnwidth]{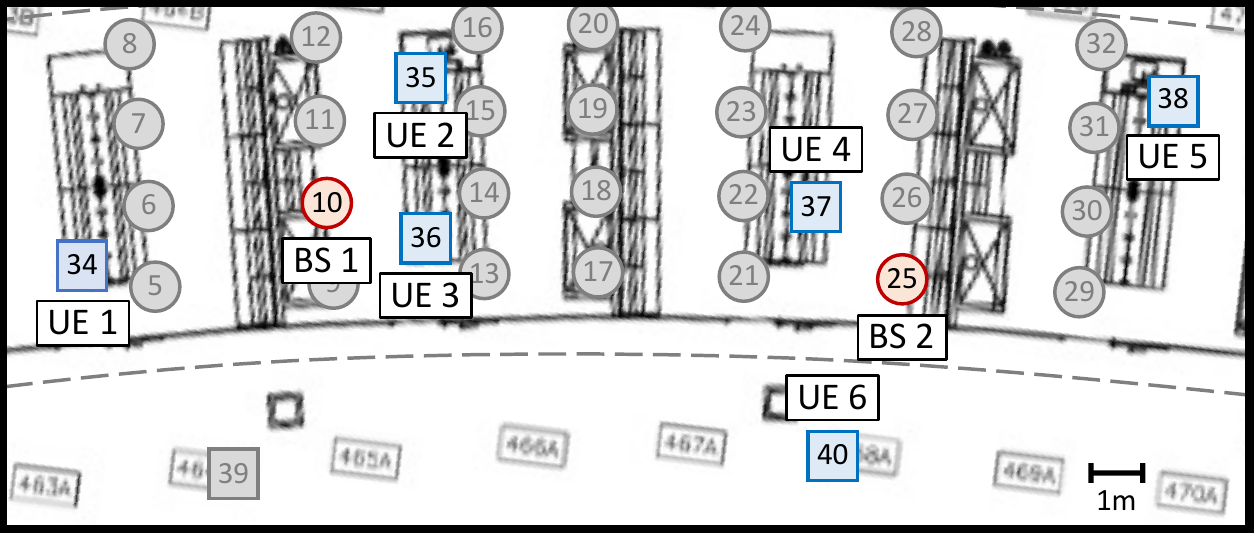}
    \caption{\blue{Location of the nodes in the multiple \glspl{bs} cellular use-case, consisting of 2 \glspl{bs} (BS 1, BS 2), and 6 static \glspl{ue}, equally assigned to each \gls{bs}: UE 1, UE 2, and UE 3 for BS 1; UE 4, UE 5, and UE 6 for BS 2.}}
    \label{fig:cellular-nodes-multiplebs}
\end{figure}

\blue{\textbf{Multiple \glspl{bs}.} In the second cellular networking use case, we use a similar setup as for the first experiment by leveraging SCOPE and Xiaomi Redmi Go phones to deploy a twinned srsRAN protocol stack with two \glspl{bs} located in positions 10 (BS 1) and 25 (BS 2), as shown in Figure~\ref{fig:cellular-nodes-multiplebs}.
We assign three \glspl{ue} to each \gls{bs}, for a total of 6~\glspl{ue} (i.e., the number of smartphones currently at our disposal).
%
Specifically, UE 1, UE 2, and UE 3 are assigned to BS 1, while UE 4, UE 5, and UE 6 are assigned to BS 2.}

\begin{figure}[ht]
    \centering
    \includegraphics[width=.99\columnwidth]{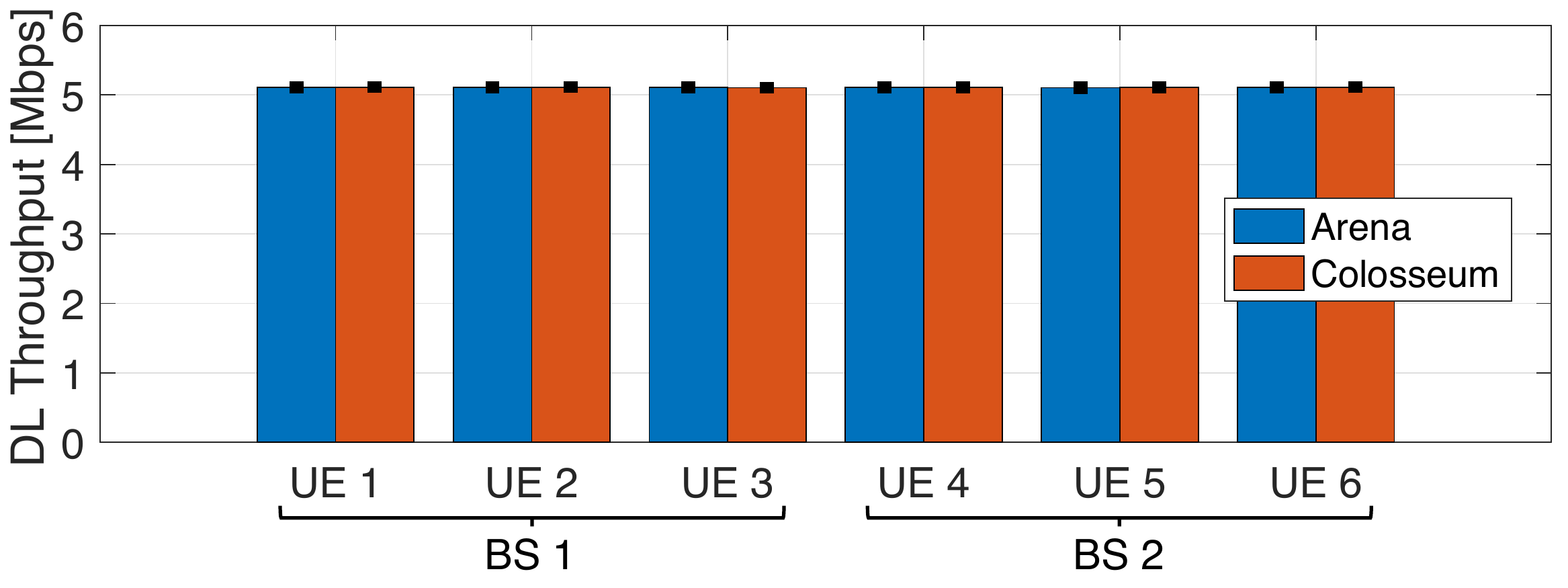}
    \caption{\blue{Downlink throughput average results and 95\% confidence interval of the second cellular networking use-case with 2 \glspl{bs} and 6 \glspl{ue}.}}
    \label{fig:th-cellular-multiplebs}
\end{figure}

\blue{Once each \gls{ue} has completed the attachment procedures, we initiate a continuous $5$\:Mbps UDP downlink traffic stream from each smartphone using iPerf towards its corresponding \gls{bs}. 
Figure~\ref{fig:th-cellular-multiplebs} shows the average downlink throughput results over a period of more than $30$\:minutes of data collection, along with 95\% confidence intervals.
These metrics are collected at the data-link layer at the \gls{bs} level through the SCOPE framework.
We observe that all \glspl{ue} are able to consistently ensure the $5$\:Mbps throughput with a very low 95\% confidence interval showing an average margin of error of $0.02$\:Mbps.
This demonstrates the accuracy of the digital twin representation, even in the presence of multiple \glspl{bs} and \glspl{ue}, showcasing the scalability of our \gls{dtmn}.
}

\textbf{\blue{Comparison.}} By analyzing the results after the \glspl{ue} have completed the attachment procedures, \blue{we can assess the similarity between the two testbeds by following}
\begin{equation}
    \label{eq:xcorr}
    \blue{
    \rho(k) = \frac{\sum_{n=1}^{N} (x(n) - \bar{x})(y(n+k) - \bar{y})}{\sqrt{\sum_{n=1}^{N} (x(n) - \bar{x})^2 \sum_{n=1}^{N} (y(n) - \bar{y})^2}},
    }
\end{equation}
\blue{which measures the normalized cross-correlation $\rho(k)$ between the Arena data sequence, denoted as $x(n)$, and shifted lagged copies of the Colosseum data sequence, denoted as $y(n)$, as a function of the lag $k$.
In this context, $\bar{x}$ and $\bar{y}$ represent the means of the input sequences and $N$ is their length.
If $x(n)$ and $y(n)$ have different lengths, zeros are appended to the end of the shorter vector to ensure both sequences have the same length $N$.}

\blue{Table~\ref{table:cellular-corr-results} presents the normalized cross-correlation results and their averages for throughput and \gls{sinr} of each \gls{ue}, considering the maximum values between 10~lags, of the \gls{udp} and \gls{tcp} \blue{single \gls{bs}} use-case experiments.
The multiple \glspl{bs} experiments would yield similar comparison results.}
\begin{table}[ht]
    \centering
    \footnotesize
    \caption{Normalized cross-correlation results, and their averages, considering the maximum values between 10 lags for the single \gls{bs} cellular experiment.}
    \label{table:cellular-corr-results}
    
    \begin{tabularx}{\columnwidth}{
        >{\raggedright\arraybackslash\hsize=1.3\hsize}X
        >{\raggedright\arraybackslash\hsize=0.5\hsize}X
        >{\raggedright\arraybackslash\hsize=0.5\hsize}X
        >{\raggedright\arraybackslash\hsize=0.5\hsize}X
        >{\raggedright\arraybackslash\hsize=0.5\hsize}X}
        \toprule
        Metric & Static\newline UE 1 & Static\newline UE 2 & Mobile\newline UE 3 & Average \\
        \midrule
        UDP Throughput & 0.986 & 0.998 & 0.999 & 0.994 \\
        TCP Throughput & 0.937 & 0.998 & 0.998 & 0.978 \\
        TCP SINR & 0.997 & 0.994 &	0.982 &	0.991 \\
        \bottomrule
    \end{tabularx}
\end{table}

\noindent
We can observe a very high similarity between the two testbeds in all use-case experiments, with individual \gls{ue} values consistently above 0.93 and an average exceeding 0.97 for each use-case metric.
It is worth noting, as observed in Figure~\ref{fig:sinr-cellular-results-tcp}, that the \gls{sinr} results in Arena are quite similar from \gls{ue} 1 and \gls{ue} 3, and they present a fixed difference of about $5$\:dB compared to Colosseum for \gls{ue} 2.
%
%
This difference can mainly be attributed to the additional and uncontrolled interference and impairments of a real-world \gls{rf} environment, as well as the different power levels between a simulated \gls{ue} and a real smartphone.
However, these variances can be compensated for in the \gls{dt} by adjusting factors such as the node gains at the transmitter and receiver, as well as by adding stochastically representative interference models to the channel that represents the real-world behavior more closely.
%
%
These findings confirm the capabilities of the \gls{dt} to perform emulated cellular experiments that closely replicate the behavior of real-world setups and environments, even in the presence of mobile nodes.

\subsubsection{\blue{Wi-Fi Jamming}}
Adversarial jamming has continuously plagued the wireless spectrum over the years with the ability to disrupt, or fully halt, communications between parties.
%
%
While there are potential solutions to specific types of jamming, due to the open nature of wireless communication, this kind of attack continues to find ways to be effective.
However, the development of new techniques to counter this attack is not always straightforward, as even experimenting with possible solutions requires complying with strict \gls{fcc} regulations~\cite{noauthor_jammer_2011}.
%
%
Even though some environments allow for jamming research, e.g., anechoic chambers or Faraday cages, these setups can hardly capture the characteristics and scale of real-world network deployments.
To bridge this gap, a \gls{dt} environment---such as the Colosseum wireless network emulator---could be fundamental in further developing techniques for jamming mitigation research 
as shown in our previous work in~\cite{robinson2023esword} where we implement jamming software within Colosseum to test the impact that jamming signals have within a cellular scenario as well as compare real-world and \gls{dt} throughput results.

\begin{figure}[ht]
    \centering
    \includegraphics[width=\columnwidth]{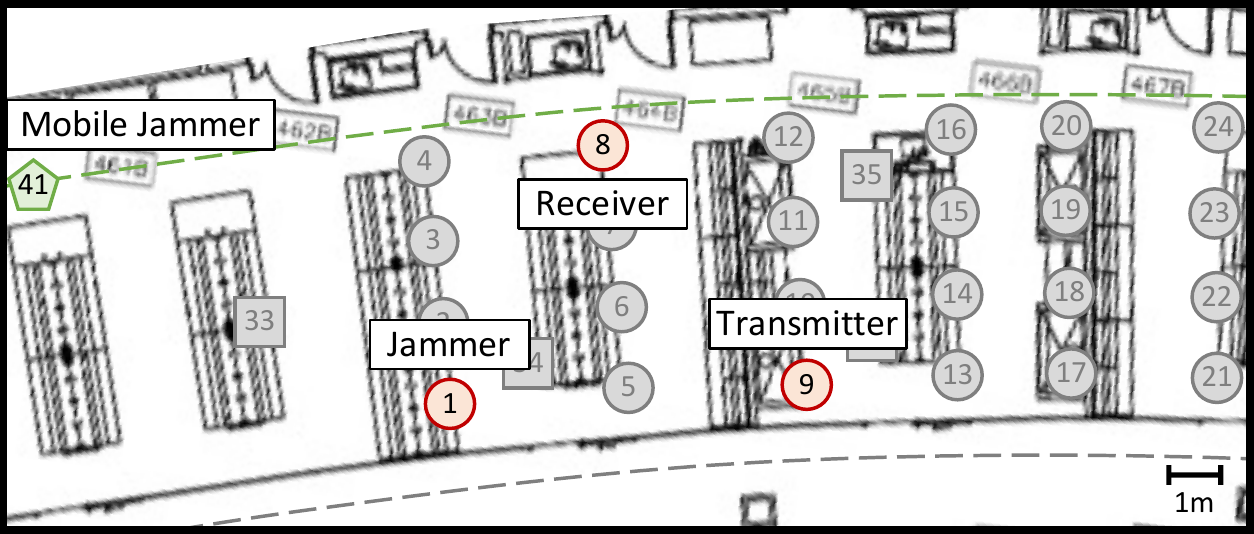}
    \caption{Location of the nodes in the jamming experiment, consisting of three static (1, 8, 9) and one mobile (41).}
    \label{fig:jamming-map}
\end{figure}

\begin{figure}[t]
    \centering
    \subfloat[Narrowband]{\label{fig:jamming-narrowband-static-results}\includegraphics[width=0.49\columnwidth]{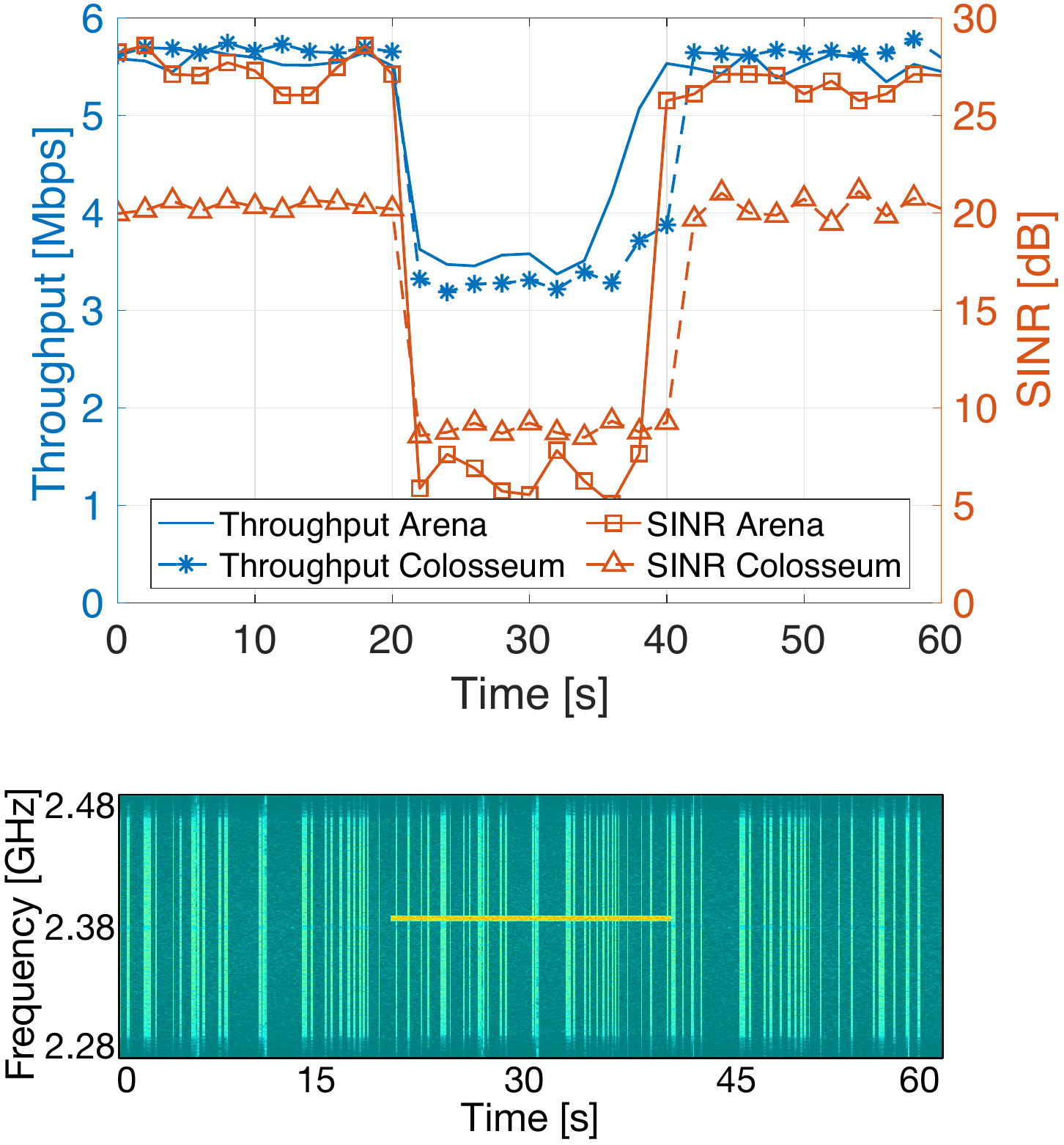}}
    \hfill
    \subfloat[Wideband]{\label{fig:jamming-wideband-static-results}\includegraphics[width=0.49\columnwidth]{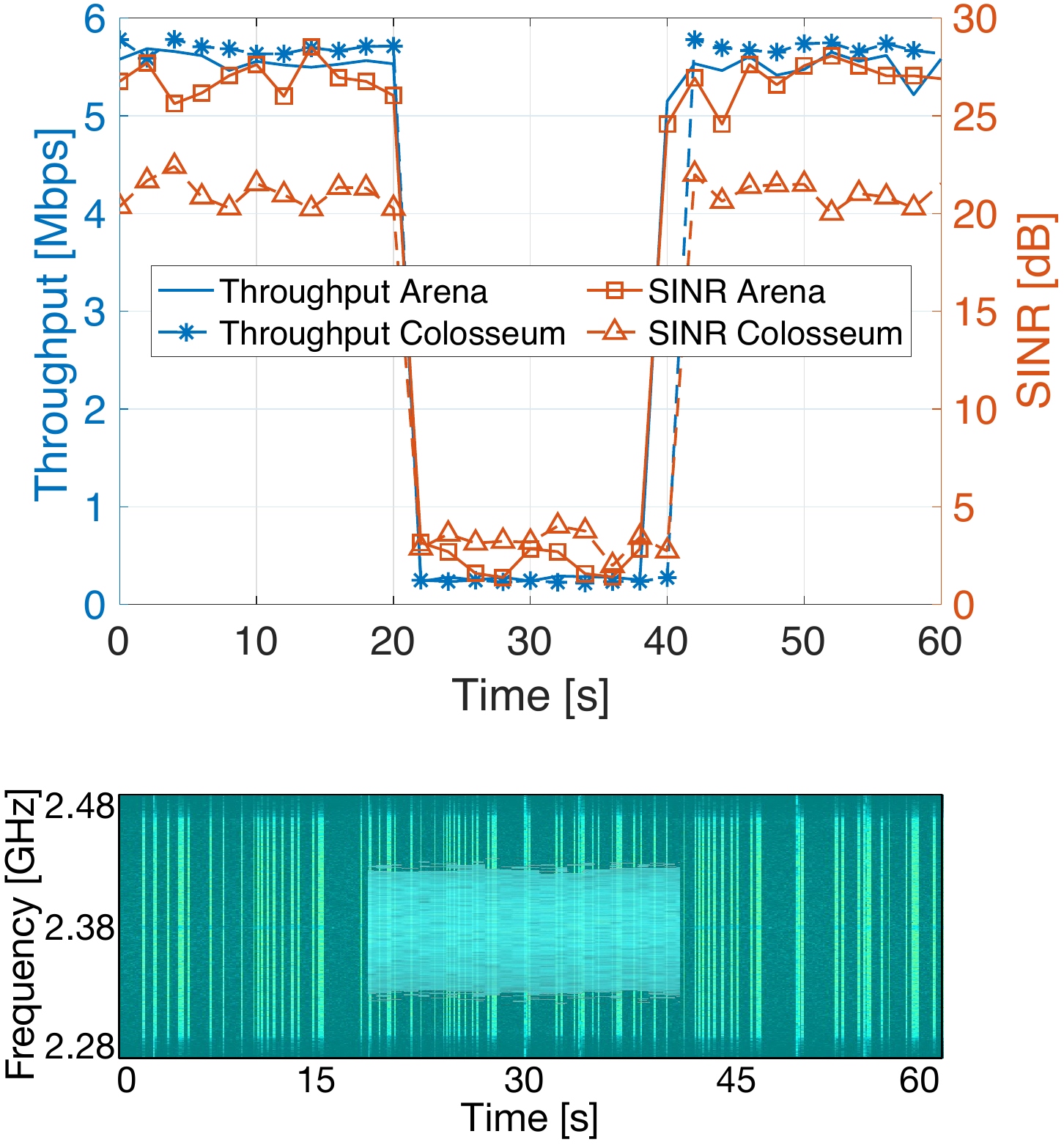}}
    \hfill
    \caption{Throughput and \gls{sinr} results on the Arena and Colosseum testbeds of the jamming experiments for the narrowband and wideband use cases. The spectrogram is shown for both forms of jamming, showing the wideband and narrowband signals over a channel.} 
    \label{fig:jamming-static-results}
    \vspace{-10pt}
\end{figure}

Here, we leverage the GNU Radio-based IEEE 802.11 implementation~\cite{bloessl_ieee_2022} to deploy two Wi-Fi nodes (\gls{tx} and \gls{rx}) communicating over a $20$\:MHz spectrum on the Arena testbed~\cite{bertizzolo2020comnet}.
Additionally, we leverage GNU Radio to deploy a jammer (both stationary and mobile) that transmits Gaussian noise signals to hamper the correct functioning of our Wi-Fi network.
Our setup can be seen in Figure~\ref{fig:jamming-map}.
For the sake of fairness in the transmitted signals, in the stationary case, we deployed our nodes so that the Wi-Fi transmitter and jammer are at the same distance from the Wi-Fi receiver.
We consider two common forms of static jamming: (i)~jamming through narrowband signals (shown in Figure~\ref{fig:jamming-narrowband-static-results}); and (ii)~jamming through wideband signals (Figure~\ref{fig:jamming-wideband-static-results}).

The first type of jamming only occupies a small portion of the Wi-Fi bandwidth (i.e., $\sim$$156$\:kHz), resulting in a minimal displacement of the Wi-Fi signals.
On the contrary, the latter covers half of the spectrum used by the Wi-Fi nodes (i.e., $10$\:MHz), causing larger disruptions in the network.

\blue{\textbf{Static Jamming.}} Figure~\ref{fig:jamming-static-results} evaluates how narrowband and wideband stationary jammers impact the throughput and \gls{sinr} of a Wi-Fi network in the real and \gls{dt}-based scenarios.
%
%
In this experiment, the Wi-Fi nodes communicate for $60$ seconds, and the jammer starts transmitting at second $20$ for a duration of $20$ seconds.
Specifically, Figure~\ref{fig:jamming-narrowband-static-results} shows Wi-Fi throughput and \gls{sinr} for the narrowband jamming experiment in both the real-world and \gls{dt}, while the wideband jamming experiment throughput and \gls{sinr} results as perceived by the Wi-Fi nodes are shown in Figure~\ref{fig:jamming-wideband-static-results}.

By looking at the narrowband jamming case, we notice that in the real-world experiment, the Wi-Fi throughput achieves between $5$ and $6$\:Mbit/s when there is no jammer (Figure~\ref{fig:jamming-narrowband-static-results}).
Once the jammer starts (at second 20), we notice a rapid decrease in the throughput (i.e., between 37\% and 43\% decrease).
%
The wideband jammer (Figure~\ref{fig:jamming-wideband-static-results}), instead, has a more severe impact on the Wi-Fi throughput, causing a performance drop between 94\% and 96\% (with the throughput achieving values between $220$ and $290$\:kbit/s).
In both narrowband and wideband cases, we notice that the behavior obtained in the \gls{dt} is consistent with that of the real-world scenario.
%
%
%
%
Analogous trends can be seen for the \gls{sinr} of both signal types, where the narrowband jammer causes an \gls{sinr} decrease of approximately $20$\:dB (i.e., $\sim$77\% decrease), while the wideband jammer of approximately $25$\:dB (i.e., $\sim$92\% decrease) in the real-world scenario.
Similarly to the previous case, results are consistent with those of the \gls{dt}.

\begin{figure}[b]
    \vspace{-15pt}
    \centering
    \subfloat[Narrowband]{\label{fig:jamming-narrowband-mobile-results}\includegraphics[width=0.49\columnwidth]{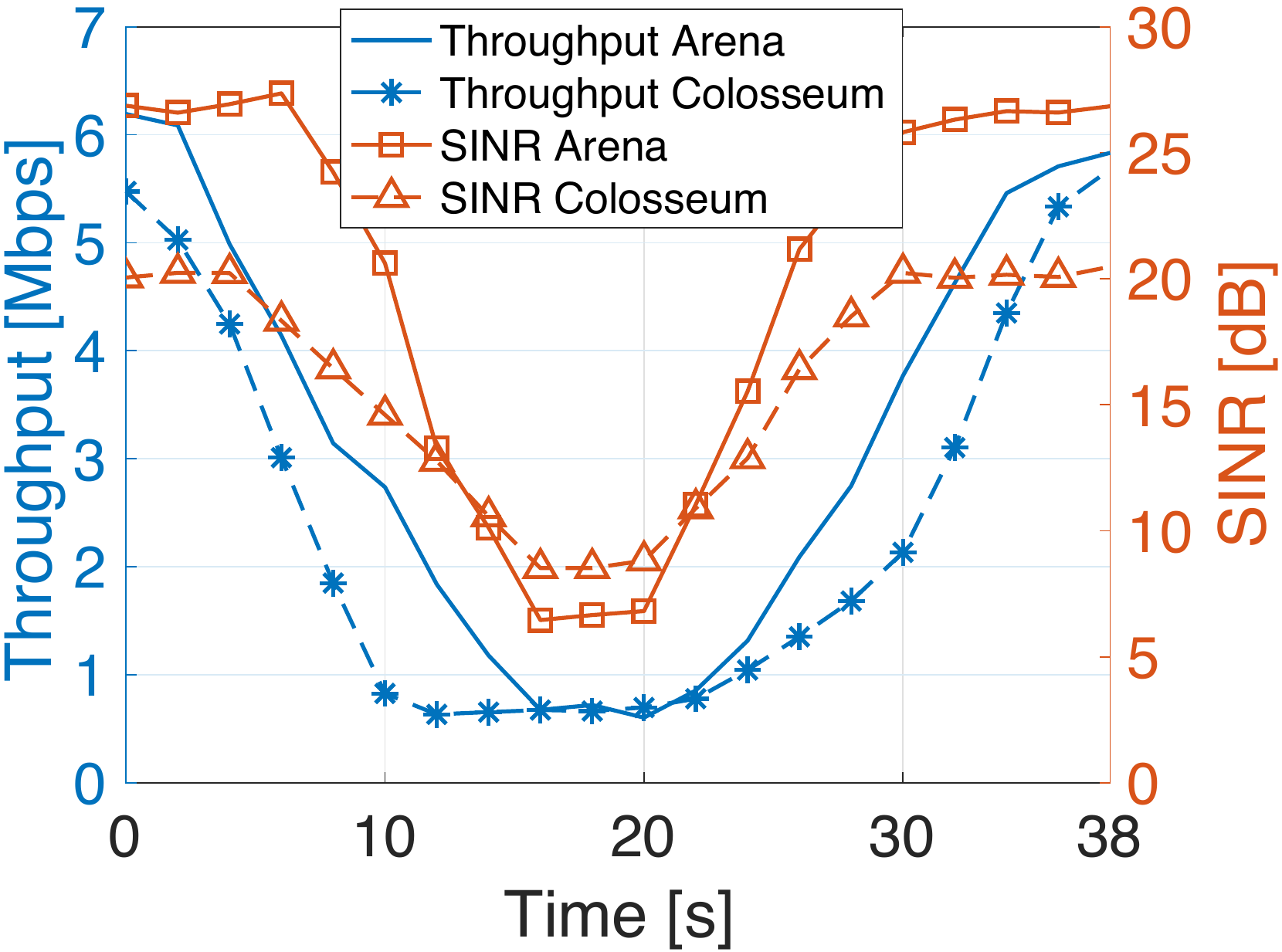}}
    \hfill
    \subfloat[Wideband]{\label{fig:jamming-wideband-mobile-results}\includegraphics[width=0.49\columnwidth]{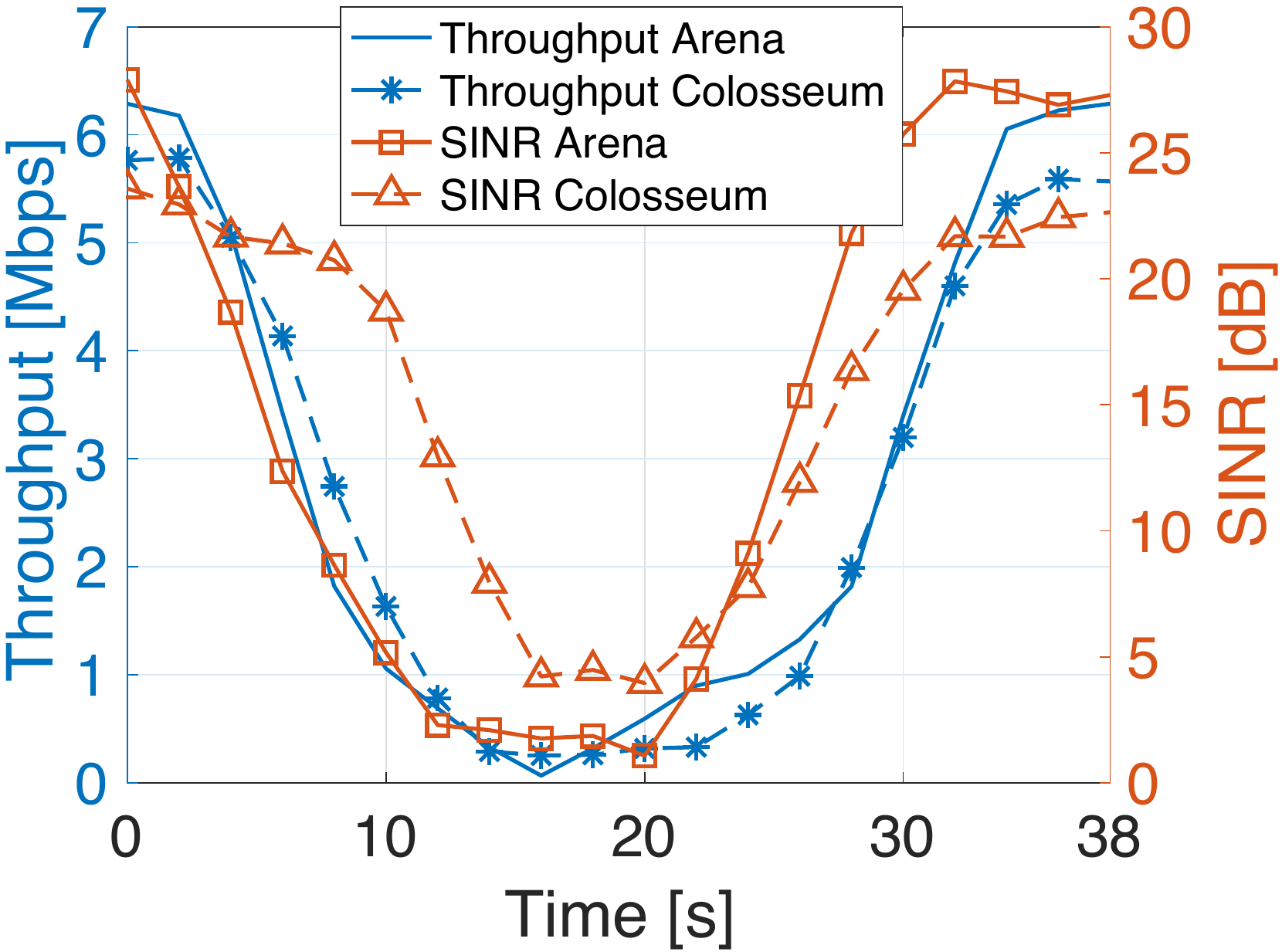}}
    \hfill
    \caption{\blue{Impact of a moving jammer for narrowband (a) and wideband (b) use cases on the throughput and \gls{sinr} of Wi-Fi nodes on Arena and Colosseum testbed.}}
    \label{fig:jamming-mobile-results}
\end{figure}

\blue{\textbf{Mobile Jamming.}} Now, we evaluate the impact that a \blue{narrowband and wideband} mobile jammer (node~41 in Figure~\ref{fig:jamming-map}) moving at pedestrian speed has on the Wi-Fi throughput.
Wi-Fi nodes are located as in the previous case, i.e., nodes~8 and~9 in the figure.
%
%
Results are shown in Figure~\ref{fig:jamming-mobile-results}.
As expected, the impact of the jamming signal on the Wi-Fi throughput varies as the jammer moves closer or farther from the Wi-Fi receiver, \blue{and it also depends on the type of jamming, i.e., narrowband vs. wideband}.
Specifically, as the jammer gets closer to the Wi-Fi nodes (i.e., seconds $5$ to $30$) \blue{in the narrowband case}, we observe a $\sim$90\% decrease in the Wi-Fi throughput in both real-world and \gls{dt} scenarios (see Figure~\ref{fig:jamming-narrowband-mobile-results}).
\blue{A comparable decrease can be observed in the \gls{sinr} as well, where we notice a clear drop in both the over-the-air Arena case and the \gls{dt} of about $\sim$65\% in a lookalike trend}.
\blue{Following a similar pattern, in the wideband case, we observe a more pronounced drop as the jammer approaches (i.e., seconds $5$ to $30$), reaching peaks near 100\% drops in throughput and \gls{sinr} in the closest locations with the nodes at around second $15$.}

\textbf{\blue{Comparison.}} As for the cellular experiment, Table~\ref{table:jamming-corr-results} shows the normalized cross-correlation results and their averages, considering the maximum values between 10 lags, for each jamming experiment.
We observe a strong similarity between the two testbeds in both static and mobile experiments, with individual values consistently above \blue{0.93}.
\begin{table}[ht]
    \centering
    \footnotesize
    \caption{Normalized cross-correlation results, and their averages, considering the maximum values between 10 lags for each jamming experiment.}
    \label{table:jamming-corr-results}
    
    \begin{tabularx}{\columnwidth}{
        >{\raggedright\arraybackslash\hsize=1.2\hsize}X
        >{\raggedright\arraybackslash\hsize=0.7\hsize}X
        >{\raggedright\arraybackslash\hsize=0.7\hsize}X
        >{\raggedright\arraybackslash\hsize=0.7\hsize}X}
        \toprule
        Metric & Narrowband & Wideband & Average \\
        \midrule
        Static Throughput & 0.996 &	0.982 &	0.989 \\
        Static \gls{sinr} & 0.986 &	0.984 &	0.985 \\
        Mobile Throughput & 0.982 & \blue{0.993} &	\blue{0.988} \\
        Mobile \gls{sinr} & 0.993 & \blue{0.935} & \blue{0.964} \\
        \bottomrule
    \end{tabularx}
\end{table}

Overall, considering all sample experiments in this work, our \gls{dtmn} is able to achieve an average similarity of \blue{0.987} in throughput and \blue{0.982} in \gls{sinr}.
These results prove the ability of our system to properly emulate various use case experiments with different protocol stacks and scenarios.

\section{Related Work}
\label{sec:relatedwork}

The concept of \gls{dt} is rapidly gaining momentum in both industry and academia.
Initial approaches showcase the use of \gls{dt}s for industry~4.0~\cite{rolle2020architecture}, and to assist design, assembly, and production operations in the manufacturing process~\cite{Tao2018dt}.
A comprehensive literature review on \gls{dt}-related applications in manufacturing is provided by Kritzinger et al.\ in~\cite{kritzinger2018dt}.

Recently, researchers and practitioners have started to apply the concept of \gls{dt} to the wireless ecosystem due to the potential of digitalization processes, and easier integration and monitoring of interconnected intelligent components, as Zeb et al.\ discuss in~\cite{zeb2022industrial}.
Nguyen et al.\ theoretically discuss how \gls{dt}s can enable swift testing and validation on real-time digital replicas of real-world 5G~cellular networks~\cite{nguyen2021dt}, while Khan et al.\ provide the architectural requirements for 5G-oriented \gls{dt}s, mentioning them as key components for the development of 6G~networks~\cite{latif2022dte}.
He et al.\ leverage the \gls{dt}s and mobile edge computing in cellular networks to enhance the creation of digital models affected by the straggler effect of user devices in a \gls{fl} process~\cite{he2022resource}.
%
Lu et al.\ incorporate \gls{dt}s into wireless networks to mitigate long and unreliable communications among users and \gls{bs} and define a permissioned blockchain-based \gls{fl} framework for edge computing~\cite{lu2021low}.
Zhao et al.\ combine 
\gls{dt}s with software-defined vehicular networks to learn, update, and verify physical environments to foresee future states of the system while improving the network performance~\cite{zhao2020idt}.

Overall, the above works agree on the potential of \gls{dt}s in: (i)~assessing the network performance; (ii)~creating realistic and accurate system models; (iii)~predicting the impact of changes in the deployment environment; and (iv)~reacting and optimizing the performance of the network.

The works most similar to our \gls{cast} toolchain in modeling and simulating channel characteristics are those of Patnaik et al.~\cite{patnaik2014implementation}, Ju and Rappaport~\cite{ju2018simulating}, Bilibashi et al.~\cite{bilibashi2020dynamic}, and Oliveira et al.~\cite{oliveria2019ray}.
Specifically, Patnaik et al.\ compare the response of \gls{fir} filters with their simulated counterpart~\cite{patnaik2014implementation}, while Ju and Rappaport devise a technique to improve the representation of channel impairments and variations for adaptive antenna algorithms in a mmWave channel simulator~\cite{ju2018simulating}.
Bilibashi et al., and Oliveira et al., instead, leverage ray-tracing approaches to include mobility in the emulated channels in~\cite{bilibashi2020dynamic} and~\cite{oliveria2019ray}, respectively.
However, these works only target specific use cases, and they cannot model generic scenarios and deployments, as instead our \gls{cast} toolchain does.

Finally, to the best of our knowledge, there are no practical works that encompass all the various building blocks of a \gls{dt} system, from channel characterization and modeling to large-scale experimentation on a \gls{dt}, to real-world validation on an over-the-air testbed, as instead, we carry out in this work.

\section{Conclusions}
\label{sec:conclusion}
In this paper, we applied the concept of \gls{dt} to the wireless communication field, and we have presented Colosseum, the World's largest wireless network emulator, as an ideal candidate for a \gls{dtmn}.
We demonstrated its capabilities by digitizing an over-the-air testbed, namely Arena, and by, first tuning, and then running various use case experiments on both testbeds.
The results showed that the \gls{dt} is able to accurately represent the real-world environment.
Thanks also to its public release, the Colosseum \gls{dt} enables the whole research community to properly run wireless experiments and to generate results as accurate as possible to the ones from real-world experimentation.



\balance
\footnotesize
\bibliographystyle{IEEEtran}
\bibliography{biblio}

\vspace{-20pt}

\begin{IEEEbiography}[{\includegraphics[width=1in,height=1.25in,clip,keepaspectratio]{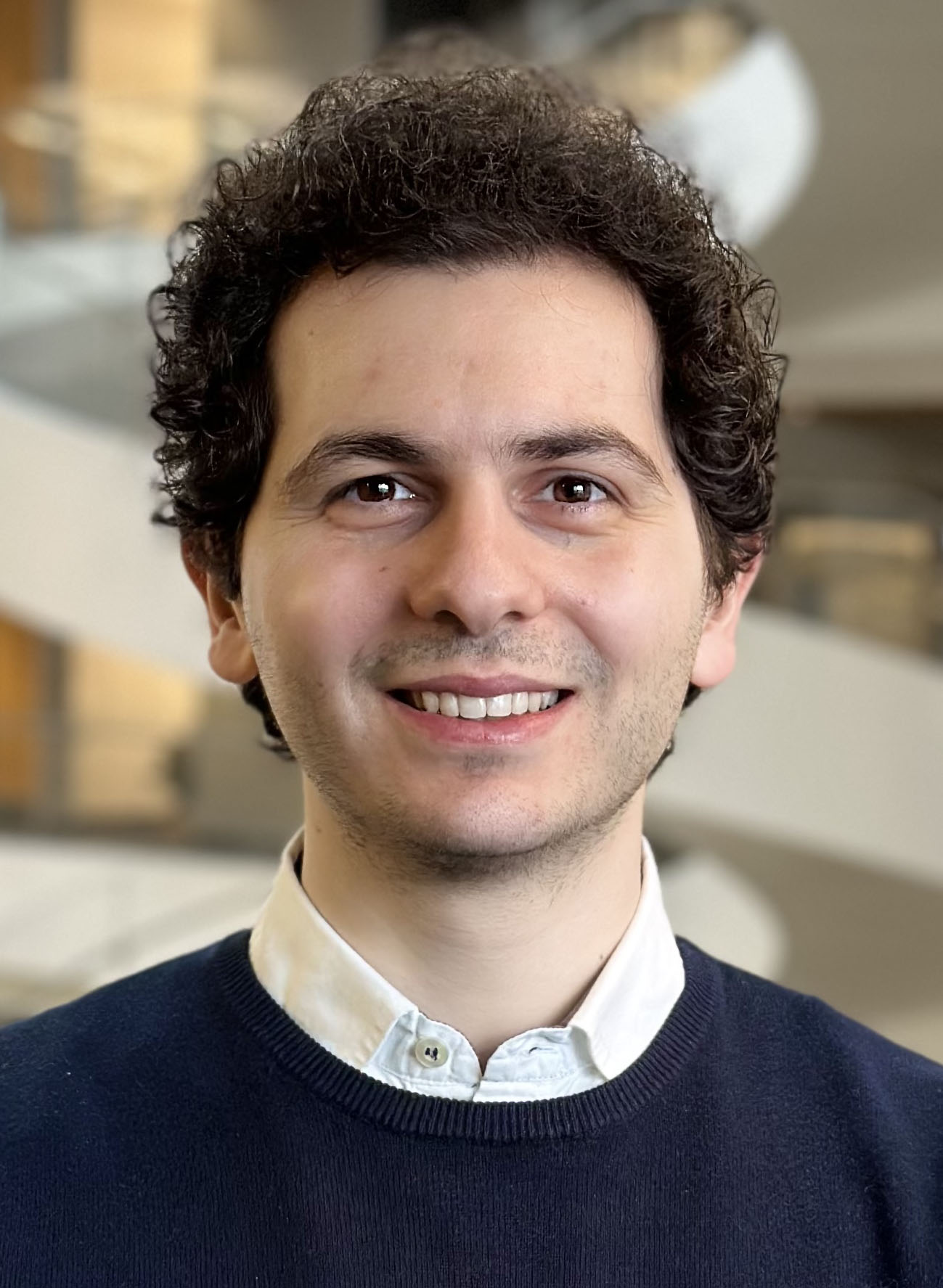}}]{Davide Villa} received his B.S. in Computer Engineering from University of Pisa, Italy, in 2015, and his M.S. in Embedded Computing Systems from Sant'Anna School of Advanced Studies and University of Pisa, Italy, in 2018. He worked as a Research Scientist in the Embedded Systems and Network Group at United Technologies Research Center in Cork, Ireland, from 2018 to 2020. He is currently pursuing a Ph.D. in Computer Engineering at Northeastern University in Boston, USA. His research interests are on 5G and beyond cellular networks, channel characterization for wireless systems, O-RAN, and software-defined networking for wireless networks.
\end{IEEEbiography}

\vspace{-10pt}

\begin{IEEEbiography}[{\includegraphics[width=1in,height=1.25in,clip,keepaspectratio]{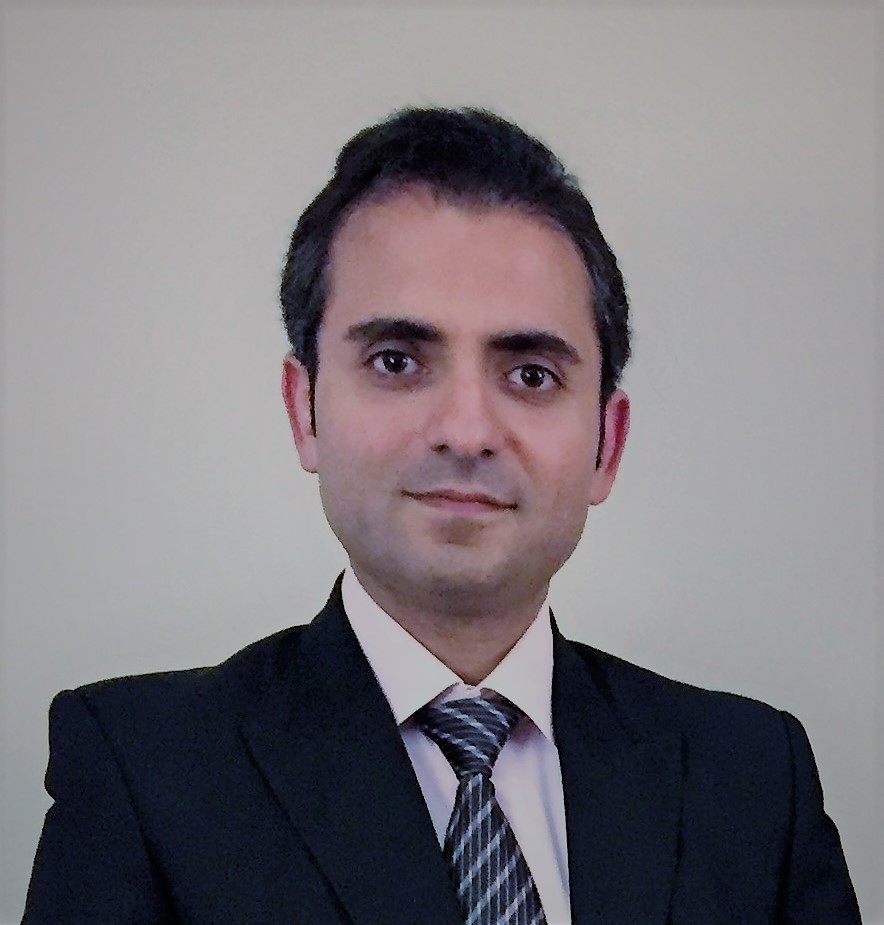}}]{Miead Tehrani-Moayyed}
is a Ph.D. candidate in Computer Engineering at Northeastern University in Boston. He is working under Prof. Stefano Basagni's supervision on RF channel models for static and mobile scenarios, from simulations to models for large-scale emulations. His research interests include applying AI/ML algorithms to wireless communication, propagation models for next-generation cellular systems, and computer networks. He received his M.S. in Computer Systems Architecture Engineering from Azad University, Iran in 2013 and his B.S. in Computer Engineering from Shomal University, Iran in 2007.
\end{IEEEbiography}

\vspace{-10pt}

\begin{IEEEbiography}[{\includegraphics[width=1in,height=1.25in,clip,keepaspectratio]{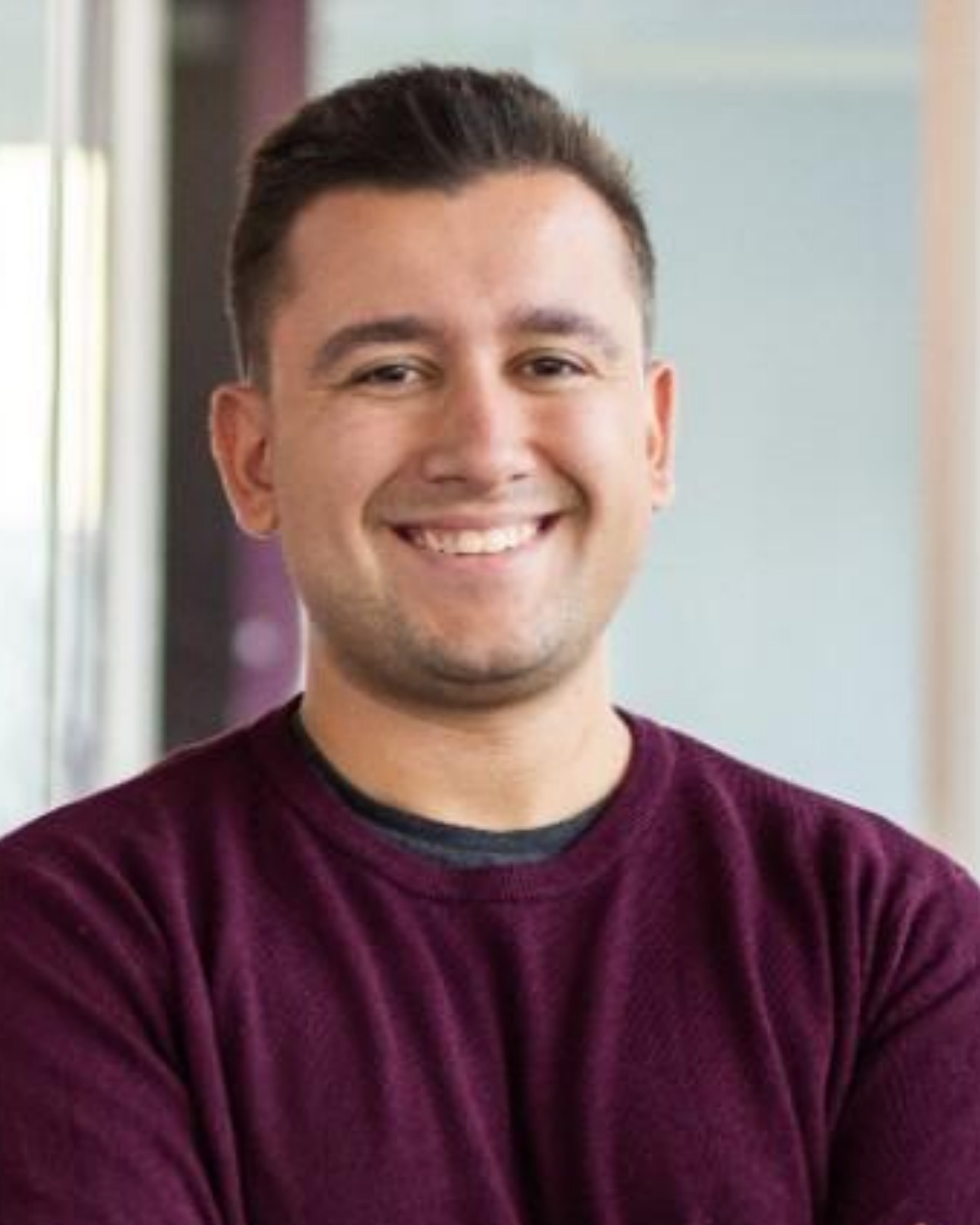}}]{Clifton Paul Robinson}
is a Cybersecurity Ph.D. candidate at Northeastern University advised by Professor Tommaso Melodia in the Wireless Networks and Embedded Systems (WiNES) Laboratory. He received his B.S. in Computer Science \& Mathematics at Bridgewater State University in 2018, and his M.S. in Cybersecurity from Northeastern University in 2020. His main area of research is in wireless network security, focusing on adversarial signals, and signal processing and detection with deep learning applications.
\end{IEEEbiography}

\vspace{-10pt}

\begin{IEEEbiography}
[{\includegraphics[width=1in,height=1.25in,keepaspectratio]{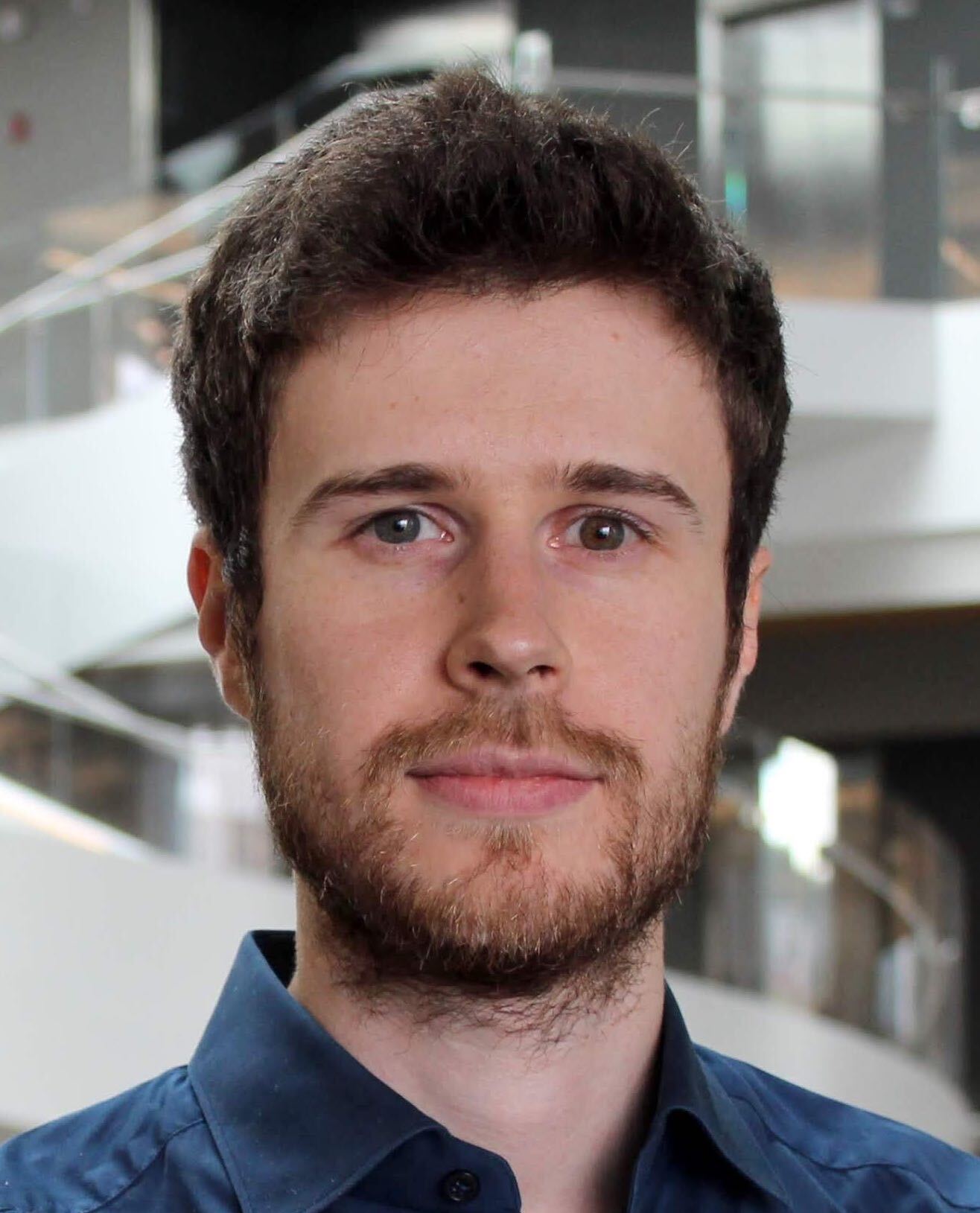}}]{Leonardo Bonati}
is an Associate Research Scientist at the Institute for the Wireless Internet of Things, Northeastern University, Boston, MA. He received a Ph.D. degree in Computer Engineering from Northeastern University in 2022. His main research focuses on softwarized approaches for the Open Radio Access Network (RAN) of the next generation of cellular networks, on O-RAN-managed networks, and on network automation and orchestration. He served multiple times on the technical program committee of the ACM Workshop WiNTECH and as guest editor of the special issue of Elsevier's Computer Networks journal on Advances in Experimental Wireless Platforms and Systems.
\end{IEEEbiography}

\vspace{-10pt}

\begin{IEEEbiography}[{\includegraphics[width=1in,height=1.25in,clip,keepaspectratio]{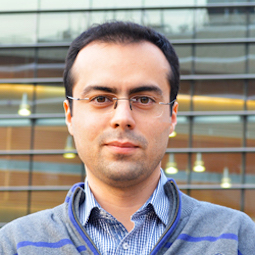}}]{Pedram Johari}
is a Principal Research Scientist at the Institute for Wireless Internet of Things at Northeastern University, Boston, MA.
Pedram received his Ph.D. in Electrical Engineering from the University at Buffalo in 2018 where he also was a Research Assistant Professor and Adjunct Lecturer in 2018 and 2019.
His research interests are in the fusion of AI and future generation of cellular networks (5G and beyond), in particular, focused on spectrum sharing, vehicular communications, full protocol wireless network emulators enabling wireless digital twins, and Internet of wireless medical things. Pedram has collaborated with several
research institutions, including University at Buffalo, Georgia Tech, Qualcomm, MathWorks, InterDigital, MITRE, and VIAVI.
\end{IEEEbiography}

\vspace{-10pt}

\begin{IEEEbiography}
[{\includegraphics[width=1in,height=1.25in,keepaspectratio]{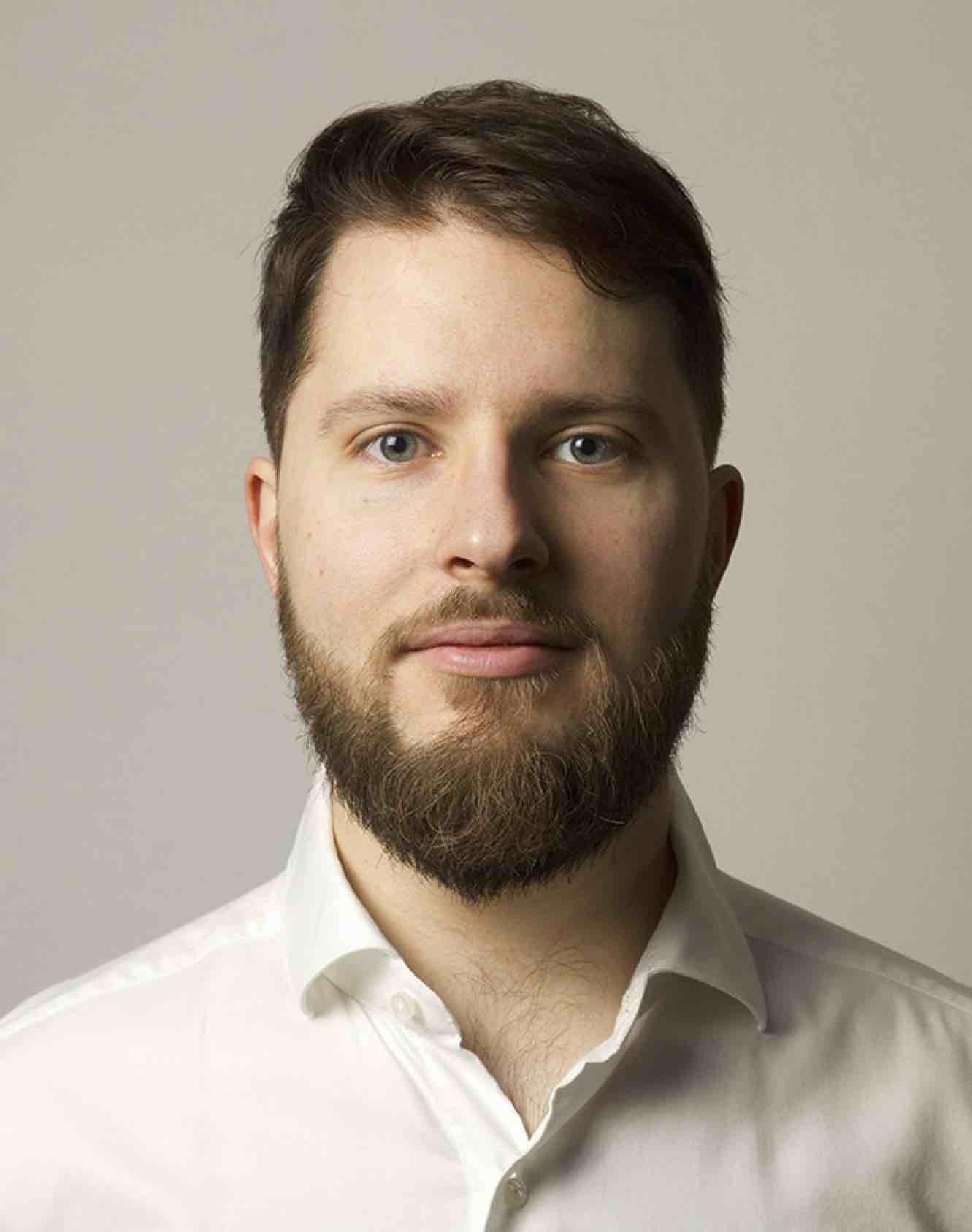}}]{Michele Polese} is a Principal Research Scientist at the Institute for the Wireless Internet of Things, Northeastern University, Boston, since March 2020. He received his Ph.D. at the Department of Information Engineering of the University of Padova in 2020. He also was an adjunct professor and postdoctoral researcher in 2019/2020 at the University of Padova, and a part-time lecturer in Fall 2020 and 2021 at Northeastern University. 
%
His research interests are in the analysis and development of protocols and architectures for future generations of cellular networks (5G and beyond).
He has contributed to O-RAN technical specifications and submitted responses to multiple FCC and NTIA notice of inquiry and requests for comments.
He received several best paper awards, is serving as TPC co-chair for WNS3 2021-2022, as an Associate Technical Editor for the IEEE Communications Magazine, and has organized the Open 5G Forum in Fall 2021.
\end{IEEEbiography}



\vspace{-10pt}

\begin{IEEEbiography}
[{\includegraphics[width=1in,height=1.25in,keepaspectratio]{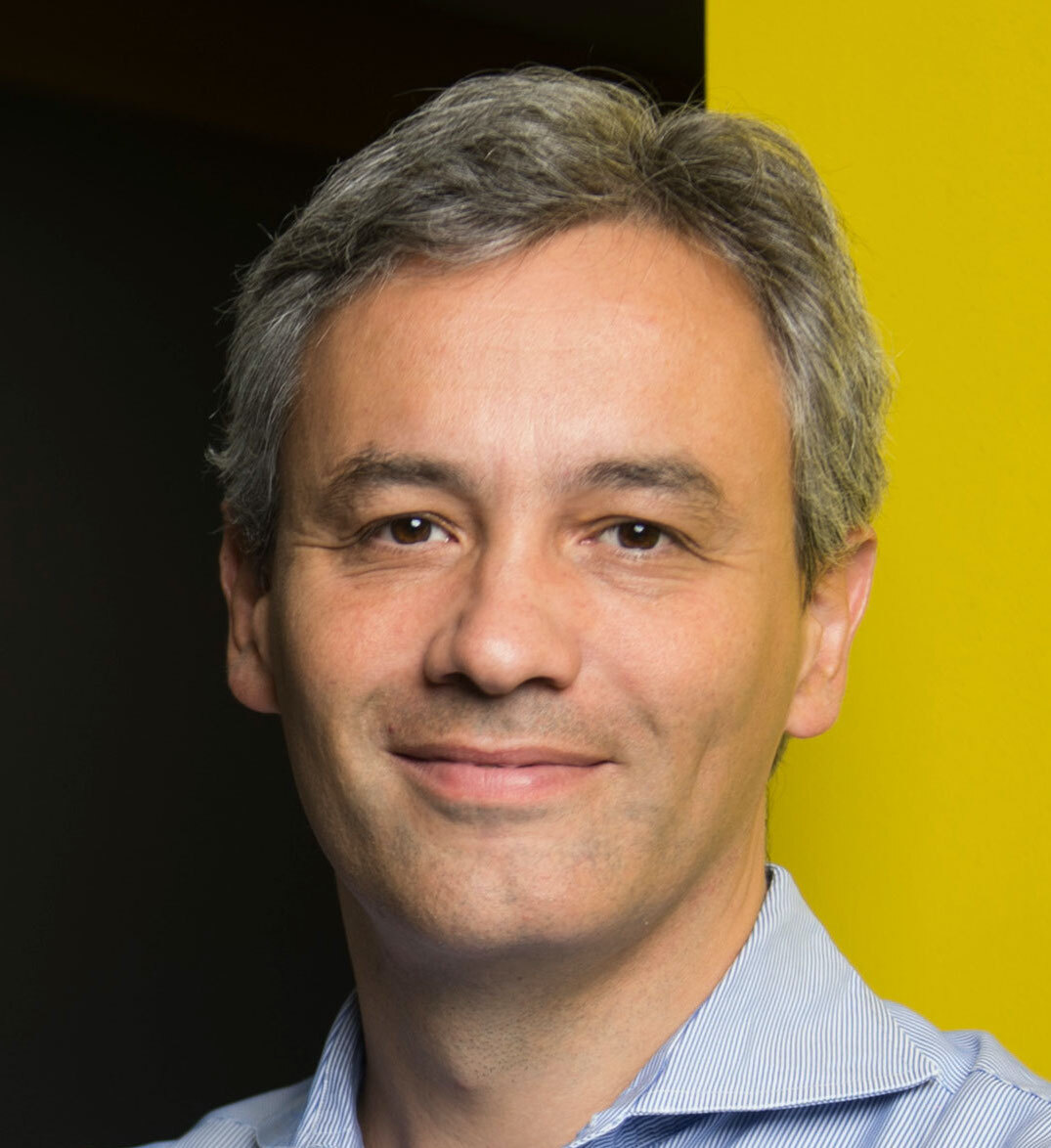}}]{Tommaso Melodia}
is the William Lincoln Smith Chair Professor with the Department of Electrical and Computer Engineering at Northeastern University in Boston. He is also the Founding Director of the Institute for the Wireless Internet of Things and the Director of Research for the PAWR Project Office. He received his Ph.D. in Electrical and Computer Engineering from the Georgia Institute of Technology in 2007. He is a recipient of the National Science Foundation CAREER award. Prof. Melodia has served as Associate Editor of IEEE Transactions on Wireless Communications, IEEE Transactions on Mobile Computing, Elsevier Computer Networks, among others. He has served as Technical Program Committee Chair for IEEE INFOCOM 2018, General Chair for IEEE SECON 2019, ACM Nanocom 2019, and ACM WUWnet 2014. Prof. Melodia is the Director of Research for the Platforms for Advanced Wireless Research (PAWR) Project Office, a \$100M public-private partnership to establish 4 city-scale platforms for wireless research to advance the US wireless ecosystem in years to come. Prof. Melodia's research on modeling, optimization, and experimental evaluation of Internet-of-Things and wireless networked systems has been funded by the National Science Foundation, the Air Force Research Laboratory the Office of Naval Research, DARPA, and the Army Research Laboratory. Prof. Melodia is a Fellow of the IEEE and a Senior Member of the ACM.
\end{IEEEbiography}

\end{document}

%% file: acronyms.tex
\newacronym{3gpp}{3GPP}{3rd Generation Partnership Project}
\newacronym{4g}{4G}{4th generation mobile network}
\newacronym{5g}{5G}{5th generation mobile network}
\newacronym{6g}{6G}{6th generation mobile network}
\newacronym{nextg}{NextG}{Next Generation}
\newacronym{5gc}{5GC}{5G Core}
\newacronym{adc}{ADC}{Analog to Digital Converter}
\newacronym{aerpaw}{AERPAW}{Aerial Experimentation and Research Platform for Advanced Wireless}
\newacronym{ai}{AI}{Artificial Intelligence}
\newacronym{aimd}{AIMD}{Additive Increase Multiplicative Decrease}
\newacronym{am}{AM}{Acknowledged Mode}
\newacronym{amc}{AMC}{Adaptive Modulation and Coding}
\newacronym{amf}{AMF}{Access and Mobility Management Function}
\newacronym{aops}{AOPS}{Adaptive Order Prediction Scheduling}
\newacronym{api}{API}{Application Programming Interface}
\newacronym{apn}{APN}{Access Point Name}
\newacronym{aqm}{AQM}{Active Queue Management}
\newacronym{ausf}{AUSF}{Authentication Server Function}
\newacronym{avc}{AVC}{Advanced Video Coding}
\newacronym{awgn}{AGWN}{Additive White Gaussian Noise}
\newacronym{balia}{BALIA}{Balanced Link Adaptation Algorithm}
\newacronym{bbu}{BBU}{Base Band Unit}
\newacronym{bdp}{BDP}{Bandwidth-Delay Product}
\newacronym{ber}{BER}{Bit Error Rate}
\newacronym{bf}{BF}{Beamforming}
\newacronym{bler}{BLER}{Block Error Rate}
\newacronym{brr}{BRR}{Bayesian Ridge Regressor}
\newacronym{bsr}{BSR}{Buffer Status Report}
\newacronym{bs}{BS}{Base Station}
\newacronym{bpsk}{BPSK}{Binary Phase-shift keying}
\newacronym{bss}{BSS}{Business Support System}
\newacronym{ca}{CA}{Carrier Aggregation}
\newacronym{caas}{CaaS}{Connectivity-as-a-Service}
\newacronym{cb}{CB}{Code Block}
\newacronym{cc}{CC}{Congestion Control}
\newacronym{ccid}{CCID}{Congestion Control ID}
\newacronym{cco}{CC}{Carrier Component}
\newacronym{cd}{CD}{Continuous Delivery}
\newacronym{cdd}{CDD}{Cyclic Delay Diversity}
\newacronym{cdf}{CDF}{Cumulative Distribution Function}
\newacronym{cdma}{CDMA}{Code-Division Multiple Access}
\newacronym{cdn}{CDN}{Content Distribution Network}
\newacronym{ci}{CI}{Continuous Integration}
\newacronym{cicd}{CI/CD}{Continuous Integration/Continuous Delivery}
\newacronym{cir}{CIR}{Channel Impulse Response}
\newacronym{cn}{CN}{Core Network}
\newacronym{codel}{CoDel}{Controlled Delay Management}
\newacronym{comac}{COMAC}{Converged Multi-Access and Core}
\newacronym{cord}{CORD}{Central Office Re-architected as a Datacenter}
\newacronym{cornet}{CORNET}{COgnitive Radio NETwork}
\newacronym{cosmos}{COSMOS}{Cloud Enhanced Open Software Defined Mobile Wireless Testbed for City-Scale Deployment}
\newacronym{cots}{COTS}{Commercial Off-the-Shelf}
\newacronym{cp}{CP}{Control Plane}
\newacronym{cpu}{CPU}{Central Processing Unit}
\newacronym{cqi}{CQI}{Channel Quality Information}
\newacronym{cr}{CR}{Cognitive Radio}
\newacronym{cran}{CRAN}{Cloud \gls{ran}}
\newacronym{crs}{CRS}{Cell Reference Signal}
\newacronym{csi}{CSI}{Channel State Information}
\newacronym{csirs}{CSI-RS}{Channel State Information - Reference Signal}
\newacronym{cu}{CU}{Central Unit}
\newacronym{d2tcp}{D$^2$TCP}{Deadline-aware Data center TCP}
\newacronym{d3}{D$^3$}{Deadline-Driven Delivery}
\newacronym{dac}{DAC}{Digital to Analog Converter}
\newacronym{dag}{DAG}{Directed Acyclic Graph}
\newacronym{darpa}{DARPA}{Defense Advanced Research Projects Agency}
\newacronym{das}{DAS}{Distributed Antenna System}
\newacronym{dash}{DASH}{Dynamic Adaptive Streaming over HTTP}
\newacronym{dc}{DC}{Dual Connectivity}
\newacronym{dccp}{DCCP}{Datagram Congestion Control Protocol}
\newacronym{dce}{DCE}{Direct Code Execution}
\newacronym{dci}{DCI}{Downlink Control Information}
\newacronym{dcl}{DCL}{Dear Colleague Letter}
\newacronym{dctcp}{DCTCP}{Data Center TCP}
\newacronym{devops}{DevOps}{Development and Operations}
\newacronym{dl}{DL}{Downlink}
\newacronym{dmr}{DMR}{Deadline Miss Ratio}
\newacronym{dmrs}{DMRS}{DeModulation Reference Signal}
\newacronym{drlcc}{DRL-CC}{Deep Reinforcement Learning Congestion Control}
\newacronym{drs}{DRS}{Discovery Reference Signal}
\newacronym{dt}{DT}{Digital Twin}
\newacronym{dtn}{DTN}{Digital Twin Network}
\newacronym{dtmn}{DTMN}{Digital Twins for Mobile Networks}
\newacronym{dtwn}{DTWN}{Digital Twin Wireless Network}
\newacronym{du}{DU}{Distributed Unit}
\newacronym{e2e}{E2E}{end-to-end}
\newacronym{ecaas}{ECaaS}{Edge-Cloud-as-a-Service}
\newacronym{ecn}{ECN}{Explicit Congestion Notification}
\newacronym{edf}{EDF}{Earliest Deadline First}
\newacronym{em}{EM}{Electro-Magnetic}
\newacronym{embb}{eMBB}{Enhanced Mobile Broadband}
\newacronym{empower}{EMPOWER}{EMpowering transatlantic PlatfOrms for advanced WirEless Research}
\newacronym{enb}{eNB}{evolved Node Base}
\newacronym{endc}{EN-DC}{E-UTRAN-\gls{nr} \gls{dc}}
\newacronym{epc}{EPC}{Evolved Packet Core}
\newacronym{eps}{EPS}{Evolved Packet System}
\newacronym{es}{ES}{Edge Server}
\newacronym{etsi}{ETSI}{European Telecommunications Standards Institute}
\newacronym[firstplural=Estimated Times of Arrival (ETAs)]{eta}{ETA}{Estimated Time of Arrival}
\newacronym{eutran}{E-UTRAN}{Evolved Universal Terrestrial Access Network}
\newacronym{faas}{FaaS}{Function-as-a-Service}
\newacronym{fapi}{FAPI}{Functional Application Platform Interface}
\newacronym{fcc}{FCC}{Federal Communications Commission}
\newacronym{fdd}{FDD}{Frequency Division Duplexing}
\newacronym{fdm}{FDM}{Frequency Division Multiplexing}
\newacronym{fdma}{FDMA}{Frequency Division Multiple Access}
\newacronym{fed4fire}{FED4FIRE+}{Federation 4 Future Internet Research and Experimentation Plus}
\newacronym{fir}{FIR}{Finite Impulse Response}
\newacronym{fit}{FIT}{Future \acrlong{iot}}
\newacronym{fl}{FL}{Federated Learning}
\newacronym{fpga}{FPGA}{Field Programmable Gate Array}
\newacronym{fr2}{FR2}{Frequency Range 2}
\newacronym{fs}{FS}{Fast Switching}
\newacronym{fscc}{FSCC}{Flow Sharing Congestion Control}
\newacronym{ftp}{FTP}{File Transfer Protocol}
\newacronym{fw}{FW}{Flow Window}
\newacronym{ga128}{Ga}{Golay Sequence type A}
\newacronym{ge}{GE}{Gaussian Elimination}
\newacronym{glfsr}{GLFSR}{Galois Linear Feedback Shift Register}
\newacronym{gnb}{gNB}{Next Generation Node Base}
\newacronym{gold}{Gold}{Gold}
\newacronym{gop}{GOP}{Group of Pictures}
\newacronym{gpr}{GPR}{Gaussian Process Regressor}
\newacronym{gpu}{GPU}{Graphics Processing Unit}
\newacronym{gtp}{GTP}{GPRS Tunneling Protocol}
\newacronym{gtpc}{GTP-C}{GPRS Tunnelling Protocol Control Plane}
\newacronym{gtpu}{GTP-U}{GPRS Tunnelling Protocol User Plane}
\newacronym{gtpv2c}{GTPv2-C}{\gls{gtp} v2 - Control}
\newacronym{gw}{GW}{Gateway}
\newacronym{harq}{HARQ}{Hybrid Automatic Repeat reQuest}
\newacronym{hetnet}{HetNet}{Heterogeneous Network}
\newacronym{hh}{HH}{Hard Handover}
\newacronym{hol}{HOL}{Head-of-Line}
\newacronym{hqf}{HQF}{Highest-quality-first}
\newacronym{hss}{HSS}{Home Subscription Server}
\newacronym{http}{HTTP}{HyperText Transfer Protocol}
\newacronym{ia}{IA}{Initial Access}
\newacronym{iab}{IAB}{Integrated Access and Backhaul}
\newacronym{ic}{IC}{Incident Command}
\newacronym{ietf}{IETF}{Internet Engineering Task Force}
\newacronym{ifw}{IFW}{Interference Free Window}
\newacronym{imsi}{IMSI}{International Mobile Subscriber Identity}
\newacronym{imt}{IMT}{International Mobile Telecommunication}
\newacronym{iot}{IoT}{Internet of Things}
\newacronym{ip}{IP}{Internet Protocol}
\newacronym{iq}{IQ}{In-phase and Quadrature}
\newacronym{isi}{ISI}{Inter-Symbol Interference}
\newacronym{itu}{ITU}{International Telecommunication Union}
\newacronym{kpi}{KPI}{Key Performance Indicator}
\newacronym{kvm}{KVM}{Kernel-based Virtual Machine}
\newacronym{lfsr}{LFSR}{Linear Feedback Shift Register}
\newacronym{los}{LOS}{Line-of-Sight}
\newacronym{ls}{LS}{Loosely Synchronised}
\newacronym{lsm}{LSM}{Link-to-System Mapping}
\newacronym{lstm}{LSTM}{Long Short Term Memory}
\newacronym{lte}{LTE}{Long Term Evolution}
\newacronym{lxc}{LXC}{Linux Container}
\newacronym{m2m}{M2M}{Machine to Machine}
\newacronym{mac}{MAC}{Medium Access Control}
\newacronym{mai}{MAI}{Multiple Access Interference}
\newacronym{manet}{MANET}{Mobile Ad Hoc Network}
\newacronym{mano}{MANO}{Management and Orchestration}
\newacronym{mc}{MC}{Multi-Connectivity}
\newacronym{mcc}{MCC}{Mobile Cloud Computing}
\newacronym{mchem}{MCHEM}{Massive Channel Emulator}
\newacronym{mcs}{MCS}{Modulation and Coding Scheme}
\newacronym{mec}{MEC}{Multi-access Edge Computing}
\newacronym{mec2}{MEC}{Mobile Edge Cloud}
\newacronym{mec3}{MEC}{Mobile Edge Computing}
\newacronym{mfc}{MFC}{Mobile Fog Computing}
\newacronym{mi}{MI}{Mutual Information}
\newacronym{mib}{MIB}{Master Information Block}
\newacronym{miesm}{MIESM}{Mutual Information Based Effective SINR}
\newacronym{mimo}{MIMO}{Multiple Input, Multiple Output}
\newacronym{mgen}{MGEN}{Multi-Generator}
\newacronym{ml}{ML}{Machine Learning}
\newacronym{mlr}{MLR}{Maximum-local-rate}
\newacronym[plural=\gls{mme}s,firstplural=Mobility Management Entities (MMEs)]{mme}{MME}{Mobility Management Entity}
\newacronym{mmtc}{mMTC}{Massive Machine-Type Communications}
\newacronym{mmwave}{mmWave}{millimeter wave}
\newacronym{mpdccp}{MP-DCCP}{Multipath Datagram Congestion Control Protocol}
\newacronym{mptcp}{MPTCP}{Multipath TCP}
\newacronym{mr}{MR}{Maximum Rate}
\newacronym{mrdc}{MR-DC}{Multi \gls{rat} \gls{dc}}
\newacronym{mse}{MSE}{Mean Square Error}
\newacronym{mss}{MSS}{Maximum Segment Size}
\newacronym{mt}{MT}{Mobile Termination}
\newacronym{mtd}{MTD}{Machine-Type Device}
\newacronym{mtu}{MTU}{Maximum Transmission Unit}
\newacronym{mumimo}{MU-MIMO}{Multi-user \gls{mimo}}
\newacronym{mvno}{MVNO}{Mobile Virtual Network Operator}
\newacronym{nalu}{NALU}{Network Abstraction Layer Unit}
\newacronym{nas}{NAS}{Network Attached Storage}
\newacronym{nbiot}{NB-IoT}{Narrow Band IoT}
\newacronym{nfv}{NFV}{Network Function Virtualization}
\newacronym{nfvi}{NFVI}{Network Function Virtualization Infrastructure}
\newacronym{nic}{NIC}{Network Interface Card}
\newacronym{nlos}{NLOS}{Non-Line-of-Sight}
\newacronym{now}{NOW}{Non Overlapping Window}
\newacronym{nrdz}{NRDZ}{National Radio Dynamic Zone}
\newacronym{nsf}{NSF}{National Science Foundation}
\newacronym{nsm}{NSM}{Network Service Mesh}
\newacronym[type=hidden]{nr}{NR}{New Radio}
\newacronym{nrf}{NRF}{Network Repository Function}
\newacronym{nsa}{NSA}{Non Stand Alone}
\newacronym{nse}{NSE}{Network Slicing Engine}
\newacronym{nssf}{NSSF}{Network Slice Selection Function}
\newacronym{ntp}{NTP}{Network Time Protocol}
\newacronym{o2i}{O2I}{Outdoor to Indoor}
\newacronym{oai}{OAI}{OpenAirInterface}
\newacronym{oaicn}{OAI-CN}{\gls{oai} \acrlong{cn}}
\newacronym{oairan}{OAI-RAN}{\acrlong{oai} \acrlong{ran}}
\newacronym{oam}{OAM}{Operations, Administration and Maintenance}
\newacronym[plural=\gls{obu}s,firstplural=Onboard Units (OBUs)]{obu}{OBU}{Onboard Unit}
\newacronym{ofdm}{OFDM}{Orthogonal Frequency Division Multiplexing}
\newacronym{olia}{OLIA}{Opportunistic Linked Increase Algorithm}
\newacronym{omec}{OMEC}{Open Mobile Evolved Core}
\newacronym{onap}{ONAP}{Open Network Automation Platform}
\newacronym{onf}{ONF}{Open Networking Foundation}
\newacronym{onos}{ONOS}{Open Networking Operating System}
\newacronym{oom}{OOM}{\gls{onap} Operations Manager}
\newacronym{opnfv}{OPNFV}{Open Platform for \gls{nfv}}
\newacronym[type=hidden]{oran}{O-RAN}{Open \gls{ran}}
\newacronym{orbit}{ORBIT}{Open-Access Research Testbed for Next-Generation Wireless Networks}
\newacronym{os}{OS}{Operating System}
\newacronym{osm}{OSM}{Open Street Map}
\newacronym{oss}{OSS}{Operations Support System}
\newacronym{pa}{PA}{Position-aware}
\newacronym{pase}{PASE}{Prioritization, Arbitration, and Self-adjusting Endpoints}
\newacronym{pawr}{PAWR}{Platforms for Advanced Wireless Research}
\newacronym{pbch}{PBCH}{Physical Broadcast Channel}
\newacronym{pcef}{PCEF}{Policy and Charging Enforcement Function}
\newacronym{pcfich}{PCFICH}{Physical Control Format Indicator Channel}
\newacronym{pcrf}{PCRF}{Policy and Charging Rules Function}
\newacronym{pdcch}{PDCCH}{Physical Downlink Control Channel}
\newacronym{pdcp}{PDCP}{Packet Data Convergence Protocol}
\newacronym{pdsch}{PDSCH}{Physical Downlink Shared Channel}
\newacronym{pdu}{PDU}{Packet Data Unit}
\newacronym{pdp}{PDP}{Power Delay Profile}
\newacronym{pf}{PF}{Proportional Fair}
\newacronym{pgw}{PGW}{Packet Gateway}
\newacronym{phich}{PHICH}{Physical Hybrid ARQ Indicator Channel}
\newacronym{phy}{PHY}{Physical}
\newacronym{pl}{PL}{Path Loss}
\newacronym{pmch}{PMCH}{Physical Multicast Channel}
\newacronym{pmi}{PMI}{Precoding Matrix Indicators}
\newacronym{powder}{POWDER}{Platform for Open Wireless Data-driven Experimental Research}
\newacronym{ppo}{PPO}{Proximal Policy Optimization}
\newacronym{ppp}{PPP}{Poisson Point Process}
\newacronym{prach}{PRACH}{Physical Random Access Channel}
\newacronym{prb}{PRB}{Physical Resource Block}
\newacronym{psnr}{PSNR}{Peak Signal to Noise Ratio}
\newacronym{pss}{PSS}{Primary Synchronization Signal}
\newacronym{pucch}{PUCCH}{Physical Uplink Control Channel}
\newacronym{pusch}{PUSCH}{Physical Uplink Shared Channel}
\newacronym{qam}{QAM}{Quadrature Amplitude Modulation}
\newacronym{qci}{QCI}{\gls{qos} Class Identifier}
\newacronym{qoe}{QoE}{Quality of Experience}
\newacronym{qos}{QoS}{Quality of Service}
\newacronym{qtgui}{QT-GUI}{QT Graphical User Interface}
\newacronym{quic}{QUIC}{Quick UDP Internet Connections}
\newacronym{rach}{RACH}{Random Access Channel}
\newacronym{ran}{RAN}{Radio Access Network}
\newacronym[firstplural=Radio Access Technologies (RATs)]{rat}{RAT}{Radio Access Technology}
\newacronym{rcn}{RCN}{Research Coordination Network}
\newacronym{rec}{REC}{Radio Edge Cloud}
\newacronym{red}{RED}{Random Early Detection}
\newacronym{renew}{RENEW}{Reconfigurable Eco-system for Next-generation End-to-end Wireless}
\newacronym{rf}{RF}{Radio Frequency}
\newacronym{rfc}{RFC}{Request for Comments}
\newacronym{rfr}{RFR}{Random Forest Regressor}
\newacronym{ric}{RIC}{\gls{ran} Intelligent Controller}
\newacronym{rlc}{RLC}{Radio Link Control}
\newacronym{rlf}{RLF}{Radio Link Failure}
\newacronym{rlnc}{RLNC}{Random Linear Network Coding}
\newacronym{rmse}{RMSE}{Root Mean Squared Error}
\newacronym{rnis}{RNIS}{Radio Network Information Service}
\newacronym{rr}{RR}{Round Robin}
\newacronym{rrc}{RRC}{Radio Resource Control}
\newacronym{rrm}{RRM}{Radio Resource Management}
\newacronym{rru}{RRU}{Remote Radio Unit}
\newacronym{rs}{RS}{Remote Server}
\newacronym{rsrp}{RSRP}{Reference Signal Received Power}
\newacronym{rsrq}{RSRQ}{Reference Signal Received Quality}
\newacronym{rss}{RSS}{Received Signal Strength}
\newacronym{rssi}{RSSI}{Received Signal Strength Indicator}
\newacronym{rsu}{RSU}{Road-Side Unit}
\newacronym{rtt}{RTT}{Round Trip Time}
\newacronym{ru}{RU}{Radio Unit}
\newacronym{rw}{RW}{Receive Window}
\newacronym{rx}{RX}{Receiver}
\newacronym{s1ap}{S1AP}{S1 Application Protocol}
\newacronym{sa}{SA}{standalone}
\newacronym{sack}{SACK}{Selective Acknowledgment}
\newacronym{sap}{SAP}{Service Access Point}
\newacronym{sc2}{SC2}{Spectrum Collaboration Challenge}
\newacronym{scef}{SCEF}{Service Capability Exposure Function}
\newacronym{sch}{SCH}{Secondary Cell Handover}
\newacronym{scoot}{SCOOT}{Split Cycle Offset Optimization Technique}
\newacronym{sctp}{SCTP}{Stream Control Transmission Protocol}
\newacronym{sdap}{SDAP}{Service Data Adaptation Protocol}
\newacronym{sd}{SD}{Standard Deviation}
\newacronym{sdk}{SDK}{Software Development Kit}
\newacronym{sdm}{SDM}{Space Division Multiplexing}
\newacronym{sdma}{SDMA}{Spatial Division Multiple Access}
\newacronym{sdn}{SDN}{Software-defined Networking}
\newacronym{sdr}{SDR}{Software-defined Radio}
\newacronym{seba}{SEBA}{SDN-Enabled Broadband Access}
\newacronym{sgsn}{SGSN}{Serving GPRS Support Node}
\newacronym{sgw}{SGW}{Service Gateway}
\newacronym{si}{SI}{Study Item}
\newacronym{sib}{SIB}{Secondary Information Block}
\newacronym{sinr}{SINR}{Signal to Interference plus Noise Ratio}
\newacronym{sip}{SIP}{Session Initiation Protocol}
\newacronym{siso}{SISO}{Single Input, Single Output}
\newacronym{sla}{SLA}{Service Level Agreement}
\newacronym{sm}{SM}{Saturation Mode}
\newacronym{smf}{SMF}{Session Management Function}
\newacronym{smo}{SMO}{Service Management and Orchestration}
\newacronym{sms}{SMS}{Short Message Service}
\newacronym{smsgmsc}{SMS-GMSC}{\gls{sms}-Gateway}
\newacronym{snr}{SNR}{Signal-to-Noise-Ratio}
\newacronym{son}{SON}{Self-Organizing Network}
\newacronym{sptcp}{SPTCP}{Single Path TCP}
\newacronym{srb}{SRB}{Service Radio Bearer}
\newacronym{srn}{SRN}{Standard Radio Node}
\newacronym{srs}{SRS}{Sounding Reference Signal}
\newacronym{ss}{SS}{Synchronization Signal}
\newacronym{sss}{SSS}{Secondary Synchronization Signal}
\newacronym{st}{ST}{Spanning Tree}
\newacronym{svc}{SVC}{Scalable Video Coding}
\newacronym{tb}{TB}{Transport Block}
\newacronym{tcp}{TCP}{Transmission Control Protocol}
\newacronym{tdd}{TDD}{Time Division Duplexing}
\newacronym{tdm}{TDM}{Time Division Multiplexing}
\newacronym{tdma}{TDMA}{Time Division Multiple Access}
\newacronym{tfl}{TfL}{Transport for London}
\newacronym{tfrc}{TFRC}{TCP-Friendly Rate Control}
\newacronym{tft}{TFT}{Traffic Flow Template}
\newacronym{tgen}{TGEN}{Traffic Generator}
\newacronym{tip}{TIP}{Telecom Infra Project}
\newacronym{tm}{TM}{Transparent Mode}
\newacronym{to}{TO}{Telco Operator}
\newacronym{toa}{ToA}{Time of Arrival}
\newacronym{tr}{TR}{Technical Report}
\newacronym{trp}{TRP}{Transmitter Receiver Pair}
\newacronym{ts}{TS}{Technical Specification}
\newacronym{tti}{TTI}{Transmission Time Interval}
\newacronym{ttt}{TTT}{Time-to-Trigger}
\newacronym{tx}{TX}{Transmitter}
\newacronym{uas}{UAS}{Unmanned Aerial System}
\newacronym{uav}{UAV}{Unmanned Aerial Vehicle}
\newacronym{udm}{UDM}{Unified Data Management}
\newacronym{udp}{UDP}{User Datagram Protocol}
\newacronym{udr}{UDR}{Unified Data Repository}
\newacronym{ue}{UE}{User Equipment}
\newacronym{uhd}{UHD}{\gls{usrp} Hardware Driver}
\newacronym{ul}{UL}{Uplink}
\newacronym{um}{UM}{Unacknowledged Mode}
\newacronym{uml}{UML}{Unified Modeling Language}
\newacronym{upa}{UPA}{Uniform Planar Array}
\newacronym{upf}{UPF}{User Plane Function}
\newacronym{urllc}{URLLC}{Ultra Reliable and Low Latency Communication}
\newacronym{usa}{U.S.}{United States}
\newacronym{usim}{USIM}{Universal Subscriber Identity Module}
\newacronym{usrp}{USRP}{Universal Software Radio Peripheral}
\newacronym{utc}{UTC}{Urban Traffic Control}
\newacronym{vim}{VIM}{Virtualization Infrastructure Manager}
\newacronym{vm}{VM}{Virtual Machine}
\newacronym{vnf}{VNF}{Virtual Network Function}
\newacronym{volte}{VoLTE}{Voice over \gls{lte}}
\newacronym{voltha}{VOLTHA}{Virtual OLT HArdware Abstraction}
\newacronym{vr}{VR}{Virtual Reality}
\newacronym{vran}{vRAN}{Virtualized \gls{ran}}
\newacronym{vss}{VSS}{Video Streaming Server}
\newacronym{wbf}{WBF}{Wired Bias Function}
\newacronym{wf}{WF}{Wired-first}
\newacronym{wi}{WI}{Wireless InSite}
\newacronym{wlan}{WLAN}{Wireless Local Area Network}
\newacronym{pnf}{PNF}{Physical Network Function}
\newacronym{drl}{DRL}{Deep Reinforcement Learning}
\newacronym{mtc}{MTC}{Machine-type Communications}
\newacronym{v2x}{V2X}{Vehicle-to-everything}
\newacronym{cast}{CaST}{Channel emulation scenario generator and Sounder Toolchain}
\newacronym{gui}{GUI}{Graphical User Interface}
\newacronym{ups}{UPS}{Uninterruptible Power Supply}
\newacronym{ota}{OTA}{Over-the-Air}
\newacronym{hitl}{HITL}{hardware-in-the-loop}
\newacronym{soc}{SoC}{System-on-Chip}